\documentclass[
 reprint,=
 amsmath,amssymb,
 aps,unsortedaddress]{revtex4-1}
\usepackage{xcolor}
\usepackage{graphicx}
\usepackage{dcolumn}
\usepackage{bm}
\usepackage{amsmath}
\begin{document}

\title{Higher order symmetry-protected topological states for interacting bosons and fermions}

\author{Yizhi You}
\affiliation{Princeton Center for Theoretical Science, Princeton University, 
NJ, 08544, USA}

\author{Trithep Devakul}
\affiliation{Department of Physics, Princeton University, 
NJ, 08544, USA}

\author{F.~J. Burnell}
\affiliation{Department of Physics, University of Minnesota Twin Cities, 
MN, 55455, USA}

\author{Titus Neupert}
\affiliation{Department of Physics, University of Zurich, Winterthurerstrasse 190, 8057 Zurich, Switzerland}

\date{\today}
\begin{abstract}
Higher-order topological insulators have a modified bulk-boundary correspondence compared to other topological phases: instead of gapless edge or surface states, they have gapped edges and surfaces, but protected modes at corners or hinges.
Here, we explore symmetry protected topological phases in strongly interacting many-body systems with this generalized bulk-boundary correspondence. We introduce several exactly solvable bosonic lattice models  as candidates for interacting higher order symmetry protected topological (HOSPT) phases protected by spatial symmetries, and develop a topological field theory that captures the non-trivial nature of the gapless corner and hinge modes.  We show how, for rotational symmetry, this field theory leads to a natural relationship between HOSPT phases and conventional SPT phases with an enlarged internal symmetry group.  We also explore the connection between bosonic and fermionic HOSPT phases in the presence of strong interactions, and comment on the implications of this connection for the classification of interacting fermionic HOSPT phases. 
Finally, we explore how gauging internal symmetries of these phases leads to topological orders characterized by nontrivial braiding statistics between topological vortex excitations and geometrical defects related to the spatial symmetry. 
\end{abstract}

\maketitle

\section{Introduction}

After a decade of intense effort, the classification and characterization of symmetry protected topological phases has been thoroughly investigated. Beginning with non-interacting topological insulators and superconductors~\cite{kane2005quantum,ryu2010topological,fu2008superconducting,bernevig2006quantum,kitaev2009periodic,roy2009topological,schnyder2009classification}, the focus has shifted to interacting systems, most of which could be classified by group cohomology, Chern-Simons theory, cobordism theory and  non-linear $\sigma$ models (NL$\sigma$M) \cite{Chen2011-et,Chen2011-et,chen2012symmetry,pollmann2012symmetry,lu2012theory,xu2013wave,bi2015classification,vishwanath2013physics,wen2015construction,PhysRevX.6.041068,kapustin2017fermionic,2014arXiv1406.3032M,2018arXiv180408628F}. 
Such a symmetry-protected topological (SPT) phase can essentially be defined by its topological bulk-boundary correspondence: if the boundary does not break the  symmetry, it must be gapless or (for 3 dimensional systems) topologically ordered. 
In parallel to the exploration of SPT phases, the concept of symmetry protection has been extended to spatial symmetries and point group symmetries in common crystalline materials~\cite{fu2011topological,hsieh2012topological,cheng2016translational,shiozaki2017many,ando2015topological,2018PhRvB..97i4508S,2017PhRvB..95t5139S,hong2017topological,qi2015anomalous,huang2017building,teo2013existence,song2017topological,cheng2018microscopic,huang2018surface,watanabe2017structure,po2017symmetry,schindler2017higher,kunst2017lattice,song2017d,cheng2016translational}. 
Topological states protected by spatial symmetries come with a much richer bulk-boundary correspondence: in addition to gapless surfaces (or edges), depending on the protecting symmetries, they also admit boundaries where the edges and surfaces are gapped,
but protected gapless modes appear at corners or hinges of the system. 
Topological crystalline phases
with this phenomenology have recently been termed  higher-order topological insulators (HOTI)~\cite{wang2018weak,ezawa2018minimal,benalcazar2017electric,benalcazar2017quantized,khalaf2018higher,matsugatani2018connecting,lin2017topological,dwivedi2018majorana,langbehn2017reflection,2018arXiv180704141Q,2018arXiv180603007E,2018arXiv180611116W,song2017d,parameswaran2017topological,yan2018majorana,zhu2018tunable,langbehn2017reflection,benalcazar2017quantized,ezawa2018minimal,dwivedi2018majorana,khalaf2018higher,kunst2017lattice,schindler2017higher,ezawa2018higher,song2017d,benalcazar2017electric,qi2017folding,peterson2017demonstration,isobe2015theory,matsugatani2018connecting,song2017interaction,benalcazar2014classification,lapa2016interaction,huang2017building}. This concept has its correspondence in the study of strongly interacting crystalline SPTs \cite{isobe2015theory,song2017topological,song2017interaction,huang2017building}, which we shall refer to as higher-order symmetry protected topological (HOSPT) phases. n nth order SPT must have some gapless modes on its boundary, but these may be confined to live on a d-n- dimensional submanifold of the boundary. In general, an $n$-th order SPT must have some gapless modes on its boundary, but these may be confined to live on a $d-n$- dimensional submanifold of the boundary. In this nomenclature, conventional SPT phases are of first order. A second order SPT phase in two spatial dimensions (2D) and three spatial dimensions (3D) then supports ungappable corner and hinge modes, respectively.

HOTIs are closely connected to the crystalline SPT phases discussed in Refs.\cite{parameswaran2017topological,fu2011topological,hsieh2012topological}. While examples of interacting systems where crystalline symmetry protects ($d-2$)-dimensional boundary modes have been discussed in the literature\cite{song2017topological,song2017interaction,isobe2015theory}, most of the efforts to date have focused on classification, rather than on exploring the higher-order nature of the bulk-boundary correspondence in these systems. In this work, we focus on this  bulk-boundary correspondence in interacting HOTI's. Several complimentary approaches can be employed. First, we anticipate that HOSPT phases can be realized in simple but exactly solvable models whose many-body spectrum, including the surface and corner or hinge modes, is exposed in a clear way. Second, low energy effective theories, either obtained in a phenomenological or microscopic way, should be able to represent the topological structure of these phases and the symmetries that protect them. In particular, as a HOSPT phase supports gapless modes at the corner/hinge, one should be able to infer from an effective theory the symmetry protection of the $(d-n)$-dimensional boundary of the $n$-th order SPT state. Third, an important feature of HOSPT phases protected by spatial symmetries (together with some internal symmetry) is the interplay between symmetry and geometric defects. In the study of conventional SPT phases, a rich phenomenology including the existence of symmetry protected zero modes was uncovered at geometrical defects, such as cross-caps, dislocations, or disclinations~\cite{song2017topological,huang2018surface,gopalakrishnan2013disclination,you2016response,teo2010topological,thorngren2018gauging}. In addition, for SPT phases protected by global unitary symmetries, gauging these symmetries reveals nontrivial braiding statistics between flux defects~\cite{wang2014braiding,levin2012braiding,jian2014layer,bi2014anyon,jiang2014generalized,chen2015anomalous}, suggesting that a similar form of braiding arises between lattice defects and symmetry fluxes when the internal symmetries of an HOSPT are gauged.

In this work, we address the above points by studying  characteristic examples of bosonic HOSPT phases, with corner or hinge modes protected by a combination of internal and lattice symmetries. In addition to mirror symmetry, for which hinge and corner modes have been discussed previously\cite{song2017topological}, we present explicit examples where $C_4$ rotational symmetry leads to second order, rather than first order, SPTs.  
We show that these examples can be described by a NL$\sigma$M with a topological term,  reminiscent of the situation for many first order SPTs~\cite{,xu2013wave,abanov2000theta,bi2015classification,wen2015construction,Senthil2015-tp}. In our HOSPT examples, however, the bulk topological term is trivial, but 
each corner (hinge) supports a 0D (1D) Wess-Zumino-Witten (WZW) term that characterizes the non-trivial topology of the bulk.  
We show explicitly how these corner (or hinge) WZW terms arise from a NL$\sigma$M description of a first-order SPT with $G \times \mathbb{Z}_2$ symmetry, where $G$ is the internal symmetry of the final HOSPT.
Based on this, we argue that in general if a first-order SPT in $d$ dimensions with symmetry group $G \times \mathbb{Z}_2$ exists and admits a non-trivial decorated domain wall construction~\cite{chen2014symmetry}, then a second-order SPT also exists on a $d$ dimensional cubic lattice, with gapless hinge or corner modes protected by a combination of $G$ symmetry and $C_4$ rotations. This provides a framework connecting conventional symmetry protected topological phases and higher order topological crystalline phases in the same dimension via symmetry reduction.

An important question that arises in our discussion is what is the appropriate definition of a higher-order topological phase.  For example, in 2D any model that realizes an internal symmetry projectively at its corners has gapless corner modes that are robust to {\it local} perturbations. However, we will show that in some cases these projective representations can be eliminated by attaching 1D SPTs to the boundary in a way that respects all relevant lattice symmetries.  Such systems are not true HOSPT, since the projective representations cannot reflect any bulk properties of the system.  We will thus take the view that two HOSPT phases are equivalent if they are related by attaching {\it any} lower dimensional system to the boundary.  This leads to a non-trivial interplay between the possible projective representations of a system's internal symmetry, and the lattice geometry required to allow a true HOSPT.  

In addition to studying bosonic HOSPT, we
also study their fermionic counterparts. The classification of fermionic TCIs is reduced by interactions, i.e., when they are considered as crystalline SPTs\cite{song2017topological,song2017interaction,isobe2015theory}. We show explicitely how this breakdown can be understood by focusing on the topological boundary states of a HOSPT phases. In the process, we demonstrate that as for conventional SPT phases, our bosonic models can be constructed from multiple copies of fermionic higher-order topological superconductors.

The remainder of the paper is structured as follows.
In Sec.~\ref{sec: bosonic}, we study two representative examples of bosonic HOSPT in 2D, and introduce an effective low-energy field theory that captures the topological nature of their boundary modes.  
We use this to draw general conclusions about HOSPT protected by $C_4$ rotation combined with an internal symmetry group $G$.  
In Sec.~\ref{3dm} we repeat this analysis for 3D systems, where the relationship between first and second order SPTs is qualitatively the same as in 2D. We also discuss an example of a third order SPT, with protected gapless corner modes in 3D.
This is followed by Sec.~\ref{sec: fermions}, in which we study fermionic HOSPTs with a $\mathbb{Z}$ classification in the noninteracting limit. We show how interaction reduce the classification to a subgroup $\mathbb{Z}_{N}$. At the same time, we demonstrate that $N$ copies of such fermionic HOSPT with interactions between them are equivalent to the bosonic HOSPT from Sec.~\ref{sec: bosonic}.
Finally, in Sec.~\ref{sec: gauging} we establish that by gauging the spatial and internal symmetry in HOSPT, the resultant gauge flux and lattice defects have nontrivial 3-loop braiding statistics. 


\section{Higher order bosonic SPT phases in 2D}
\label{sec: bosonic}

While the understanding of higher order topological phases of fermions is rather complete~\cite{parameswaran2017topological,langbehn2017reflection,benalcazar2017quantized,yan2018majorana,ezawa2018minimal,dwivedi2018majorana,khalaf2018higher,kunst2017lattice,schindler2017higher,ezawa2018higher,song2017d,benalcazar2017electric,qi2017folding,peterson2017demonstration,isobe2015theory,matsugatani2018connecting,song2017interaction,benalcazar2014classification,lapa2016interaction,huang2017building}, the corresponding focus on sub-dimensional gapless boundary modes in bosonic systems with explicit models is less known.~\cite{benalcazar2014classification,teo2013existence}. The only constructive approach is that of Ref.~\cite{song2017d}, where the authors build an HOSPT state from spin degrees of freedom via a coupled wire construction.

In this section, we introduce two types of exactly solvable 2D bosonic models realizing interacting HOSPT phases with gapless corner modes.  We also introduce a field theoretic description of the underlying physics, and discuss its relation to the paradigm developped for higher order fermionic SPT phases, and its implications for a classification of interacting bosonic HOSPT. 

As for all SPT phases, the corner modes we construct are gapless only in the presence of certain symmetries.  Here, the relevant symmetries are an on-site symmetry (which will be either time-reversal $\mathcal{T}$ or $\mathbb{Z}_m \times \mathbb{Z}_n$  symmetry, where $m$ and $n$ are not mutually prime), together with a lattice symmetry. In this work, we consider the latter to be either reflections or $C_4$ rotations as examples.

\subsection{Exactly solvable model with local $\mathbb{Z}_2\times \mathbb{Z}_2$ and $C_4$ rotation or reflection symmetry}\label{2dz}

\begin{figure}[t]
  \centering
      \includegraphics[width=0.25\textwidth]{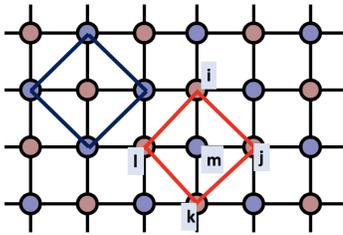}
  \caption{The topological plaquette paramagnet model on checkerboard lattice. The Pauli spins $\tau,\sigma$ live on the red/blue sites. The interaction $ \sigma^z_i \sigma^z_j \sigma^z_k \sigma^z_l  \tau^x_m$ involves the four $\sigma_z$ spin operators on the blue plaquette and the $\tau_x$ in the middle of plaquette.}
  \label{one}
\end{figure}

Our first model for a bosonic HOSPT has global $\mathbb{Z}_2\times \mathbb{Z}_2$ symmetry and gapless corner modes that transform projectively under this $\mathbb{Z}_2\times \mathbb{Z}_2$ symmetry.  We show that the resulting degeneracy is  protected by a combination of the global $\mathbb{Z}_2\times \mathbb{Z}_2$ symmetry together with either $C_4$ rotation or reflection symmetry.    We will also describe how very similar models can be constructed for global $\mathbb{Z}_m\times \mathbb{Z}_n$ symmetry, and discuss under what conditions these harbor protected gapless corner modes.

The model was introduced in Ref.~\onlinecite{2018arXiv180302369Y}, where another aspect of it was studied: it can be viewed  as  an SPT phase in which the gapless boundary modes are protected by subsystem symmetry (meaning that rotating spins independently along different one-dimensional (1D) lines constitutes a symmetry). 
Here, we show that if we relax the condition that such subsystem spin rotations be symmetries of the full Hamiltonian, but retain {\it global} $\mathbb{Z}_2 \times \mathbb{Z}_2$ symmetry, the edges can be gapped -- but that there remain gapless corner modes that transform projectively under the global symmetry.

The model is defined on the checkerboard lattice with two flavors of spins, $\sigma$ and $\tau$, residing on one of the sublattices each. The Hamiltonian is
\begin{align} 
H=-\sum_{i\in a} \tau^x_i \prod_{j\in P_i}\sigma^z_j  -\sum_{i\in b} \sigma^x_i \prod_{j\in P_i}\tau^z_j.
\label{topo1}
\end{align}
Here, $a(b)$ refers to the red(blue) sublattice as shown in Fig.~\ref{one}, and
$P_i$ refers to the four spins neighboring $i$ as shown in Fig.~\ref{one}. 
As all of these terms commute, the Hamiltonian is exactly solvable. 
\begin{figure}[t]
  \centering
      \includegraphics[width=0.3\textwidth]{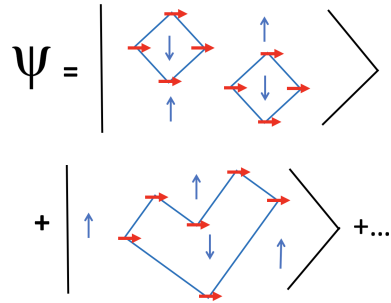}
  \caption{Ground state of the topological plaquette paramagnet defined in Eq.~\eqref{topo1}. The blue blocks illustrate the domain wall for $\sigma$ spin where $\sigma_z=\pm 1$ inside/outside the domain wall. The corner of the blue block contains a $\tau$ (red) spin polarized at $\tau_x=-1$. The ground state is a superposition of all domain wall block configurations where the corners of the blocks are decorated with $\tau_x=-1$.}
  \label{ten}
\end{figure}

The ground state of Eq.~\eqref{topo1} may be understood in the $\sigma^z$ and $\tau^x$ basis.  
Each flavor of spin forms a tilted square lattice, where $P_i$ then refers to four spins on a tilted plaquette.
The first term in Eq.~\eqref{topo1} may be understood as decorating plaquettes $P_i$ in which $\prod_{j\in P_i}\sigma^z_i=-1$ (the domain wall \emph{corners} in $\sigma^z$) with a $\tau^x=-1$ as illustrated in Fig.~\ref{ten}.
The second term then flips between domain wall configurations of $\sigma^z$, while maintaining the decoration of $\tau^x$.
The ground state is therefore a coherent equal superposition of all configurations of $\sigma^z$, decorated in this way.


\subsubsection{Edge and corner modes}
\begin{figure}[t]
  \centering
      \includegraphics[width=0.3\textwidth]{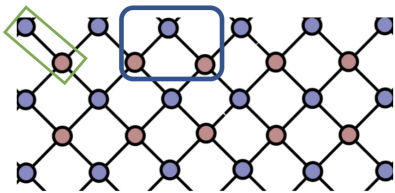}
  \caption{Boundary degrees of freedom of Eq.~\ref{topo1}. The three-spins cluster in the blue rectangle refers to the effective edge spin, while the corner two-spin cluster in the green rectangle is the corner effective spin.} 
  \label{boson2}
\end{figure}

Although the ground state of topological plaquette paramagnet on a closed manifold is unique, the ground state becomes highly degenerate in the presence of a boundary~\cite{2018arXiv180302369Y}, if one simply excludes terms in the Hamiltonian that are not fully supported in the system.
To illustrate this, we consider the horizontal/vertical edge of a checkerboard lattice that is tilted by $\pi/2$, as shown in Fig.~\ref{boson2}.
One can define a set of three anticommuting Pauli operators for each cluster of three sites along the edge (blue rectangle in Fig.~\ref{boson2}), which commute with the Hamiltonian and all operators associated with different boundary clusters. They are given by
\begin{align} 
\pi^x_\text{edge} =\tau^z \sigma^x \tau^z,\quad \pi^y_\text{edge}=\tau^z \sigma^y \tau^z,\quad  \pi^z_\text{edge}= \sigma^z.
\label{edge3}
\end{align}
We will call $\pi^{x,y,z}_\text{edge}$ the edge spin operators. (Here, we omit the site indices for convenience).
At the corner, the spin operators that commute with the Hamiltonian and with the boundary cluster operators from the adjacent edges are defined using a 2-site cluster (green rectangle in Fig.~\ref{boson2}), via
\begin{align} 
\pi^x_\text{corner}= \sigma^x\tau^z,\quad \pi^y_\text{corner}=\sigma^y\tau^z,\quad \pi^z_\text{corner}= \sigma^z
\end{align}
In total, including edges and corners, there is a $2^{N_B}$ dimensional Hilbert space (where $N_B$ denotes the number of boundary spins) associated with these boundary spin operators.  As discussed in Ref.~\cite{2018arXiv180302369Y}, the Hamiltonian~(\ref{topo1}) has a subsystem symmetry that protects this large degeneracy. 

Here, we will instead require only global $\mathbb{Z}_2 \times \mathbb{Z}_2$ symmetry, generated by $\prod_i \sigma^x_i$ and $\prod_i \tau^x_i$. 
In this case each edge degree of freedom along the vertical or horizontal edge can be gapped out in a symmetric way by adding a uniform polarization term $-h \pi^x_{\text{edge}}$ to the Hamiltonian, which commutes with both generators of the global $\mathbb{Z}_2 \times \mathbb{Z}_2$ symmetry.
However, at the corners, none of the three edge Pauli operators commute with both $\mathbb{Z}_2$ symmetries, and such a term cannot be added. 

To see more generally that these corner modes cannot be gapped out in any symmetric (and local) way, let us consider the possible action of an arbitrary symmetry respecting perturbation on the ground state manifold.  
The boundary spins may be viewed as a 1D Ising chain described by the spin operators $\pi_{\text{edge}}$ and $\pi_{\text{corner}}$ going around the perimeter.
The first $\mathbb{Z}_2$ symmetry acts on this Ising chain as a global $\mathbb{Z}_2$ symmetry, $\prod \sigma^x \rightarrow \prod \pi^x_\text{edge}\prod\pi^x_\text{corner}$ on both edge and corner spins.
Meanwhile, the second one acts on the Ising chain only on the four sites at each corner as $\prod \tau^x \rightarrow \prod \pi^z_\text{corner}$.
Considering a square sample, the $\mathbb{Z}_2\times \mathbb{Z}_2$ symmetry is therefore realized projectively within each quadrant (as the generators anticommute locally when acting on the boundary degrees of freedom).
The argument then proceeds as follows.  $\pi^z_{\text{corner}}$ remains a good quantum number under any local interaction term that may be added (as such terms must commute with the symmetry $\prod_\text{corners}\pi^z_{\text{corner}}$).
The energy must also be independent of the $\pi^z_{\text{corner}}$ quantum number, as it is flipped by the global $\prod \sigma^x$ symmetry which commutes with the Hamiltonian (as long as the corners are separated by much larger than the correlation length and the edge is not spontaneously symmetry broken).
Therefore, there is a two-fold degeneracy at each corner.
Without $C_4$ or reflection symmetry, one may simply move the gapless mode from one corner to the another via local terms, where they can be gapped by a local coupling.
This possibility is not allowed when $C_4$ or reflection symmetry is imposed on the system. We thus conclude that $C_4$ or reflection symmetry together with the local $\mathbb{Z}_2\times \mathbb{Z}_2$ symmetry protect this HOSPT phase.



\subsubsection{$\mathbb{Z}_n\times \mathbb{Z}_m$ generalization}

Finally, we describe the more general case with $\mathbb{Z}_n\times \mathbb{Z}_m$ and reflection symmetry, where we require $\text{gcd}(n,m)\neq 1$, as in Ref.~\onlinecite{2018arXiv180302369Y}.
We replace the $\sigma$ Pauli operators by $\mathbb{Z}_n$ operators $Z$, $X$, satisfying the algebra
\begin{subequations}
\begin{eqnarray}
Z^n &=& X^n = 1,\\
XZ &=& \omega ZX,\label{eq:znpauli}
\end{eqnarray}
\end{subequations}
with $\omega=e^{2\pi i / n}$, and also define $Y=-i ZX$.
Similarly, we replace the $\tau$ Pauli operators by $\mathbb{Z}_m$ operators $\widetilde{Z}$, $\widetilde{X}$, $\widetilde{Y}$, which satisfy the same algebra but with $n$ replaced by $m$ and $\widetilde{\omega}=e^{2\pi i/m}$.
Let $q=\text{gcd}(n,m)$.
We may define the following generalization of Eq.~\eqref{topo1}
\begin{equation}
    \begin{split}
    H = -\sum_{\substack{m\in a\\(ijkl\in P_m)}} (\widetilde{Z}_i^\dagger \widetilde{Z}_j\widetilde{Z}_k^\dagger\widetilde{Z}_l)^{\frac{mz}{q}} X_m + \mathrm{h.c.}\\
    -\sum_{\substack{m\in b\\(ijkl\in P_m)}} ({Z}_i^\dagger {Z}_j{Z}_k^\dagger {Z}_l)^{\frac{nz}{q}} \widetilde{X}_m + \mathrm{h.c.}.
\end{split}
\label{eq: generalization spin H}
\end{equation}
Here, the sums are over all sites $m$ in the $a$ or $b$ sublattice, and $ijkl\in P_m$ are the four spins surrounding site $m$, in the order labeled in Fig.~\ref{one}.
 Hamiltonian~\eqref{eq: generalization spin H} consists of mutually commuting terms, where each $z=1\dots q$  corresponds to a distinct phase (which we characterize here by a distinct projective representations of $\mathbb{Z}_n\times\mathbb{Z}_m$ at the corners).
This model possesses the $\mathbb{Z}_n$ symmetry $\prod_i X_i$, and the $\mathbb{Z}_m$ symmetry $\prod_i \widetilde{X}_i$.
Note that this model does not possess $C_4$ rotation symmetry, as such a rotation maps $z\rightarrow q-z$ and does not leave the Hamiltonian invariant.
However, it does have reflection symmetry along the vertical and horizontal axes in Fig.~\ref{one}.
We define this model on the $45^\circ$ rotated square lattice shown in Fig.~\ref{boson2}, with diagonal reflection symmetries which terminate at the corners.

We may similarly define $\mathbb{Z}_n$ degrees of freedom along the edges (see Fig.~\ref{boson2}),
\begin{align} 
\pi^x_\mathrm{edge} = \widetilde{Z}^{\frac{mz}{q}} X \widetilde{Z}^{\dagger{\frac{mz}{q}}},\ \ 
\pi^y_\mathrm{edge} = \widetilde{Z}^{\frac{mz}{q}} Y \widetilde{Z}^{\dagger{\frac{mz}{q}}},\ \ 
\pi^z_\mathrm{edge} =  Z ,
\end{align}
as well as at the corners, which, depending on the corner orientation, are either given by
\begin{equation} 
\pi^x_\mathrm{corner} = X \widetilde{Z}^{\dagger{\frac{mz}{q}}},\ \ 
\pi^y_\mathrm{corner} = Y \widetilde{Z}^{\dagger{\frac{mz}{q}}},\ \
\pi^z_\mathrm{corner} = Z.
\end{equation}
or with $\widetilde{Z}^\dagger\leftrightarrow\widetilde{Z}$.
As in the previous section, if we simply exclude terms in the Hamiltonian that are not fully fully supported in the system, the ground state manifold is highly degenerate.  However, the degeneracy associated with the $\pi_{\text{edge}}$ spins can be locally removed without breaking the $\mathbb{Z}_m\times\mathbb{Z}_n$ symmetries, while the corner degeneracy cannot.

As before, this is because the $\mathbb{Z}_n$ symmetry acts on the ground state manifold as $\prod \pi^x_\mathrm{edge}\prod\pi^x_\mathrm{corner}$, while the $\mathbb{Z}_m$ symmetry acts as $\prod (\pi^{z\dagger}_\mathrm{corner})^{\frac{nz}{q}}$.
Thus, the edge can be gapped out in a symmetric way by adding $-h\pi^x_\mathrm{edge}+\mathrm{h.c.}$ along the edge.  
At the corner, we can add $-(\pi^x_\mathrm{corner})^k+\mathrm{h.c.}$, which commutes with both $\mathbb{Z}_m$ and $\mathbb{Z}_n$ symmetry, if $k$ is chosen such that $zk/q$ is integer.  
The smallest non-zero $k$ one can add is $k = q/\text{gcd}(z,q)$.
This term lifts some of the ground state degeneracy associated locally with the corner -- but not all of it.
Specifically, all eigenstates of $\pi^z_\mathrm{corner}$ with eigenvalue $\omega^a$ remain ground states if $\omega^{ka}=1$.  
There are $\text{gcd}(n,k)$ such eigenvalues. Thus, $\text{gcd}(n,k)$ is the protected ground state degeneracy per corner.
In particular, if we choose $z=1$, then the ground state degeneracy per corner is simply $q=\text{gcd}(n,m)$.

However, for some choices of $m$ and $n$ these corner modes are not fully protected by the spatial symmetries.
Consider $m=n=3$ and $z=1$ as an example.  
The onsite symmetries protect a 3-fold degeneracy at the corners stable to small local perturbations, due to the fact that the corners transform as a non-trivial projective representation of $\mathbb{Z}_3\times\mathbb{Z}_3$.
However, strong interaction terms along the edges can gap this out.  
To see this, notice that the projective representations of this group have a $\mathbb{Z}_3=\{1,\nu,\nu^2\}$ classification, and we may define the representation at the top left and bottom right to be of the class $\nu$, and those at the remaining corners of class $\nu^{-1}=\nu^2$.  
Now, consider lining 1D $\mathbb{Z}_3\times\mathbb{Z}_3$ SPTs along each edge, which can be done in a reflection symmetric way such that at each corner we have a total of three copies of $\nu$ or $\nu^2$ projective representations, which thus leads to a completely trivial representation at the corners which can be gapped out.

More generally, projective representations of $\mathbb{Z}_n\times\mathbb{Z}_m$ are given by the second cohomology group $\mathcal{H}^2[\mathbb{Z}_n\times\mathbb{Z}_m,U(1)] = \mathbb{Z}_q$.  
Let $\nu$ be the generator of $\mathbb{Z}_q$, then,
 our model with general $z$ has the projective representations $\nu^{z}$ at the corner.
Lining 1D $\mathbb{Z}_n\times\mathbb{Z}_m$ SPT chains along the edge, we may modify this projective representation as $\nu^z \rightarrow \nu^{z+2c}$, for any integer $c$.  
Thus if $q$ is odd, it is always possible to gap out the corners completely by adding 1D systems to the boundary.
However, if $q$ is even, this is not possible.  
Allowing the symmetric stacking of 1D systems along the boundary into our phase equivalence relation, this implies a $\mathbb{Z}_2$ classification for even $q$, and trivial ($\mathbb{Z}_1$) for odd $q$.
This suggests that the existence of projective representations alone is not sufficient to ensure an interacting HOSPT phase, and that the lattice geometry imposes non-trivial conditions on which projective representations can lead to HOSPT phase.  



\subsection{Exactly solvable model with time-reversal and reflection symmetry}\label{2dt}

In addition to unitary symmetries like $\mathbb{Z}_n \times \mathbb{Z}_m$, it is instructive to consider time reversal symmetry $\mathcal{T}$.  Here we propose a lattice model where $\mathcal{T}$, combined with either reflections or $C_4$ rotation symmetry, can also lead to bosonic HOSPT phase.  We will use this construction to formulate a field theory that describes both this HOSPT phase and that of the previous subsection.  We note that according to the group cohomology classification~\cite{Chen2011-et,chen2011two}, there exists no nontrivial usual bosonic SPT in 2D protected only by $\mathcal{T}$ symmetry. Instead, we show that these gapless corner modes arise from a non-trivial bulk topological term that is precisely that of the 1D Haldane or the Affleck-Lieb-Kennedy-Tasaki (AKLT) chain, which requires only $\mathcal{T}$ symmetry to be topological. This is consistent with the construction of crystallne SPTs from lower-dimensional systems, pioneered in Ref.~\cite{song2017interaction,huang2017building,song2017interaction,isobe2015theory}.

Our construction is similar in spirit to the 1D AKLT chain, in which neighbouring spins are coupled in such a way that in the ground state each boundary has an effective free spin-$1/2$.  In 2D, we instead use interactions which entangle the spins between different sites on the same plaquette, such that in the ground state, each corner of the lattice contains an odd number of free spin-$1/2$ degrees of freedom, which are decoupled from the bulk.

\begin{figure}[t]
  \centering
      \includegraphics[width=0.35\textwidth]{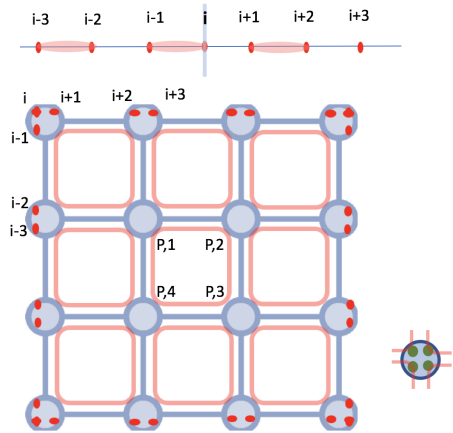}
  \caption{Spin model on square lattice with four spin-1/2 per site. The red square denotes the spin cluster between four spins on the corner of the plaquette. The edge site contains two free spin-1/2 (red dot) while the corner site contains three. One can dimerize the edge spin via inter/intra-bond singlet.
  The edge dimerization is odd under the reflection symmetry that leaves the diagonal invariant.} 
  \label{edge1}
\end{figure}

More precisely, we begin with spins arranged on the square lattice as shown in Fig.~\ref{edge1}. There are four spins (green dots) per site (blue circle) and each spin independently interacts with the one of the four plaquette clusters (red) adjacent to the site. The plaquette cluster interaction involves the four spins coming from the four corners of the plaquette and the interaction projects the four spin into a unique state.

The Hamiltonian of this model is the sum of all plaquette clusters with no mutual overlap, rendering it exactly solvable
 \begin{equation}
 \begin{split} 
H=&\,\sum_{P}  |\alpha_P \rangle \langle \alpha_P |, \\
|\alpha_P \rangle=&\,
\frac{1}{\sqrt{2}}
\bigl(|0 \rangle_{P,1}|0 \rangle_{P,2} |0 \rangle_{P,3} |0 \rangle_{P,4}\\
&\quad+|1 \rangle_{P,1}|1 \rangle_{P,2} |1 \rangle_{P,3}|1 \rangle_{P,4}\bigr),\\
\label{TM}
\end{split}
\end{equation}
where $P$ labels a plaquette, shown in red in Fig.~\ref{edge1}, which contains four spins at its corners, labeled by the pairs $(P,1)$, $(P,2)$, $(P,3)$, and $(P,4)$.
The plaquette cluster interaction $|\alpha_P \rangle \langle \alpha_P |$ for plaquette $P$ projects the four spins into a unique ground state $|\alpha \rangle_P$. Here, $|0 \rangle_{P,j}$ and  $|1 \rangle_{P,j}$ denote the spin up and down state for the $j=1,\dots,4$ spin-1/2 belonging to plaquette $P$.

Hamiltonian~\eqref{TM} is invariant under the antiunitary time-reversal symmetry defined for any $P$ and $j=1,\dots,4$ by
\begin{equation}
\mathcal{T}: \quad |0 \rangle_{P,j} \rightarrow |1 \rangle_{P,j},
\quad |1 \rangle_{P,j} \rightarrow  -|0 \rangle_{P,j}
\end{equation}
and under the mirror symmetries $M_1$ and $M_2$
\begin{equation}
\begin{split}
M_1:& \quad 
|\lambda \rangle_{(x,y)}\rightarrow 
|\lambda \rangle_{(y,x)},
\\
M_2:& \quad 
|\lambda \rangle_{(x,y)}\rightarrow 
|\lambda \rangle_{(-y,-x)},
\end{split}
\end{equation}
for $\lambda=0,1$. For notational convenience we have used a second way of labeling the spins $|\lambda \rangle_{(x,y)}$, using the coordinates of 2D space in which the lattice is embedded. Our choice of origin coincides with the center of a blue site. The Hamiltonian preserves a number of additional symmetries which are not crucial to our discussion.

Each site along the edge contains two unpaired spins, which may be gapped locally into an onsite singlet. 
However, doing this in a way that preserves the reflections $M_1$ and $M_2$, necessarily leaves an odd number of unpaired spins at each corner. 
Since these are separated by a distance $L_x$ or $L_y$, they cannot be coupled by any local interaction; thus the resulting Kramers degeneracy at each corner can be lifted only by breaking time reversal.  In other words, these corner zero modes cannot be gapped without breaking either reflection or $\mathcal{T}$ symmetry. 

\begin{figure}[t]
  \centering
      \includegraphics[width=0.25\textwidth]{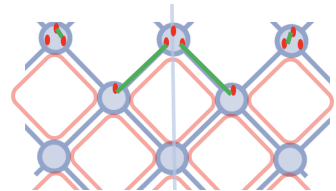}
  \caption{Take a zigzag edge with the reflection plane in the middle of the edge. The edge spins could be gapped in a symmetric way while the reflection points contains a zero mode due to the free spin-1/2.} 
  \label{edge11}
\end{figure}

Though we will often call these unpaired spin-1/2 ``corner modes", in reality they are located at the intersection of the boundary with the reflection plane along the square diagonals. For example, if we instead take a zigzag edge as shown in Fig.~\ref{edge11} and let the reflection axis hit the middle of the edge, the reflection symmetry ensures that the unpaired spin-1/2 sits in the center of the edge, where it is bisected by the reflection plane.  

It is worth pointing out that the zero modes of the Hamiltonian \eqref{TM} can also be protected by $\mathcal{T}$ and $C_4$ rotation symmetry.  
In this case the system consists of four identical quadrants, each containing an odd number of  spin-$1/2$ on the boundary (including the corner).  
Rotational symmetry ensures that the unpaired spin-$1/2$ are separated by a distance on the order of the linear system size, and hence cannot be coupled by any local interaction.
Insisting that the edges along the $x,y$ directions are fully gapped pins these zero modes to the corners---while in general they may reside anywhere on the boundary, provided they are arranged in a rotationally symmetric way.

The robustness of these zero modes can be made apparent using a low-energy effective field theory of the edge. 
We denote the vector of Pauli matrices acting on an edge spin $j$ by $\vec{s}_j=(\sigma^x_j,\sigma^y_j,\sigma^z_j)$ (consult Fig.~\ref{edge1} for the definition of the labels $j$). 
Each spin-1/2 can be written in terms of the vector boson $\vec{n}_j=(n_{1,j},n_{2,j},n_{3,j})$.  Let  $n_{4,j}=\langle \vec{s}_j\cdot\vec{s}_{j+1}- \vec{s}_{j}\cdot \vec{s}_{j-1} \rangle$  denote the valence bond order parameter which forms an on-site or inter-site singlet. 
Time-reversal transforms the fields as
\begin{equation} 
\begin{split}
\mathcal{T}:\qquad
& n_{k,j}(t) \rightarrow -n_{k,j}(-t),\ \ k=1,2,3,\\
& n_{4,j}(t)  \rightarrow n_{4,j}(-t).
\end{split}
\end{equation}
For the boundary shown in  Fig.~\ref{edge1}, the diagonal reflection symmetry $M_1$  acts on $n_{4,i}$ via 
\begin{equation}
    \begin{split}
M_1:\ & n_{4,i+n}= \langle \vec{s}_{i+n} \vec{s}_{i+n+1}- \vec{s}_{i+n} \vec{s}_{i+n-1}\rangle\\
&\rightarrow  \langle\vec{s}_{i-n} \vec{s}_{i-n-1}- \vec{s}_{i-n} \vec{s}_{i-n+1}\rangle=-n_{4,i-n},
    \end{split}
\end{equation} 
where $i$ is the central spin-$1/2$ at the corner, as shown in Fig.~\ref{edge1}.
Note that here we are treating the entire boundary (including the corners) as a single 1D system.  In this 1D system an onsite singlet along the $x$ edge is mapped by reflection to an inter-site singlet on the $y$ edge due to the odd number of spins associated with the corner (see Fig.~\ref{edge1}).  This in turn implies that for a reflection-invariant lattice configuration $n_4$ is odd under diagonal reflections.  

Taking a continuum limit by trading the discrete index $j$ for a continuous co-ordinate $w$ gives an $O(4)_1$ Wess-Zumino-Witten (WZW) model in $(1+1)d$~\cite{chen2013critical,xu2013nonperturbative,xu2013wave,bi2015classification,you2014symmetry},
\begin{equation} \label{BdyQFT}
\mathcal{L}_{\text{edge}}=\frac{1}{g}(\partial_{\mu} \vec{n})^2+\frac{2\pi}{\Omega^3} \int_0^1 du~ \epsilon^{ijkl}  n_i\partial_w n_j \partial_t n_k\partial_u n_l,
\end{equation}
where
$\Omega^N$ is the surface of the N dimensional unit sphere and 
we extended the field $(n_1,n_2,n_3,n_4)(w,t)$ to an extra dimension $u$ to express the $O(4)$ WZW term. The boundary conditions are
\begin{equation} 
\begin{split}
   & (n_1,n_2,n_3,n_4)(w,t,u=0)=(1,0,0,0),\\
   & (n_1,n_2,n_3,n_4)(w,t,u=1)=(n_1,n_2,n_3,n_4)(w,t).
\end{split}
\end{equation}
  $\mathcal{T}$ invariance  ensures that the $O(3)$ rotor $\vec{n}$ is disordered. 
Based on the definitions above, we expect that a domain wall of $n_4$, which is at the interface between an edge region of onsite-singlets and an edge region of inter-bond singlets, contains a free spin-1/2 degree of freedom. 
This is captured by the $O(4)$ WZW term, which indicates that a domain wall of $n_4$ is associated with an $O(3)_1$ WZW term in $(0+1)d$---exactly the field theory of the free spin-1/2.

Since reflection invariance requires
$n_4(w,t) = -n_4(-w,t)$, where $w=0$ corresponds to the corner site $i$ (see Fig.~\ref{edge1}),
$n_4$ must have a domain wall at the corner, resulting in a spin-1/2 zero mode that cannot be gapped unless we break $\mathcal{T}$ or $M_1$. 
This implies our model supports a higher order topological phase protected by $\mathcal{T}$ and reflection $M_1$.

A similar field theoretic picture applies both to the case of reflections about the centers of the zig-zag edges shown in Fig.~\ref{edge11}, with the domain wall of $n_4$ being pinned to the intersection of the mirror plane with the boundary.  A similar argument can be used to show that four-fold rotation symmetry, rather than reflection symmetry can be used to stabilize the HOSPT corner modes, because $n_4$ must change sign under $C_4$ rotations as well.


\subsection{Bulk field theory} \label{subsec:qft1}

The effective edge field theory described in Eq.~\eqref{BdyQFT} suggests a connection between HOSPT and conventional SPT phases, which we now explore in more detail.  This gives a general picture of the relationship between a subset of the possible HOSPT phases, and conventional SPT phases that can be described within the NL$\sigma$M  framework, which applies to most SPTs within the group cohomology classification~\cite{bi2015classification}.

To see how such a connection arises, let us first scrutinize the effective edge theory of Eq.~\eqref{BdyQFT} in more detail.  
Suppose that our bulk gap is large compared to the edge gap, such that it is natural to begin with a theory that has a gapless boundary and consider which symmetry-allowed terms can be used to gap it.
In this case, we can think of our edge theory as a WZW theory emerging from a 2D 
bulk 
described by the $O(4)$ NL$\sigma$M :
\begin{equation} 
\mathcal{L}=\frac{1}{g}(\partial_{\mu} \vec{n})^2+\frac{\Theta}{\Omega^3} \epsilon^{ijkl}  n_i\partial_x n_j \partial_t n_k\partial_y n_l
\label{omg}
\end{equation}
with $\Theta=2\pi$, and $\Omega^3 = 2 \pi^2$ is the volume of the unit 3-sphere. Time reversal symmetry acts according to
\begin{equation}
\mathcal{T} :\quad (n_1,n_2,n_3,n_4) \rightarrow (-n_1,-n_2,-n_3,n_4), 
\label{SymsT}
\end{equation}
The remaining spatial symmetry also acts non-trivially on the bosonic fields $\vec{n}_i$.  Here we will focus on the case of $C_4$ rotation symmetry, for which
\begin{equation}
C_4:\quad (n_1,n_2,n_3,n_4) \rightarrow (n_1,n_2,-n_3,-n_4). \\
\label{SymsZ2}
\end{equation}
where we have suppressed the arguments of each field, which transform in the usual way under $C_4$.  For reflection symmetry, the analogous transformation is $M_1:\quad (n_1,n_2,n_3,n_4) \rightarrow (n_1,n_2,n_3,-n_4)$; in this case an odd number of fields must change sign under the reflection symmetry in order for the Theta term to remain invariant.  

The  gapless edge of this bulk theory is described by an $O(4)_1$ WZW model,
\begin{equation} 
\mathcal{L}_{\text{edge}}=\frac{1}{g}(\partial_{\mu} \vec{n})^2+\frac{2\pi}{\Omega^3} \int_0^1 du~ \epsilon^{ijkl}  n_i\partial_w n_j \partial_t n_k\partial_u n_l,
\end{equation}
with the full $O(4)$ symmetry.
Here $w$ parameterizes the direction along the boundary and $u$ is an extra dimension, with boundary conditions
\begin{equation}
\vec{n}(w,t,u=0)=(1,0,0,0),\quad\vec{n}(w,t,u=1)=\vec{n}(w,t).
\end{equation}

\begin{figure}[t]
  \centering
      \includegraphics[width=0.25\textwidth]{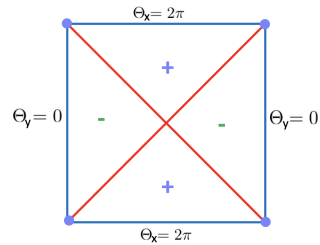}
  \caption{Distribution of $n_4$ in space in a $C_4$ symmetric way. Red lines are domain walls for the $n_4$ scalar field.}
  \label{c4}
\end{figure}

However, this gapless boundary is not protected, since we can polarize $n_4$ in a way that preserves both time reversal and the relevant lattice symmetry.  
This can for instance be achieved by choosing the $n_4$ polarization pattern shown in Fig.~\ref{c4}.  
The result is precisely the field theory in Eq.~(\ref{BdyQFT}), where the lattice symmetry, together with time-reversal symmetry, protects the HOSPT phase. 


The above description relies on a picture in which the bulk gap of the parent $\mathbb{Z}_2\times \mathcal{T}$ SPT is large compared to the symmetry-breaking terms near the boundary, such that we can understand the corner modes starting from the WZW description of the SPT's boundary. As a complementary view, we now ask whether one can obtain a description of the corner modes directly from a bulk description. We will separately discuss the cases of reflection and rotation symmetry, as the corresponding field theories are qualitatively different.
%
%

\subsubsection{Rotation Symmetry}

For an HOSPT protected by rotation symmetry, the NL$\sigma$ model in Eq. ~\eqref{omg} can be viewed as the end result of a 2-step process in which a 2 $d$ first-order SPT with both $C_4$ rotation and an internal $\mathbb{Z}_2$ symmetry is reduced to a second order SPT with only a modified version of the lattice symmetry.  
To see this, we begin with Eq.~\eqref{omg}, with the symmetry actions:
\begin{equation}
\begin{split}
\mathcal{T} :&\quad (n_1,n_2,n_3,n_4) \rightarrow (-n_1,-n_2,-n_3,n_4), \\
\mathbb{Z}_2:&\quad (n_1,n_2,n_3,n_4) \rightarrow (n_1,n_2,-n_3,-n_4), \\
C_4:&\quad (n_1,n_2,n_3,n_4) \rightarrow (n_1,n_2,n_3,n_4). \\
\end{split}
\label{SymsBulk}
\end{equation}
We then break the $\mathbb{Z}_2$ symmetry by ordering $n_4$ in the bulk. However, we chose to do this such that the product of $C_4$ rotation and $\mathbb{Z}_2$ symmetry is preserved.
By this, we mean 
 we preserve the symmetry group  generated by $g=g_1 g_2$, where $g_1$, $g_2$ are the generators of the original $\mathbb{Z}_2$ and $C_4$ respectively.
The product of $C_4$ and $\mathbb{Z}_2$ generators defines a new rotation symmetry  (which we will denote from here on by $\tilde{C}_4 = \mathbb{Z}_2 C_4$), under which $n_4$  is odd. 

Let us now study what the topological Theta term of our original SPT tells us about the boundary of the resulting system.  To do this we begin with Eq.~\eqref{omg} in polar coordinates $(r,\phi)$, and take  $n_4 \equiv \langle n_4 \rangle=\cos(2\phi)$, which preserves both $C_4$ rotations and diagonal reflections.  We define a new $O(3)$ vector boson field $\vec{N}$ normalized as $\sum_i N_i^2 = 1$, via
\begin{equation} 
\begin{split}
n_i=&N_i \sin(2\phi),\quad i=1,2,3,\\
n_4=&\cos(2\phi)
\end{split}
\end{equation}
We further let $\Theta$ be spatially depended as
\begin{equation}
\Theta(r)=\left[1-\mathrm{sgn}(r-R)\right]\pi,
\end{equation}
where $R$ is the radius of the system, which is assumed to be a disk.

The resulting topological term has the form:

\begin{widetext} 
 \begin{equation} 
 \begin{split}
L_{\Theta}=& 
\int  dr \int_0^{2\pi} d \phi~\frac{\Theta(r)}{\Omega^3} \epsilon^{ijk} \left [ 2\sin^2(2\phi) N_i \partial_r N_j \partial_t N_k 
+\cos(2\phi)\sin^3(2\phi) \partial_{\phi} N_i \partial_r N_j \partial_t N_k \right].
 \end{split}
\end{equation}
  Because $n_4$ is ordered, the bulk topological term is trivial, and we can integrate over $r$.  To do this, first note that
up to boundary terms the second term in parentheses is a total derivative in $r$. 
The first term is not a total derivative, but can be made to be one by introducing an extra dimension $u$, and exploiting the fact that 
\begin{align}
\partial_u \left( \epsilon^{ijk}  N_i \partial_r N_j \partial_t N_k \right ) =  \epsilon^{ijk} \partial_u N_i \partial_r N_j \partial_t N_k 
\end{align}
to write this term as an integral over $u$.
After doing so we can integrate both terms by parts in $r$, to obtain
\begin{equation}
\begin{split}
\label{LtopFinal1}
L_{\Theta}=&\int  dr \int_0^{2\pi} d \phi~ \epsilon^{ijk}~ \frac{\delta(r-R)2\pi}{\Omega^3} \left [\int_0^1 du~  2\sin^2(2\phi) N_i \partial_u N_j \partial_t N_k+\cos(2\phi)\sin^3(2\phi) N_i \partial_{\phi} N_j \partial_t N_k \right]\\
=&\int_0^{2\pi} d \phi~ \left [\int_0^1 du~\frac{2\pi}{\Omega^3} \epsilon^{ijk}~ 2\sin^2(2\phi) N_i \partial_u N_j \partial_t N_k+\frac{2\pi \cos(2\phi)\sin^3(2\phi)}{\Omega^3} \epsilon^{ijk}~ N_i \partial_{\phi} N_j \partial_t N_k \right ],
\end{split}
\end{equation}
with the boundary conditions $\vec{N}(\phi,t,u=0)=(1,0,0)$ and $\vec{N}(\phi,t,u=1)=\vec{N}(\phi,t)$.
\end{widetext}

The second term in Eq.~\eqref{LtopFinal1} is precisely the $O(3)$ Theta term in $(1+1)d$ that we encountered in Eq.~(\ref{BdyQFT}). 
However, its coefficient $\Theta = \cos(2\phi)\sin^3(2\phi)2\pi$ is not quantized. In the infrared limit of the renormalization group, $\Theta$ will flow to one of the discrete stable fixed points $\Theta=2\pi K,\ K\in \mathbb{Z}$~\cite{xu2013nonperturbative}, depending on its microscopic magnitude~\cite{xu2013nonperturbative}. In our case this magnitude is small, and we expect $\Theta$ to flow to 0 in the infrared, corresponding to two topologically trivial boundaries.  However choosing a slightly different ordering configuration for $n_4$ (for example, with an abrupt sign change, as in Fig.~\ref{c4}), we could equally arrive at the conclusion that $\Theta$ flows to $2 \pi$ along each boundary. Both cases are topologically equivalent.

The first term in Eq.~(\ref{LtopFinal1}) resembles a $(0+1)d$ WZW term, delocalized along the edge.  To make this more precise, consider integrating along one quarter of the integration domain, from $\phi = 0$ to $\phi = \pi/2$, such that the coefficient is non-vanishing at both ends of the range of integration.  
Ignoring the $\phi$-dependence of the O(3) rotor $\vec{N}$, we obtain
\begin{equation} 
\begin{split}
    \mathcal{L}_{\Theta}=&\int_0^{\pi/2} d \phi \int_0^1 du~\frac{2\pi}{\Omega^3} \epsilon^{ijk}~ 2\sin^2(2\phi) N_i \partial_u N_j \partial_t N_k \\
=& \int_0^1 du~\frac{2\pi}{\Omega^2} \epsilon^{ijk} N_i \partial_u N_j \partial_t N_k.
\end{split}
\end{equation}
where $\Omega^2 = 4 \pi$ is the area of the unit sphere.  Thus each quarter of the system contains an $O(3)_1$ WZW term in $(0+1)d$, which describes a free spin-1/2 zero mode coming from the bulk.\footnote{Here we have ignored the $\phi$ dependence of the fields $N_i$.  More generally, we could expand the topological term into modes of different angular momenta.  For the $l=0$ mode the calculation described here applies.  For modes with $l>0$ we should consider the integral over the entire boundary rather than over a single quadrant; doing so reveals that the net topological term for these modes is $0$.}

\subsubsection{Reflection Symmetry}

If we replace $C_4$ rotational symmetry with reflection in the discussion above, a somewhat different field theory applies.  This is because the topological Theta term is odd under mirror symmetry, which changes the sign of only one of the derivatives.  As a consequence, in this case we cannot begin with a conventional SPT and reduce the symmetry.  Instead, for reflection symmetry the appropriate field theoretic treatment indicates a relationship between HOSPT phases and lower-dimensional SPTs, similar to what has been previously discussed for topological crystalline insulators~\cite{song2017interaction,isobe2015theory}.

We illustrate the general framework by considering phases protected by $\mathcal{T}$ and reflection symmetry.  In any even spatial dimension $d=N$, 
we begin with
an $O(N+2)$ NL$\sigma$M in $(d+1)$ dimensional spacetime, with an internal $\mathbb{Z}_2$ symmetry under which an odd number of the vector components change sign.  The corresponding topological $\Theta$ term is therefore odd under both reflection and $\mathbb{Z}_2$ symmetry, but even under their combination.  As $N$ is even, the $\Theta$ term is also invariant under $\mathcal{T}$ symmetry.  In this case, our starting point is a topological field theory that respects $\mathcal{T}$ symmetry and a reflection symmetry under which an odd number of components of $\vec{n}$ change sign:  
\begin{align}  \label{Eq:Ltopref}
\mathcal{L}=\frac{1}{g}(\partial_{\mu} \vec{n})^2+\frac{\Theta}{\Omega^{N+1}} ~ \epsilon^{ijkl..}  n_i\partial_z n_j \partial_t n_k\partial_u n_l...\ .
\end{align}
The symmetries act as
\begin{equation}
    \begin{split}
        \mathcal{T}:&\quad n_i(t) \rightarrow -n_i(-t),\quad i=1,\dots, N+1\\
        &\quad n_{N+2}(t) \rightarrow n_{N+2}(-t)
        \\
R_a: &\quad n_i(r_a) \rightarrow n_i(-r_a),\quad i=1,\dots, N+1\\
        &\quad n_{N+2}(r_a) \rightarrow -n_{N+2}(-r_a),
    \end{split}
\end{equation}
where we omitted all spatio-temporal variables of the field that are unchanged under the respective symmetry operation.

Naively, the action in Eq.~\ref{Eq:Ltopref} suggests a topological crystalline phase with gapless boundaries.  However, the last component $n_{N+2}$ in the $O(N+2)$ vector field can be ordered in a reflection-symmetric way, provided $n_{N+2}(r_a)=-n_{N+2}(-r_a)$. This renders the bulk $\theta$ term trivial without breaking any symmetries.  But as we have seen, this trivialized bulk topological term can generate non-trivial lower dimensional topological terms associated with the boundary.  In particular, $n_{N+2}$ must contain a domain wall at the reflection symmetric plane ($r_a=0$). 
As discussed above, the intersection of this reflection plane with the boundary can be viewed as an intersection of two domain walls, where simultaneously $\langle n_{N+2}(r_a) \rangle$ switches sign and $\Theta$ jumps from $0$ to $2\pi$. A calculation similar to the one described above shows that this point is described by an $O(N+1)$ WZW theory in  $(d-1)$ spacetime dimensions, which  cannot be trivially gapped as long as the $\mathcal{T}$ symmetry is unbroken.  


\subsubsection{Generalizations}

Finally, let us turn to the question of which 2D HOSPT phases  admit a field theoretic description similar to that presented here.  Evidently, the field theory presented above can equally be applied to the case of $\mathbb{Z}_2 \times \mathbb{Z}_2 $ symmetry, with the nature of the field theory itself unchanged. More generally, for 
bosonic SPT phases which have a NL$\sigma$M description, we can make a connection between the conventional SPT state with gapless boundary modes protected by $\mathbb{Z}_2$ and $G$ symmetry, and the higher order SPT protected by $C_4$ rotation and $G$ symmetry, similar to that proposed by Ref.~\onlinecite{song2017topological}. Specifically, $\mathbb{Z}_2 \times G$ SPT phases (with $G$ satisfying certain compatibility conditions) may contain a decorated domain wall structure where the $\mathbb{Z}_2$ domain wall is equipped with a lower dimensional SPT with $G$ symmetry~\cite{you2016decorated,chen2014symmetry}. If we break the $\mathbb{Z}_2$ symmetry while keeping the combination of $\mathbb{Z}_2$ and $C_4$ ($\tilde{C}_4=\mathbb{Z}_2 C_4$) invariant,
the system contains perpendicular planes with $\mathbb{Z}_2$ domain walls, each of which contains a lower dimensional SPT with $G$ symmetry.
The boundary of these planes thus carries a gapless mode protected by $G$ symmetry, which will appear in a field theoretic description as a zero-dimensional WZW term associated with each corner.  

More generally, these arguments can be extended to any $C_m \times G$ symmetric topological crystalline phases, which contain $m$ copies of gapless $d-2$ modes on a boundary that respects $C_m$ symmetry, if they support a HOSPT phase.

\section{Higher-order bosonic SPT phases in 3D}\label{3dm}

In 3D, fermionic systems admit two distinct classes of HOSPT: phases with gapless corner modes similar to those of our 2D examples, and phases with ``hinge" states - 1D gapless modes that are confined to live at the boundary between two distinct surfaces of the crystal.  
Systems with protected gapless modes at a boundary of co-dimension $n$ are referred to as $n$-th order topological phases. Following this terminology, Ref.~\onlinecite{schindler2017higher} introduced a second order TI in 3D which exhibits protected hinge states with (spectral) flow between the valence and conduction bands. Subsequent work has systematically classified the
Higher order TIs and topological superconductors (TSCs) for non-interacting fermion systems.\cite{PhysRevB.95.235143,wang2018weak,song2017d,langbehn2017reflection,dwivedi2018majorana,watanabe2017structure,watanabe2017topological,khalaf2018higher,peng2017boundary}.  

We now discuss how 3D second order SPT phases can also exist in strongly interacting bosonic systems, using a combination of exactly solvable models and field theoretic descriptions.  We begin by discussing a model for which the symmetry required to protect the hinges is $\mathbb{Z}_2$ combined with reflection or $C_4$ rotation symmetry.  Since the hinge modes are described by the same topological field theory as the edges of a 2D bosonic SPT,  we must have an internal $\mathbb{Z}_2$ symmetry, rather than $\mathcal{T}$ (which does not yield a non-trivial bosonic SPT in 2D). In this case, there is a bulk NL$\sigma$M description that relates the phase with hinge modes protected by $C_4$ rotation to a 3D bosonic SPT with $Z_2 \times Z_2$ symmetry.

We will then discuss third order interacting SPT phases in 3D, presenting a model with $\mathbb{Z}_2 \times \mathbb{Z}_2$ and rotation symmetry that appears to have protected gapless corner modes. To best of our knowledge, this is the first explicit third order HOSPT state in interacting boson systems. 

\subsection{Second order SPT phase with gapless hinge modes protected by $Z_2$ and $C_4$ rotation or reflection symmetry}\label{3dz}


 The first 3D model we consider is a cubic-lattice version of the CZX model introduced by Ref.~\onlinecite{chen2011two}. There it was shown that the square lattice CZX model realizes a non-trivial 2D $\mathbb{Z}_2$ SPT phase, with symmetry-protected gapless edge modes.  Here we will show that the cubic lattice version is a second order SPT protected by $\mathbb{Z}_2$ symmetry and reflections. The surfaces of of this model can be gapped in a symmetric way, but the hinge separating two faces of a cubic system harbors a gapless mode which is exactly the edge of the 2D $\mathbb{Z}_2$ SPT~\cite{chen2011two,levin2012braiding}.

\begin{figure}[t]
  \centering
      \includegraphics[width=0.5\textwidth]{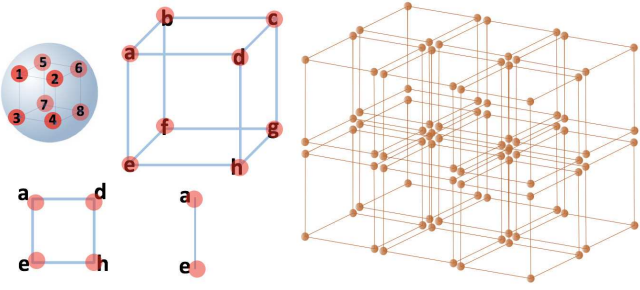}
  \caption{Lattice for the CZX model defined in Eq.~\eqref{eq: CZX Ham}. Each site contains eight spin-1/2 degrees of freedom (labeled as 1,2,...,8). Each spin interacts with one of the eight spin-1/2 cubic clusters adjacent to the site. The Hamiltonian projects the eight spins beonging to a cube into a unique state.}
  \label{xie3d2}
\end{figure}

The Hilbert space for the cubic CZX model is a cubic lattice with eight spin-1/2 per site, each of which interacts independently with a cluster of eight spin-1/2 from sites at the corners of an elementary cube, as shown in Fig.~\ref{xie3d2}. 
The Hamiltonian is a sum of projectors, each of which projects eight spins on the corners of a cube into the entangled state. It is given by

\begin{equation} 
\label{eq: CZX Ham}
\begin{split}
H=&-\sum_{C} \left(H_{C}\otimes \prod_{P \in \mathrm{plaq}(C)} H_{P}\otimes \prod_{B \in \mathrm{bond}(C)} H_{B}\right),\\
H_{C}
=&
|0 0 0 0 0 0 0 0\rangle_C
\langle 0 0 0 0 0 0 0 0|_C
\\
&+
|1 1 1 1 1 1 1 1 \rangle_C
\langle 0 0 0 0 0 0 0 0|_C
\\
&+
|0 0 0 0 0 0 0 0\rangle_C
\langle 1 1 1 1 1 1 1 1 |_C
\\
&+
|1 1 1 1 1 1 1 1 \rangle_C\langle 1 1 1 1 1 1 1 1 |_C,\\
H_{P}=&|0 0 0 0 \rangle_P\langle  0 0 0 0|_P+|1 1 1 1 \rangle_P \langle  1 1 1 1 |_P,\\
H_{B}=&|0 0 \rangle_B \langle  0 0|_B +| 1 1 \rangle_B \langle  1 1 |_B.
\end{split}
\end{equation}

The Hamiltonian sums over all cubes, denoted by $C$, of the cubic lattice. 
Each cube $C$ consists of eight sites at its corners, and recall that each site further consists of eight spin-1/2.
$H_C$ acts on the eight inner spins forming a cube.
Labeling each spin in the cube $C$ by the pair $(\alpha,i)$, for $\alpha\in\{a,\dots h\}$, and $i\in\{1,\dots,8\}$, as in Fig.~\ref{xie3d2}, $H_C$ acts on the eight spins:
\begin{equation}
\left\{(a,8),(b,4),(c,3),(d,7),(e,6),(f,2),(g,1),(h,5)\right\}.
\end{equation}
There are six plaquette terms $H_P$ for $P\in \mathrm{plaq}(C)$, each of which act on the outer four spins on a plaquette of the cube $C$.  
Explicitly, $H_P$ for the top plaquette acts on the four spins
\begin{equation}
\left\{(a,6),(b,2),(c,1),(d,5)\right\}
\end{equation}
and similarly for the other plaquettes.
Finally, there are 12 bond terms $H_B$ for $B\in \mathrm{bond}(C)$, which each act on the two outer spins along a bond.  For instance, along the bond $B$ connecting sites $a$ and $b$, $H_B$ acts on the spins $\{(a,5),(b,1)\}$.

The Hamiltonian $H_C$ projects the eight inner spins of every cube into a unique state.
Meanwhile, $H_P$ and $H_B$ project the spins on each plaquette or bond onto a two-level subspace.
Note that all of these terms commute.

This Hamiltonian has the on-site $\mathbb{Z}_2$ symmetry $U_{czx}$,
\begin{equation}
    \begin{split}
        U_{czx}=&\,U_x U_{cz},\\
U_x=&\,\sigma_x^1\sigma_x^2\sigma_x^3\sigma_x^4\sigma_x^5\sigma_x^6\sigma_x^7\sigma_x^8,\\
U_{cz}=&\,\text{CZ}_{13}~\text{CZ}_{24}~\text{CZ}_{57}~\text{CZ}_{68} ~\text{CZ}_{56}~\text{CZ}_{12}\\\
&\times\text{CZ}_{34}~\text{CZ}_{78}~\text{CZ}_{15}~\text{CZ}_{26}~\text{CZ}_{37}~\text{CZ}_{48},\\
\text{CZ}_{ij}=&\,|0 0 \rangle_{ij}\langle  0 0|_{ij}+| 1 0 \rangle_{ij} \langle   1 0 |_{ij}\\
&+|0 1 \rangle_{ij}\langle  01|_{ij}
-| 1 1 \rangle_{ij} \langle   1 1 |_{ij},
\label{cx}
    \end{split}
\end{equation} 
where the labels $1,\dots,8$ correspond to the eight spin-1/2 on a given site (we suppress the site index).
In this representation the $\mathbb{Z}_2$ symmetry operator $U_{czx}$ is decomposed into the spin flip operator $U_x$ which flips all the spins together with a product of control-Z ($\text{CZ}$) quantum gate operators. Each $\text{CZ}$ operator acts on two spins and imposes a minus sign if they are in the $| 1 1 \rangle$ state. The eight spins on the same site form a mini cube as depicted in Fig.~\ref{xie3d2} and the $\text{CZ}$ operators act on all links of the mini cube.
The ground state is invariant under the
$U_x$ operation. Meanwhile, as the eight spins entangled via the cube operator $H_C$ across the corner sites have all-up-all-down configurations, the $\text{CZ}$ operation finally creates even number of $(-1)$ factors so the state is always invariant under $U_{czx}$. Thus, the Hamiltonian in Eq.~\eqref{cx} commutes with this $\mathbb{Z}_2$ symmetry.

We now consider the degrees of freedom on a surface. For any of the surfaces parallel to the $x$-$y$, $z$-$y$, or $z$-$x$ plane, a surface plaquette involving four corner spins is projected to the two-level system $|0 0 0 0 \rangle, |1 1 1 1 \rangle $. 
We now consider surface pertrubations that preserve the $U_{cyx}$ symmetry.
Each plaquette on the surface can be gapped out with
\begin{equation}
    \begin{split}
        H_{\text{surface}}=&-\sum_{P\in\text{surface}} \Bigl(
        |0 0 0 0 \rangle_P
        \langle 0 0 0 0 |_P
        +        |0 0 0 0 \rangle_P
        \langle 1111 |_P
        \\
        &\ \ 
        +|1 1 1 1  \rangle_P
            \langle 0000 |_P
        +|1 1 1 1  \rangle_P
            \langle 1 1 1 1 |_P\Bigr).
    \end{split}
\end{equation} 

\begin{figure}[t]
  \centering
      \includegraphics[width=0.35\textwidth]{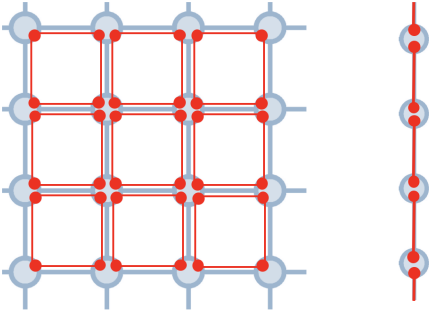}
  \caption{Boundary degrees of freedom of the CZX model defined in Eq.~\eqref{eq: CZX Ham}. Left: On the surface, one can perform a plaquette projection to gap out the surface without breaking the symmetries that define the HOSPT phase.  Right: Along the hinge between two surfaces, each bond represents a two-level system and $\mathbb{Z}_2$ symmetry acts on the hinge in a non-local way.}
  \label{xie3d3}
\end{figure}

This surface interaction does not break $\mathbb{Z}_2$ symmetry or $C^i_4$ rotation symmetry along the $i=x,y,z$ axes, where
\begin{align} 
&C_4^z: s(x,y,z) \rightarrow  s(y,-x,z),\nonumber\\
&C_4^x: s(x,y,z) \rightarrow  s(x,z,-y),\nonumber\\
&C_4^y: s(x,y,z) \rightarrow  s(z,y,-x).
\end{align}
However, the hinge along the $j-$axis between the $i$-$j$ and $j$-$k$ surface planes (with $i,j,k$ and permutation of $x,y,z$) carries an extra 2-fold degeneracy at each bond $B$ because the operator
\begin{align} 
&H_{B}=|0 0 \rangle_B\langle  0 0|_B+| 1 1 \rangle_B \langle   1 1 |_B
\label{twolevel}
\end{align}
projects on a two-dimensional subspace.
One can redefine the basis of this subspace on the hinge bond as
\begin{align} 
&|\tilde{0} \rangle=|0 0 \rangle,\quad |\tilde{1}  \rangle=|11 \rangle,
\label{twolevel2}
\end{align}
where we omit the bound label. 
In this new basis, the corresponding $\mathbb{Z}_2$ symmetry defined in Eq.~\eqref{cx} acts on the hinge spin in a nontrivial way given by
\begin{align} 
&\tilde{U}_{czx}=\prod_i \tilde{U}^i_x. \tilde{U}^{i,i+1}_{cz}
\end{align}
The $\mathbb{Z}_2$ symmetry does not act as an on-site symmetry along this hinge,
and in fact forms a nontrivial 3-cocycle~\cite{chen2011two}, which cannot be realized by on-site symmetries in 1D. 
From this we conclude that 
the hinge modes cannot be completely gapped without breaking $\mathbb{Z}_2$ symmetry locally.  
Indeed this gapless hinge mode is described by the same topological field theory as the edge of the 2D Levin-Gu model or CZX model, which both realize the non-trivial 2D $\mathbb{Z}_2$ SPT phase~\cite{bi2014anyon,xu2013wave,levin2012braiding,chen2013critical}.  The relevant field theory is a $(1+1)d$ $O(4)_1$ WZW theory,
\begin{align} 
&\mathcal{L}_{\text{edge}}=\frac{1}{g}(\partial_{\mu} \vec{n})^2+\frac{2\pi}{\Omega^3} \int_0^1 du~ \epsilon^{ijkl}  n_i\partial_z n_j \partial_t n_k\partial_u n_l,\nonumber\\
 &\vec{n}(x,t,u=0)=(1,0,0,0),\quad
 \vec{n}(x,t,u=1)=\vec{n}(x,t).
\label{col}
\end{align}
The $\mathbb{Z}_2$ symmetry flips all four components of the $O(4)$ vector boson
\begin{equation}
\mathbb{Z}_2: \vec{n}(x,t) \rightarrow -\vec{n}(x,t)
\end{equation}
and the topological term ensures that the theory must be gapless unless the $\mathbb{Z}_2$ symmetry is broken~\cite{xu2013nonperturbative}.

Note that we have gapped the bulk and surface in a $C_4$ rotation invariant way. We can therefore decompose the boundary containing the $x$-$z$ and $y$-$z$ surfaces into four equivalent quadrants (assuming periodic boundary conditions in $z$ direction and $C_4^z$ symmetry). Our analysis above shows that each quadrant contains an odd number of 1D gapless modes in Eq.~\eqref{twolevel}.  Due to the $\mathbb{Z}_2$ classification of Levin-Gu model, an odd number of gapless chains in each quadrant  cannot be fully gapped, and the  gapless hinge modes cannot annihilate each other without breaking rotational symmetry.

As in our 2D examples, reflections through the planes that leave the hinges invariant and interchange adjacent faces, combined with the global symmetry, are also sufficient to protect the gapless hinge modes.  As before, reflections fix the gapless modes to lie on the intersection of the reflection plane with the surface, whereas rotational invariance merely ensures that the hinge modes cannot be brought together and annihilated.

In summary, the combination of $C_4$ rotations or diagonal reflections with the $\mathbb{Z}_2$ CZX symmetry protects the gapless hinge modes and with it a second order 3D SPT phase. 
Breaking both of these lattice symmetries allows quadruples of gapless hinges states to be moved to the same place and annihilated with each other.


\begin{figure}[t]
  \centering
      \includegraphics[width=0.4\textwidth]{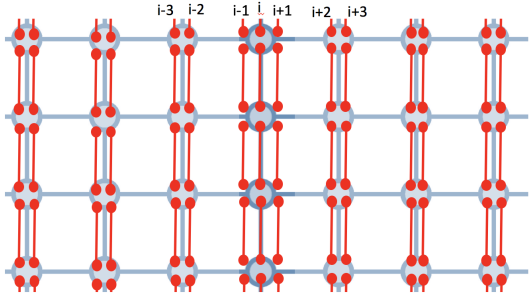}
  \caption{Geometry for hinge modes of the CZX model defined in Eq.~\eqref{eq: CZX Ham}. 2D representation of the surfaces corresponding to the $x$-$z$ plane (left) and $y$-$z$ plane (right). The reflection axis acts on the $i$-th column or surface sites and takes column $i-n$ to $i+n$.} 
  \label{edge22}
\end{figure}


Finally, we emphasize that the second order SPT with gapless hinge states cannot be constructed in an effectively 2D system, for instance by adding 2D SPT layers on the side surfaces of a trivial phase. For example, adding a 2D $\mathbb{Z}_2$ SPT phase in a $C_4$ symmetric way by stacking the CZX mode on the $x$-$z$ and $y$-$z$ planes adds two additional copies of the gapless edge modes described by Eq.~(\ref{col}), which can be trivialized.

\subsection{Field theory of second order 3D SPT phases}

Next, we turn to the question of how second order 3D SPT phases can be described field theoretically.  As in 2D, one approach is via an effective description of the boundary.  Consider the part of the boundary containing the $x$-$z$ and the $y$-$z$ surfaces, as illustrated in Fig.~\ref{edge22}.  Reflection maps the $x$-$z$ surface to the $y$-$z$ surface, by taking the $(i+n)$-th column to the $(i-n)$-th column as shown in the figure, where $i$ labels the coordinate of the hinge column. Each column is a spin-$1/2$ chain described by the $O(4)_1$ WZW theory in Eq.~(\ref{col}).  We define a new scalar field $n_5$ to characterize the coupling between columns 
($n_a^{j}$ refers to the scalar field $n_a$ on the $j$-th column in Fig.~\ref{edge22})
\begin{equation} 
 n_{5}^j=\sum_{a=1}^{4} \langle n_a^{j}n_a^{j+1}-n_a^{j}n_a^{j-1}\rangle,
\end{equation}
which transforms under the mirror symmetry $M_1$ that leaves a given hinge invariant as
\begin{equation}
\begin{split}
    M_1: n_5^{i+n} \rightarrow  -n_5^{i-n}.
\end{split}
\end{equation}
Taking a continuum limit in which the discrete variable $j$ is replaced by a continuous variable $w$, the domain wall of $n_5$ is encoded with an $O(4)_1$ WZW theory in $(1+1)d$. Following the same argument as Sec.~\ref{2dt}, the column coupling on the side surface creates a $(2+1)d$ $O(5)_1$ WZW theory~\cite{chen2013critical},
\begin{align} 
&\mathcal{L}_{\text{edge}}=\frac{1}{g}(\partial_{\mu}\vec{n})^2+\frac{2\pi}{\Omega^4} \int_0^1 du~ \epsilon^{ijklm}  n_i\partial_w n_j \partial_t n_k\partial_u n_l \partial_z n_m 
\end{align}
with the boundary conditions
\begin{equation} 
\begin{split}
\vec{n}(w,z,t,u=0)=&\,(1,0,0,0,0),\\
\vec{n}(w,z,t,u=1)=&\vec{n}(w,z,t).
\end{split}
\end{equation}
The symmetries act on the fields as
\begin{equation}
\begin{split}
    \mathbb{Z}_2:\quad &n_a(w,z,t) \rightarrow -n_a(w,z,t),\quad a=1,\dots,4,\\
    &n_5(w,z,t) \rightarrow n_5(w,z,t),\\
M_1:\quad &n_a(w,z,t) \rightarrow n_a(-w,z,t),\quad a=1,\dots,4,\\
    &n_5(w,z,t) \rightarrow -n_5(-w,z,t).
\end{split}
\end{equation}

The local $\mathbb{Z}_2$ symmetry prevents $(n_1,n_2,n_3,n_4)$ from ordering. 
However, when $n_5$ is ordered on the $x$-$z$ or $y$-$z$ side surfaces, the $O(5)$ WZW theory on the surface is reduced to the $O(4)$ Theta term with either $\Theta=0$ or $\Theta=2\pi$, which gives a gapped surface~\cite{xu2013nonperturbative}. However, since $n_5$ is odd under reflection symmetry, the effective value of $\Theta$ changes from $0$ to $2\pi$ at the hinge along the $z$-axis, which is a domain wall of $n_5$. 
This 1D domain wall is described by a $(1+1)d$ $O(4)_1$ WZW theory, which cannot be gapped in the presence of $\mathbb{Z}_2$ symmetry.

This picture suggests that, for the HOSPT protected by $C_4$ rotation symmetry, a bulk field theory can be obtained by starting with a $(3+1)d$ NL$\sigma$M at $\Theta=2\pi$:
\begin{equation} 
\mathcal{L}=\frac{1}{g}(\partial_{\mu} \vec{n})^2+\frac{\Theta}{\Omega^4} \epsilon^{ijklm} 
n_i\partial_x n_j \partial_t n_k\partial_z n_l \partial_y n_m \ \ .
\end{equation}
with $\mathbb{Z}_2\times \mathbb{Z}_2$ symmetry acting on the fields via\cite{bi2015classification}
\begin{equation}
    \begin{split}
\mathbb{Z}_2^a: &(n_1,n_2, n_3,n_4,n_5) \rightarrow (-n_1,-n_2,-n_3,-n_4,n_5), \\
\mathbb{Z}_2^b: &(n_1,n_2, n_3,n_4,n_5) \rightarrow (n_1,n_2,n_3,-n_4,-n_5).
    \end{split}
\end{equation}
This describes a 3D SPT phase with gapless boundary modes protected by the global $\mathbb{Z}_2\times \mathbb{Z}_2$ symmetry.
We now consider breaking the $\mathbb{Z}_2^b$ symmetry by polarizing $n_5$ as shown in Fig.~\ref{c4}.  This breaks also the $C_4^z$ rotation symmetry, but preserves the combination of $C^z_4$ and $\mathbb{Z}_2^b$ symmetry (which we define to be the new rotation symmetry $\tilde{C}^z_4=C^z_4\mathbb{Z}_2^b$). The corresponding state contains a gapped bulk and surface. The four hinges along the $z$ direction each carry an $O(4)_1$ WZW term, resulting in a gapless spectrum protected by $\mathbb{Z}_2$ symmetry. (A detailed derivation, along the lines of that presented in Sec.~\ref{subsec:qft1}, is included in Appendix \ref{Appendix:FieldTheory}.)

\subsection{Third order SPTs with gapless corner modes}

\begin{figure}[t]
  \centering
      \includegraphics[width=0.3\textwidth]{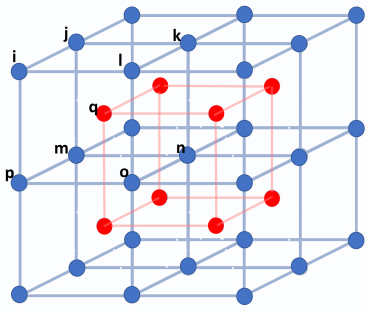}
  \caption{Lattice structure for the third order SPT defined in Eq.~\eqref{cubic}. The spin interaction involves the red spin on the eight corners of a cube together with the blue spin in the middle of the cube and vice versa.
  } 
  \label{six}
\end{figure}

We now turn to the construction of a third order SPT in 3D.   Let us begin by considering the 3D generalization of the $\mathbb{Z}_n\times \mathbb{Z}_m$-symmetric model discussed in Sec.~\ref{2dz}, focusing on the case with $n=m=3$ and the $z=1$ phase. This model lives on inter-penetrating cubic lattices, as shown in Fig.~\ref{six} (which form a body-centered cubic lattice). 
Each site $i$ contains a $\mathbb{Z}_3$ degree of freedom which is acted on label by the $\mathbb{Z}_3$ generalization of Pauli operators $X_i$, $Y_i$, $Z_i$, and $\widetilde{X}_i$, $\widetilde{Y}_i$, $\widetilde{Z}_i$ for the two sublattices $a$ or $b$, respectively.
These operators obey the algebra given in Eq.~\eqref{eq:znpauli}.

The Hamiltonian consists of the 3D cluster interaction

\begin{equation}
    \begin{split}
    H = -\sum_{\substack{q\in a\\(ijklmnop \in C_q)}} (\widetilde{Z}_i^\dagger \widetilde{Z}_j\widetilde{Z}_k^\dagger\widetilde{Z}_l
    \widetilde{Z}_m^\dagger \widetilde{Z}_n\widetilde{Z}_o^\dagger\widetilde{Z}_p
    ) X_q + \mathrm{h.c.}\\
    -\sum_{\substack{q\in b\\(ijklmnop \in C_q)}} ({Z}_i^\dagger {Z}_j {Z}_k^\dagger {Z}_l
    {Z}_m^\dagger {Z}_n {Z}_o^\dagger {Z}_p
    ) \widetilde{X}_q + \mathrm{h.c.}\,.\\
\end{split}
\label{cubic}
\end{equation}
Here $C_q$ refers to the cube of eight nearest neighbor sites of site $q$, which belong to the opposite sublattice of site $q$, as depicted in Fig.~\ref{six}.  
Hamiltonian~\eqref{cubic} is exactly solvable, in close analogy of the model defined in Eq.~\eqref{topo1} and Fig.~\ref{ten}.
Our model has two types of symmetries relevant to the HOSPT: an onsite  $\mathbb{Z}_3 \times \mathbb{Z}_3$ symmetry, generated by $\prod_{i\in a} X_i$ and $\prod_{i\in b} \widetilde{X}_i$, and
 spatial symmetries corresponding to $2\pi/3$ rotation about each the four axes $\hat{x}\pm_1 \hat{y}\pm_2 \hat{z}$, where each choice of $\pm_1$,$\pm_2$ corresponds to a separate axis and $C_3$ symmetry.

\begin{figure}[t]
  \centering
      \includegraphics[width=0.27\textwidth]{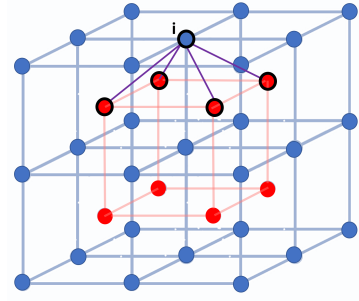}
  \caption{Definition of the surface operators $\pi_i^x$ and $\pi_i^y$ from Eq.~\eqref{edge3}, which have support on the site labelled $i$ and the four sites connected to it by lines.} 
  \label{12}
\end{figure}

We now study the surface degrees of freedom of this model by considering the model with a surface as shown in Fig.~\ref{12}. 
If we exclude all Hamiltonian terms that are not fully supported in the bulk the ground state manifold is massively degenerate.
At each site on the surface there is an effective $\mathbb{Z}_3$ degree of freedom, comprising of one $a$ site on the surface and a cluster of four $b$ spins belonging to the plaquette $P_i$ underneath, as shown in Fig.~\ref{12}. 
The spin operators associated with the surface degrees of freedom are schematically written
\begin{align} 
\begin{split}
\pi^x_\mathrm{surface} &= X \widetilde{Z} \widetilde{Z}^\dagger \widetilde{Z} \widetilde{Z}^\dagger,
\quad
\pi^y_\mathrm{surface} = Y  \widetilde{Z} \widetilde{Z}^\dagger \widetilde{Z} \widetilde{Z}^\dagger,
\\
\pi^z_\mathrm{surface} &= Z,
\end{split}
\label{edge3}
\end{align}
They are defined in such a way that they commute with the bulk Hamiltonian, as well as with all spin operators on neighbouring sites.
Similarly, along a hinge we can construct effective spin degrees of freedom associated with a hinge site and its two nearest neighbors schematically written as
\begin{align} 
\pi^x_\mathrm{hinge}=
X \widetilde{Z} \widetilde{Z}^\dagger,
\quad
\pi^y_\mathrm{hinge} =
X \widetilde{Z} \widetilde{Z}^\dagger,
\quad
\pi^z_\mathrm{hinge}= Z.
\label{edge2}
\end{align}
Finally, at a corner, we may construct operators that commute with the bulk Hamiltonian and all of the surrounding spin operators associated with faces and hinges as
\begin{align} 
\pi^x_\mathrm{corner} = X \widetilde{Z},\quad
\pi^y_\mathrm{corner} = Y \widetilde{Z},\quad
\pi^z_\mathrm{corner} = Z
\end{align}
or with $\widetilde{Z}\leftrightarrow\widetilde{Z}^\dagger$, depending on the corner orientation.

The symmetry $\prod_{i\in a} X_i$ acts as $\prod \pi^x$ on all surface/hinger/corner degrees of freedom. 
However, under the symmetry $\prod_{i \in b} \widetilde{X}$, both surface and hinge degrees of freedom transform trivially, while it acts as $\prod \pi^z_\mathrm{corner}$ on the corners.  
Thus the surface and hinge degrees of freedom are charged only under one of the two generators of the  $\mathbb{Z}_3 \times \mathbb{Z}_3$ symmetry.  
Consequently the surface and hinge spins can be gapped by adding a term $\pi^x_\mathrm{surface}$ or $\pi^x_\mathrm{hinge}$ for each site on the surfaces and hinges; this does not break any symmetry.  
The corners, on the other hand, form a projective representation of $\mathbb{Z}_3 \times \mathbb{Z}_3$. 
The resulting degeneracy cannot be lifted without breaking the symmetry.  
As in the 2D case, such corner modes can be eliminated in the absence of lattice symmetries, by adding a term that effectively moves them together such that they can be locally coupled into singlets; however, $C_3$ rotations about the corner point prevent this.

The reason we focus on $\mathbb{Z}_3\times\mathbb{Z}_3$, rather than $\mathbb{Z}_2\times\mathbb{Z}_2$ is that the latter does not possess gapless corner modes protected by the set of $C_3$ symmetries.  
For example, we may line 1D $\mathbb{Z}_2\times\mathbb{Z}_2$ non-trivial SPTs along both diagonals of each face: this does not break any of our spatial symmetries.  
However, at the corners, we now have three projective representations of $\mathbb{Z}_2\times\mathbb{Z}_2$, along with one more from the corner mode.  
However, as projective representations of $\mathbb{Z}_2\times\mathbb{Z}_2$ have a $\mathbb{Z}_2$ classification, four copies must be trivial, and thus the corner may be gapped out in a symmetric way.

\section{Connection between bosonic and fermionic HOSPT }
\label{sec: fermions}


Current interest in higher-order topology is driven primarily by the possibility of realizing it in fermionic systems\cite{benalcazar2017quantized,schindler2018higher,PhysRevB.97.035138,schindler2017higher,po2017symmetry,2018arXiv180502598T}. This raises the question of whether interacting fermionic HOSPT are related to the bosonic HOSPT studied in the prefious sections. Noninteracting fermionic HOTIs have been characterized using quantized response functions, such as electric multipole moments from Wilson loop spectra~\cite{benalcazar2017quantized} and the magneto-electric coupling~\cite{schindler2018higher}, by studying elementary band representations~\cite{PhysRevB.97.035138,schindler2017higher} and symmetry-indicators in band structures\cite{po2017symmetry}, as well as from the representation theory of Dirac electrons~\cite{2018arXiv180502598T,PhysRevX.7.041069,slager2013space}. 
Most of these concepts cannot be readily translated to interacting systems, which require a fundamentally different approach. 

For conventional SPT states, one way to determine which fermionic systems are stable to interactions is to exploit the equivalence between bosonic phases and a subset of their fermionic counterparts.  
The connection is established by coupling several copies of a non-interacting fermionic SPT phases to a fluctuating bosonic vector field~\cite{you2015bridging,bi2015classification,song2017interaction,you2014symmetry, morimoto2015breakdown,yoshida2015bosonic}.  This bosonic field generates dynamical mass terms for the fermions at the boundary, as well as introducing interactions.
In general, symmetry prevents the vector boson from ordering, and the resulting interacting theory is an SPT (trivial) phase provided  the NL$\sigma$M describing the bosonic degrees of freedom has (does not have) a topological term.  
This approach has been used to study the effect of interactions on the classification of topological insulators and superconductors\cite{bernevig2006quantum,kane2005quantum,ryu2010topological,kitaev2009periodic}, as well as topological crystalline insulators and superconductors \cite{song2017interaction,isobe2015theory}.  

Here, we apply this method to higher-order topological superconductors, where the relevant topological term is a topological Theta term for the {\it boundary} theory, which in turn produces a WZW term at a corner or hinge, as discussed in Secs.~\ref{subsec:qft1} and~\ref{3dm}.
This NL$\sigma$M description  will allow us to answer the following crucial questions about HOSPTs: 
\begin{enumerate}
    \item 
How do interactions affect the classification of HOSPT starting from  fermionic HOTIs? For the conventional SPT, strong interactions can either collapse some $\mathbb{Z}$ classified SPT into a $\mathbb{Z}_n$ classification or generate an interaction enabled SPT state in a symmetry class that is topologically trivial without interactions~\cite{you2015bridging,bi2015classification,song2017interaction,you2014symmetry,wang2014interacting,Senthil2015-tp,khalaf2017symmetry,2014arXiv1406.3032M}.  
This phenomenon has also been studied in topological crystalline insulators, in which the gapless boundary modes are protected by lattice symmetries~\cite{song2017topological,isobe2015theory}.  Because HOSPT phases also require lattice symmetries to protect their gapless boundary modes, the classification of Refs.~\onlinecite{song2017topological,isobe2015theory} also applies to these systems.  Here we make this connection explicit, studying the case where the boundary is gapped leaving only gapless corners or hinges.  In particular, we clarify the classification of interacting fermionic systems for which rotational symmetry protects gapless hinge or corner modes, which has not been discussed explicitly in the context of crystalline SPT phases.
\item
Is there a way to bridge the connection between fermionic and bosonic HOSPT states, and can  the bosonic candidates be obtained from interacting fermion models?  
\end{enumerate}

Our main general result is that, in both 2D and 3D, the following holds for HOSPT phases protected by a spatial symmetry and an internal symmetry $G$: Consider a system for which the noninteracting classification for fermions is $\mathbb{Z}$ while interactions reduce it to $\mathbb{Z}_N$. Then we show that $M$ copies of such a fermionic system with strong inter-copy interaction are akin to a bosonic HOSPT with a $\mathbb{Z}_{N/M}$ classification.

In the process, we will uncover  an alternative route to the bosonic HOSPT models discussed in the previous section: we will show how they can be obtained in systems whose fundamental degrees of freedom are fermionic. 

Our discussion will also highlight another similarity between fermionic HOSPT phases and their bosonic counterparts. In Sec.~\ref{subsec:qft1}, we discussed how bosonic HOSPT phases $C_4$ rotations can be obtained from an SPT with an enlarged symmetry, by breaking part of this symmetry in a particular way.
Similarly, it is believed that most fermionic HOTIs can be obtained from a first order SPT phase with an artificially enlarged symmetry. By breaking the enlarged symmetry, and only focusing on the boundary degrees of freedom, one can obtain the higher-order topological boundary modes at hinges or corners. An example is the construction of Ref.~\onlinecite{schindler2017higher}, which starts with at time-reversal and $C_4$ symmetric 3D TI with Dirac cones. By breaking time-reversal on the surface, while preserving the combination $\mathcal{T}C_4$, one obtains a HOTI with chiral hinge modes protected by this symmetry. We will mostly adopt an analogous procedure in what follows to construct HOSPT phases. It is important to stress that, while the construction is focused on boundary modes, it cannot be realized without the higher-dimensional bulk, since the topological boundary modes generically realize the symmetries in an anomalous way.

\subsection{Reflection symmetry in HOSPT}

In this section, we focus on a HOSPT phase protected by a pair of reflection symmetries $M_x$ and  $M_y$ as well as some internal symmetry $G$ in an interacting  system with fermionic and bosonic degrees of freedom. 
We will show that a HOSPT phase stabilized by the $G \times M_x  \times M_y$ symmetry in $d$ dimensions can be reduced to a usual SPT in $(d-1)$ dimensions with $G$ symmetry alone~\cite{song2017interaction,isobe2015theory,qi2017folding}.

Here and onwards, we will use the notation 
\begin{equation}
\sigma^{ijkl\dots} = \sigma^i \otimes \sigma^j \otimes \sigma^k \otimes \sigma^l \otimes \dots
\end{equation}
with $i,j,k,l\in\{0,1,2,3\}$
to represent tensor products of Pauli matrices,
where $\sigma^0$ is the $2\times2$ unit matrix while $\sigma^{1}$, $\sigma^{2}$, and $\sigma^{3}$ are the $x$, $y$, and $z$ Pauli matrices, respectively.

\subsubsection{HOSPT with $\mathcal{T}$ and reflection symmetry in 2D}
We start with a 2D topological crystalline superconductor (TCSC) with spinless $\mathcal{T}$ and $M_x$ as well as $M_y$ reflection symmetry
\begin{equation}
\begin{split}
 H_{\mathrm{SC}}=
 \int_{\Omega}\mathrm{d}x\,\mathrm{d}y\,
 \chi^T(x,y)\bigl[&i \partial_x \sigma^{30}+i \partial_y\sigma^{13}+m \sigma^{20}\bigr]\chi(x,y),
 \end{split}
\end{equation} 
where $\Omega$ is the area on which the system is supported.
The symmetries are realized 
as
\begin{equation}
    \begin{split}
\mathcal{T}: &\quad\chi(x,y) \rightarrow \mathcal{K}\sigma^{11} \chi(x,y) ,
\\
M_{x}: &\quad\chi(x,y) \rightarrow \sigma^{22}\chi(-x,y),
\\
M_{y}: &\quad\chi(x,y) \rightarrow \sigma^{01}\chi(x,-y).      
    \end{split}
\end{equation}


We now consider an edge parallel to the $x$ axis, which is invariant under $M_x$ symmetry.
The edges is gapless as long as only $m$ is considered.
However, we can include in the  low-energy theory of the edge
\begin{equation}
    H_{\mathrm{edge}}=
    \int\mathrm{d}x\,
    \chi^T(x)\left[i \partial_x \sigma^{3}+f(x)\sigma^{2}\right]\chi(x),
\end{equation}
with the symmetries implemented as
\begin{equation}
\begin{split}
    \mathcal{T}:&\quad \chi \rightarrow \mathcal{K}\sigma^{1} \chi ~\nonumber\\
M_{x}:&\quad\chi(x) \rightarrow \sigma^{1}\chi (-x),
\end{split}    
\end{equation}
a term proportional to $f(x)$ that 
opens a gap almost everywhere (the other mass term $\sigma_1$ is prevented by the particle-hole symmetry). To comply with $M_x$, we choose $f(x)=-f(-x)$ to be an odd function. 

At $x=0$, the fermion mass $f(x)$ changes sign and hence traps a single Majorana zero mode, which we denote by $\chi^0$ at the reflection symmetric point. (We could alternatively have arrived at a Majorana zero mode trapped at a corner by considering  two edges along the two diagonal directions meeting at the reflection symmetric corner.) 

To understand the topological stability of the Majorana mode, we now consider $N$ superimposed copies of this model. 
When uncoupled, each copy contributes one localized Majorana mode.
The symmetries act on each Majorana zero mode at the symmetric point as
\begin{equation}
    \begin{split}
        \mathcal{T}:&\quad\chi^0 \rightarrow \mathcal{K}\chi^0,\\
M_{x}:&\quad\chi^0(x) \rightarrow \chi^0(-x).
\end{split}
\end{equation} 
As a consequence of $\mathcal{T}$, it is not possible to perturbativley lift the degeneracy associated with the $N$ Majorana corner modes with a noninteracting Hamiltonian, i.e., one that is bilinear in the $\chi$ fields.
In addition, one cannot add a purely 1D system to the edge that adds Majoranas with only one mirror eigenvalue at the corners. The simplest nontrivial edge phase transition that respects both $M_x$ and $\mathcal{T}$ would add two Majorana states to the corner, one with each mirror eigenvalue. The difference $(n_+-n_-)$ between the numbers of Majoranas with mirror eigenvalues $+1$ and $-1$, denoted by $n_+$ and $n_-$, respectively, is thus invariant.
We conclude that the type of higher-order TSC constructed here has a noninteracting $\mathbb{Z}$ classification.

When considering interactions, the stability arguments now proceed in close similarity as for the case of end-modes onf a 1D system studied in Ref.~\onlinecite{fidkowski2010effects}.
If $N=2$, the only coupling term between two Majorana modes is $i\chi^0_1 \chi^0_2$, which breaks $\mathcal{T}$. For $N=4$, one can add a symmetry-allowed quartet term $\chi^0_1 \chi^0_2 \chi^0_3 \chi^0_4$, which still retains a two-fold ground state degeneracy. 
Finally, for $N=8$ copies one can locally gap out all the degrees of freedom and obtain a nondegenerate ground state.
Thus, the fermion HOSPT with $\mathcal{T}$ and reflection symmetry has a $\mathbb{Z}_8$ classification.
While the mirror symmetry seems to be irrelevant for this consideration, it is nevertheless important, becasue it prevents that the Majorana modes hybridze with those from another corner. 

Let us return to discuss the  $N=4$ case in more detail. The four Majoranas on each corner cannot be gapped, but their degeneracy can be lifted into a spin-1/2 degree of freedom. This double degeneracy locally transforms as a Kramers doublet under $\mathcal{T}$, a fact that suggests the $N=4$ system is akin to a bosonic HOSPT discussed in Sec.~\ref{2dt}. To make this correspondence explicit, we  add a bulk four-fermion interaction term to the $N=4$ TCSC. We then perform a Hubbard-Stratonovich transformation, which transforms the four-fermion interaction into  fermion bilinear terms coupled to  a fluctuating  O(3) rotor field $\vec{n}=(n_1,n_2,n_3)$,
\begin{equation}
    \begin{split}
 H_{\mathrm{SC}}=
 &\,\int_\Omega \mathrm{d}x\mathrm{d}y\,
 \chi^T(x,y)\Bigl[i \partial_x\sigma^{3000}+i \partial_y\sigma^{1300}+m\sigma^{2000}\\
&
+n_1(x,y)\sigma^{1120}
+n_2(x,y)\sigma^{1132}
\\
&+
n_3(x,y)\sigma^{1112}
+f(x,y)\sigma^{1200}\Bigr]
\chi(x,y),
    \end{split}
\end{equation}
with the symmetries realized as
\begin{equation}
    \begin{split}
\mathcal{T}:&\quad \chi(x,y) \rightarrow \mathcal{K}\sigma^{1100} \chi(x,y),\\
&\quad\vec{n}(x,y)\rightarrow-\vec{n}(x,y) \\
M_{x}:&\quad\chi(x,y) \rightarrow \sigma^{2200}\chi(-x,y),   \\
&\quad\vec{n}(x,y) \rightarrow \vec{n}(-x,y),   \\
M_{y}:&\quad\chi(x,y)  \rightarrow \sigma^{0100}\chi(x,-y), \\
&\quad\vec{n}(x,y) \rightarrow \vec{n}(x,-y),   \\
    \end{split}
\end{equation}
which implies $f(x,y)=-f(-x,y)=-f(x,-y)$ for the Hamiltonian to be mirror symmetric.

The interaction generates three dynamical masses $\vec{n}=(n_1,n_2,n_3)$ for the fermions. When this $O(3)$ rotor is in the ordered phase, the $\mathcal{T}$ symmetry is broken and the fermion acquires a band mass. In the disordered phase, the bulk and boundary are both gapped. If we integrate out the gapped fermion to obtain the effective theory for the $O(3)$ rotor, the bulk theory contains a trivial $O(3)$ NL$\sigma$M with no topological term.

 As we discussed, at the reflection symmetry point on the edge, $x=0$, $f(x)$ changes its sign and thus generates a domain wall.
 If we focus on the dynamics of the $O(3)$ rotor at the edge, provided that $f(x) \ll m$, the edge theory between the reflection symmetric point can be described as a NL$\sigma$M with $\Theta=0$ or $\Theta=2\pi$ depending on the sign of $f(x)$
 \begin{equation}
 \begin{split}
     \mathcal{L}_{\mathrm{edge}}=&\frac{1}{g}(\partial_i \vec{n})^2+\frac{\Theta}{\Omega^2} \epsilon^{ijk} n_i\partial_x n_j \partial_t n_k,\\
\Theta=&\pi\left[1+\mathrm{sgn}\,f(x)\right],
\label{nlsm}
 \end{split}
 \end{equation}
where the symmetries act as
\begin{equation}
    \begin{split}
\mathcal{T}:&\quad \vec{n}(x,t)\rightarrow-\vec{n}(x,-t) \\
M_{x}:&\quad\vec{n}(x,t)  \rightarrow \vec{n}(-x,t).   
    \end{split}
\end{equation}
Due to $\mathcal{T}$ and reflection symmetry constraint, there is no way remove such a Theta term by polarizing the vector boson, unless we break the symmetry.

Edges along the diagonal/off-diagonal direction, $x=\pm y$, each exhibit a NL$\sigma$M with either $\Theta=0$ or $\Theta=2\pi$. The reflection symmetric corner is the domain wall between $\Theta=2\pi$ and $\Theta=0$. 
The corner supports a $(0+1)d$ $O(3)_1$ WZW term which exactly incorporates a spin-1/2 degree of freedom. 
As a result, these four copies of higher-order TSC manifest the same bulk-boundary correspondence as the bosonic lattice model discussed in Sec.~\ref{2dt}---a HOSPT with $\mathcal{T}$ and reflection symmetry.

\subsubsection{HOSPT with $\mathbb{Z}_2$ and reflection symmetry in 3D}

Now we move on to discuss 3D second order SPT phases with gapless hinge modes. Such phases, with chiral or helical hinge modes, were first discussed for noninteracting fermions in Refs.~\onlinecite{schindler2017higher,langbehn2017reflection,song2017d,song2017interaction,isobe2015theory}. 
Here, we discuss a TCSC in 3D 
defined by the Hamiltonian
\begin{equation}
\begin{split}
        H_{\mathrm{SC}}=
 &\,\int_\Omega \mathrm{d}x\mathrm{d}y\mathrm{d}z\,
 \chi^T(x,y,z)\Bigl[
 i \partial_x \sigma^{330}+
 i \partial_y \sigma^{100}\\
&
+
 i \partial_z \sigma^{310}+M\sigma^{200}
\Bigr]
\chi(x,y,z).
\end{split}
\label{eq: Hamiltonian 3D TSC}
\end{equation} 
We will show that this system has hinge Majorana states protected by a local $\mathbb{Z}_2$ symmetry if in addition the hinge is invariant under one of the mirror symmetries $M_x$ or $M_y$. The symmetry actions are defined as follows
\begin{equation}
    \begin{split}
 \mathbb{Z}_2:&\quad\chi(x,y,z)  \rightarrow \sigma^{001}\chi(x,y,z),  \\      
M_{x}:&\quad\chi(x,y,z) \rightarrow \sigma^{011}\chi(-x,y,z),  \\
M_{y}:&\quad\chi(x,y,z) \rightarrow \sigma^{221}\chi(x,-y,z). 
    \end{split}
\end{equation}
The system and vacuum are are differentiated by the sign of the mass $M$, which without loss of generality we can choose to be $M>0$ in the vacuum and $M<0$ in the TCSC. 

We now solve for the hinge mode between surfaces parallel to the $z$ axis and demonstrate its topological stability.
For concreteness, consider a surface parallel to the $x$-$z$ plane. The surface theory induced by the sign change of $M$ is a $4\times4$ massless Dirac equation
\begin{equation}\label{Eq:3DRefFer}
    \begin{split} 
H_{\text{$x$-$z$}}=
\int \mathrm{d}x\mathrm{d}z\,
 \tilde{\chi}^T(x,z)\Bigl[
 i \partial_x \sigma^{30}
 +i \partial_z \sigma^{10}
\Bigr]
\tilde{\chi}(x,z)
    \end{split}
\end{equation} 
with the symmetries
\begin{equation}
\begin{split}
\mathbb{Z}_2:&\quad\tilde{\chi}(x,z)  \rightarrow \sigma^{01}\tilde{\chi}(x,z),  \\
M_{x}:&\quad\tilde{\chi}(x,z) \rightarrow \sigma^{11}\tilde{\chi}(-x,z).
\end{split}
\end{equation}
There are only two mass terms that can be added to this surface Hamiltonian and are $\mathbb{Z}_2$ symmetric:  $\sigma^{21}$ and $\sigma^{20}$. Both of these masses are odd under the $M_x$ mirror symmetry. A term in the surface Hamiltonian that includes these two mass terms,
\begin{equation}
f_1(x,z)\sigma^{21} 
+
f_2(x,z)\sigma^{20} 
\end{equation}
would have a domain wall at $x=0$, since to comply with mirror symmetry, we have to impose
\begin{equation}
    f_1(x,z)=-f_1(-x,z),\qquad
    f_2(x,z)=-f_2(-x,z).
\end{equation}
The domain wall binds a pair of 1D helical modes propagating along the $z$ direction. 
(1) If $f_1$ dominates, we obtain two 
\emph{counter-propagating} 
modes with the \emph{same} $M_x$ eigenvalue, but \emph{opposite} $\mathbb{Z}_2$ eigenvalues.
(2) If $f_2$ dominates, we obtain two \emph{co-propagating} modes with 
\emph{opposite} $M_x$ eigenvalue, and \emph{opposite} $\mathbb{Z}_2$.
This already points to a rich topological classification of this higher-order TSC. We can now imagine introducing a kink (hinge) in the surface that runs along the $x=0$ line, while maintaining the mirror symmetry. The domain wall mode would then become the hinge mode, while the surfaces on either side of the kink become side surfaces of the system.

The two cases (1) and (2) can be shown to be consistent with one another. Note that a purely 2D phase exists that complies with $M_x$ and $\mathbb{Z}_2$ symmetry and adds two co-propagating modes in one $\mathbb{Z}_2$ subspace with \emph{opposite} $M_x$ eigenvalue~\cite{langbehn2017reflection,schindler2017higher}.

We denote with $n_{\lambda, \rho}$ the net number (up- minus down-movers) of modes in the $\mathbb{Z}_2$-subspace $\lambda=\pm1$ with mirror eigenvalue $\rho=\pm1$. We find that $n_{\lambda+}-n_{\lambda-}$ is invariant against symmetry-preserving surface manipulations for both $\lambda=\pm1$.
This represents a $\mathbb{Z}\times\mathbb{Z}$ classification of robust hinge modes in the absence of interactions. The Hamiltonian~\eqref{eq: Hamiltonian 3D TSC}
 is
a representative system with $n_{++}-n_{+-}=-(n_{-+}-n_{--})=1$, while we do not provide an explicit example of the second generator of this class of topological states here. 

We  now argue that the classification reduces to $\mathbb{Z}\times\mathbb{Z}_8$ if interactions are included.
 For concreteness, consider the hinge states from case (1).
There are two anti-commuting mass matrices $\sigma^{21}$ and  $\sigma^{23}$ on the $x$-$z$ surface. They are odd under the $M_x$ and $\mathbb{Z}_2$ symmetry, respectively. 
Adding these masses clearly breaks symmetry; however following the approach of~\cite{you2014symmetry,morimoto2015breakdown,song2017interaction} we can imagine adding such mass terms and then restoring symmetry by proliferating topological defects.  In the present case the two masses may form a vortex. However, proliferating such vortices necessarily creates a gapless surface, since each vortex core contains a gapless fermionic mode.  Thus this gapless surface is robust to interactions.
If we take two copies of such a  surface, there are three such anti-commuting mass matrices $\sigma^{213},\sigma^{233},\sigma^{203}$. In this case the relevant defect is a monopole in spacetime, which carries zero modes~\cite{song2017interaction}.
If we take four copies of such surface, there are five anti-commuting mass matrices forming a WZW defect. Such a surface theory can be mapped to the surface of a bosonic HOSPT with $\mathbb{Z}_2$ classification.
Once we take eight copies of the model~(\ref{Eq:3DRefFer}), the side surface  can be fully gapped. 

Returning to the more general case, we see that an interacting HOSPT with $\mathbb{Z}_2$ and reflection symmetry in 3D has a $\mathbb{Z}\times\mathbb{Z}_8$ classification, where the factor $\mathbb{Z}$ is related to the net chirality (per mirror subspace), while the factor $\mathbb{Z}_8$ is related to the $\mathbb{Z}_2$-graded chirality of the hinge modes (per mirror subspace).

Let us return to the case of four copies of the gapless hinge, where interactions mix four pairs of counter-propagating Majorana modes into a gapless boson mode [case (1) above]. To verify this, we take four copies of such a fermionic theory and couple them via a fluctuating $O(4)$ rotor $\vec{n}=(n_1,n_2,n_3,n_4)$,
\begin{equation}
    \begin{split}
 H_{\mathrm{SC}}=
 &\,\int_\Omega \mathrm{d}x\mathrm{d}y\mathrm{d}z\,
 \chi^T(x,y,z)\Bigl[
  i \partial_x\sigma^{33000}+i \partial_y \sigma^{10000}
  \\
&\quad  +i \partial_z\sigma^{31000}
+M\sigma^{20000}
+f(x,y,z)\sigma^{32100}\\
&\quad+n_1\sigma^{32212}+n_2\sigma^{32220}+n_3\sigma^{32232}\\
&\quad +n_4\sigma^{32300}\Bigr] \chi(x,y,z)  
    \end{split}
\end{equation}
with the symmetry action
\begin{equation}
    \begin{split}
\mathbb{Z}_2:&\quad
\chi(x,y,z) \rightarrow \sigma^{00100}\chi (x,y,z),\\
&\quad
\vec{n}(x,y,z) \rightarrow -\vec{n} (x,y,z),
\\
M_{x}:&\quad\chi(x,y,z) \rightarrow \sigma^{01100}\chi(-x,y,z), 
\\
M_{y}:&\quad\chi(x,y,z) \rightarrow \sigma^{22100}\chi(x,-y,z).    
    \end{split}
\end{equation} 
The corresponding fermionic theory is fully gapped but leaves a gapless bosonic mode at the hinge.  If we integrate out the gapped fermionic degrees of freedom, the corresponding bosonic theory is trivial in the bulk. However, on the side surface, there exists a NL$\sigma$M with $\Theta=0$ or $\Theta=2\pi$ depending on the sign of $f_1(x,z)$, that is,
\begin{equation}
    \begin{split}
\mathcal{L}_{\mathrm{edge}}=&
\frac{1}{g}(\partial_i \vec{n})^2+\frac{\Theta}{\Omega^3} \epsilon^{ijkl}  n_i\partial_x n_j \partial_t n_k\partial_z n_l
\\
\Theta=&\pi\left[1+\mathrm{sgn}\,f(x,z)\right].
    \end{split}
\end{equation}
The symmetry transformations are given by
\begin{equation}
    \begin{split}
        \mathbb{Z}_2:&\quad \vec{n}(x,z) \rightarrow -\vec{n}(x,z) \\
M_{x}:&\quad\vec{n}(x,z) \rightarrow \vec{n}(-x,z),
    \end{split}
\end{equation}
and $f(x,z)=-f(-x,z)$.

At $x=0$, there is a domain wall interfacing $\Theta=2\pi$ and $\Theta=0$ which contains $O(4)_1$ WZW term, marking the hinge between two side surfaces. Such an $O(4)_1$ WZW term can be mapped onto an $SU(2)_1$ WZW theory akin to the Levin-Gu edge state. 
As a result, the hinge at the reflection symmetric line supports gapless modes which cannot be trivially gapped unless we break the $\mathbb{Z}_2$ symmetry. 
This provides a fermionic construction for the bosonic lattice model from Sec.~\ref{3dz}.


\subsection{HOSPT protected by $C_4$ rotation and $G$ symmetry}


In Sec.~\ref{subsec:qft1}, we showed how a NL$\sigma$M description~\cite{bi2014anyon,you2015bridging,bi2015classification,you2014symmetry} can be used to relate a bosonic SPT in $d$ spacetime dimensions protected by $\mathbb{Z}_2$ and $G$ symmetry to a HOSPT protected by $G$ symmetry and $C_4$ rotations, provided that the original SPT admits a decorated domain wall construction \cite{chen2014symmetry}.  This is done by  polarizing one of the components $n_i$ of the vector field in a spatially dependent way, which breaks both $\mathbb{Z}_2$ and $C_4$ individually, but preserves the combination $\mathbb{Z}_2 C_4$.  The $C_4$ symmetric corner/hinge connecting two boundary components then becomes a domain wall for $n_i$, which supports a WZW theory in $(d-2)$ dimensions.

We now consider the analogue of this approach for fermionic systems, and show how this picture is connected to our earlier discussion of bosonic HOSPT.  The connection between interacting fermionic and bosonic SPT's is well-understood~\cite{you2015bridging,you2014symmetry}. Here, we show how a similar connection applies to certain higher-order topological phases.   Specifically, we will show how a 2D topological superconductor, with gapless boundary modes protected by $\mathbb{Z}_2\times G$ symmetry, can be used to construct a fermionic HOSPT with gapless corner modes protected by $C_4 \times G$ symmetry.  We also use general field theoretic arguments to show that, since multiple copies of the fermionic SPT are equivalent to a bosonic SPT, it follows that multiple copies of the fermionic HOSPT are equivalent to a bosonic HOSPT of the type described in Sec.~\ref{sec: bosonic}.

\subsubsection{HOSPT protected by $ C_4 \times \mathcal{T}$ symmetry in 2D}

As a concrete example, we begin with a 2D topological superconductor
\begin{equation}
 \label{Eq:TSC1}
    \begin{split}
&H_{\mathrm{SC}}=
\int_{\Omega}\mathrm{d}x\,\mathrm{d}y\,
 \chi^T(x,y)\bigl[
\chi^T(i\partial_x \sigma^{300}+i\partial_y\sigma^{130}
\\
&\qquad \qquad+m \sigma^{200} \bigr]\chi(x,y)\\ 
    \end{split}
\end{equation}
with a $\mathbb{Z}_2$, $\mathcal{T}$, and $C_4$ symmetry defined as
\begin{equation}
    \begin{split}
\mathcal{T}:&\quad\chi(x,y) \rightarrow   \mathcal{K} \sigma^{110}  \chi(x,y) \\
C_4:&\quad \chi(x,y) \rightarrow e^{i \frac{\pi}{4} \sigma^{230}}  \chi(-y,x) , \\
\mathbb{Z}_2:&\quad \chi(x,y) \rightarrow \sigma^{032} \chi(x,y).
    \end{split}
\end{equation}

The mass $m$ ensures that the fermions are gapped in the bulk, but as above leaves gapless modes at the boundary.  In this case, the boundary hosts two pairs of helical Majorana modes, protected by $\mathbb{Z}_2$ and $ \mathcal{T}$ symmetry.  
Following the logic of Refs.~\onlinecite{isobe2015theory,you2015bridging}, one can show that with four copies of this model, interactions can generate dynamical mass terms at the boundary that preserve all symmetries; therefore 
such an SPT has a $\mathbb{Z}_4$ classification. 

As we did in the bosonic case, we may take this fermionic model and add a mass term
that breaks $\mathbb{Z}_2$ and rotational symmetries individually, but preserves their combination.  To do this, we may take
\begin{equation}
    \Delta H =
    \int_{\Omega}\mathrm{d}x\,\mathrm{d}y\,
    f(x,y) \chi^T(x,y)  \sigma^{120}  \chi(x,y),
\end{equation}
where $f(x,y) = - f(y, -x)$, for example through $f(x,y)=\cos(x)-\cos(y)$.  It is easy to check that this term is odd under both symmetries individually, but even under the combination of $\mathbb{Z}_2$ and a $C_4$ rotation.  
This creates a new rotation symmetry $\tilde{C}_4 = \mathbb{Z}_2 C_4$, generated by the product of the $\mathbb{Z}_2$ and $C_4$ generators.
Since we have broken the $\mathbb{Z}_2$ symmetry necessary to protect the SPT phase, this perturbation gaps the $x$ and $y$ edges of our system.  However, the function $f$ changes sign on the diagonals $x = \pm y$. Each sign change binds a pair of Majorana zero modes, whose two-fold degeneracy is protected by $\mathcal{T}$ symmetry.  Thus we obtain a higher-order TSC in essentially the same way as we obtained bosonic HOSPT from their SPT cousins in Sec.~\ref{sec: bosonic}.  
As was the case there, this approach is easily generalized to other symmetry groups of the form $\mathbb{Z}_2 \times G$.

Next, we show that two copies of the model~(\ref{Eq:TSC1}) are equivalent to the bosonic model discussed in Sec.~\ref{2dt}.  
There are two ways to do this: as in the previous discussions, we could couple two copies of the higher-order TSC to directly obtain the bosonic HOSPT. Instead, here we will show that two coupled copies of the model~(\ref{Eq:TSC1}) yield a bosonic SPT, and then introduce an appropriate $\mathbb{Z}_2$ breaking mass term into our bosonic model to obtain the bosonic HOSPT.  
This illustrates the fact that the relationship between SPTs and HOSPTs is analogous for bosonic and fermionic systems.

To see this, we couple two copies of the fermion theory in Eq.~(\ref{Eq:TSC1}) with a vector boson field $(n_1,n_2,n_3, n_4)$,
\begin{equation}
    \begin{split}
H_{\mathrm{SC}}=&
\int_{\Omega}\mathrm{d}x\,\mathrm{d}y\,
 \chi^T(x,y)\bigl[
i\partial_x \sigma^{3000}+i\partial_y \sigma^{1300}+m \sigma^{2000}  \\
& +n_1 \sigma^{1120}+n_2 \sigma^{1132}+n_3 \sigma^{1112}+n_4 \sigma^{1200}\bigr] \chi(x,y) 
    \end{split}
\end{equation}
with the symmetries acting as
\begin{equation}
    \begin{split}
\mathcal{T}:&\quad  \chi(x,y) \rightarrow \mathcal{K} \sigma^{1100} \chi(x,y) ,
\\
&\quad(n_1,n_2,n_3,n_4)(x,y) \rightarrow (-n_1,-n_2,-n_3,n_4)(x,y)   \\
C_4:&\quad \chi(x,y)  \rightarrow  e^{i  \frac{\pi}{4} \sigma^{2300}}\chi(-y,x) , \\
&\quad(n_1,n_2,n_3,n_4)(x,y) \rightarrow (n_1,n_2,n_3,n_4)(-y,x)   \\
\mathbb{Z}_2:&\quad  \chi(x,y) \rightarrow \sigma^{0320}\chi(x,y) ,\\
&\quad (n_1,n_2,n_3,n_4)(x,y) \rightarrow (n_1,n_2,-n_3,-n_4)(x,y).
    \end{split}
\end{equation}
Integrating out the gapped fermions in the bulk yields an $O(4)$ NL$\sigma$M. Due to the nontrivial Berry curvature in the fermion band, this NL$\sigma$M has a topological Theta term with $\Theta = 2\pi$~\cite{you2015bridging}.

To obtain an HOSPT, we take $\langle n_4\rangle=f(x,y)$, 
with $f(x,y)=-f(y,-x)$ so that the theory still preserves the combination of $\mathbb{Z}_2$ and $C_4$ symmetry, $\tilde{C}_4=\mathbb{Z}_2C_4$. 
Our model then becomes
\begin{equation}
    \begin{split}
H_{\mathrm{SC}}=&
\int_{\Omega}\mathrm{d}x\,\mathrm{d}y\,
 \chi^T(x,y)\bigl[
i\partial_x \sigma^{3000}+i\partial_y \sigma^{1300}+m \sigma^{2000}  \\
& +n_1 \sigma^{1120}+n_2 \sigma^{1132}+n_3 \sigma^{1112}+f \sigma^{1200}\bigr] \chi(x,y) 
    \end{split}
\end{equation}
with $\tilde{C}_4$ acting as
\begin{equation}
    \begin{split}
\tilde{C}_4:&\quad \chi(x,y) \rightarrow e^{i  \frac{\pi}{4} \sigma^{2300}} \sigma^{0320} \chi(-y,x), \\
&\quad (n_1,n_2,n_3,n_4)(x,y) \rightarrow (n_1,n_2,-n_3,-n_4)(-y,x).
    \end{split}
\end{equation}
Integrating over the fermion leads to an effective theory for the vector boson which is gapped in the bulk and on the edge.   The resulting effective theory of the edge is a NL$\sigma$M with $\Theta=2\pi$ or $\Theta=0$ term depending on the sign of $f$
\begin{equation}
    \begin{split}
\mathcal{L}_{\mathrm{edge}}=&\frac{1}{g}(\partial_i \vec{n})^2+\frac{\Theta}{\Omega^2} \epsilon^{ijk} n_i\partial_x n_j \partial_t n_k\\
\Theta=&\pi\left[1+\text{sgn}\,f\right]
    \end{split}
\end{equation}
with time-reversal acting as
\begin{equation}
\mathcal{T}: \vec{n}(x,t)\rightarrow-\vec{n} (x,-t). 
\end{equation}
In this description, the corner appears as a domain wall between regions where $\Theta=2\pi$ and $\Theta=0$.  Note, however, that as discussed in Sec.~\ref{subsec:qft1}, this effective field theory of the edge is equivalent to one in which all boundaries have the same value of $\Theta$, and the bulk contributes an extra topological term at the corners.  In either case the result is an $O(3)_1$ WZW term at each corner protected by $\mathcal{T}$ symmetry. This exactly reproduces the bosonic lattice model SPT from Sec.~\ref{2dt} with protected corner modes.  

The approach sketched here can be replicated in a straightforward way for other symmetry groups $G$. 

\section{Symmetry gauging for $C_4$ symmetric HOSPT phases}\label{sec: gauging}

In this final section we extend the scope of this work by considering phases with intrinsic topological order from the perspective of higher-order topology~\cite{dwivedi2018majorana}, which are generated from HOSPT phases. The types of topological order that we discuss can be characterized by the universal properties of their low-energy excitations. They can be constructed from SPT phases by promoting the global symmetry to a local one that is enforced by a dynamical gauge field, a procedure termed `gauging the symmetry'. 
The nontriviality of the parent  SPT phase is then reflected in the corresponding deconfined gauge theory~\cite{bi2014anyon,levin2012braiding,thorngren2018gauging,wang2014braiding,jian2014layer,jiang2014generalized,chen2015anomalous}, in particular through the propertes of its elementary excitations such as flux lines or quasiparticles. If one gauges a spatial symmetry, which is necessarily part of the symmetry group of HOSPTs, the procedure involves coupling the degrees of freedom to the background geometry~\cite{you2016geometry,thorngren2018gauging}. The dynamical excitations of the gauged theory can then be probed via their response to geometrical defects. For instance, dislocations and disclinations can act as a symmetry flux line for excitations braided around them. In this section, we discuss such geometrical lattice defects in our models of HOSPT phases, giving an explicit construction of the gapless boundary modes.  We also comment on the distinctive topological signatures of the gauged theory.

\subsection{Construction of the disclination in HOSPT}

\begin{figure}[t]
  \centering
      \includegraphics[width=0.35\textwidth]{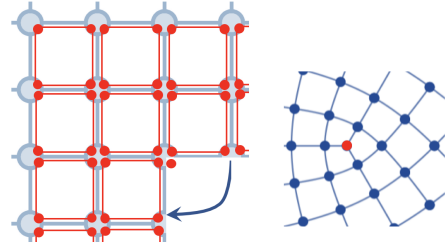}
  \caption{A $\pi/2$ disclination point is created by cutting out a quadrant of the square lattice and reconnecting the edges. Inside the disclination core, there exists a spin-1/2 zero mode protected by $\mathcal{T}$.}
  \label{dis2}
\end{figure}

We first discuss disclinations and their associated bound states in the parent HOSPT phases.
We start with the example of the 2D HOSPT phase protected by $\mathcal{T} \times C_4$ symmetry from Sec.~\ref{2dt}. A $\pi/2$ disclination can be understood as the gauge flux for $C_4$ rotation symmetry.  
On the square or cubic lattices with $C_4$ symmetry, the $\pi/2$ disclination can be generated by taking away the quadrant of sites covered by the $\pi/2$  angle and connect the residual boundary as shown in Fig.~\ref{dis2}. The $\pi/2$ angle that is removed includes a corner on the boundary with a symmetry protected spin-1/2 zero mode. As a result, the disclination core contains an unpaired spin-1/2 zero mode, which is necessarily gapless.

\begin{figure}[t]
  \centering
      \includegraphics[width=0.25\textwidth]{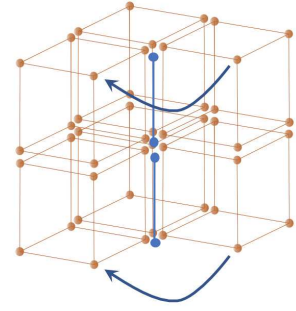}
  \caption{A $\pi/2$ disclination line is created by cutting out the quadrant of the cubic lattice and reconnect the side faces. Inside the disclination line, there exists a dispersing bosonic gapless degree of freedom akin the the hinge mode.}
  \label{dis}
\end{figure}

A similar construction applies to the 3D second order SPT we introduced in Sec.~\ref{3dz}. Removing a quadrant of the cubic lattice as shown in Fig.~\ref{dis}, the surfaces parallel to the $x$-$z$ plane and the one parallel to the $y$-$z$ plane are gapped and can be joined. The corner hinge supports a gapless mode which is preserved as the disclination line is formed by reconnecting the gapped surfaces.

A phenomenological long-wavelength field theory description for such a gapless mode inside the disclination\cite{thorngren2018gauging} is
\begin{equation}
\begin{split} 
\mathcal{L}_{2D}=&\frac{2 (\partial_x \omega_y-\partial_y \omega_x)}{\pi} \frac{2\pi}{\Omega^2} \int_0^1 du \epsilon^{ijk}  n_i \partial_u n_j \partial_t n_k \\
\mathcal{L}_{3D}=&\frac{2 (\partial_x \omega_y-\partial_y \omega_x)}{\pi}\frac{2\pi}{\Omega^3} \int_0^1 du \epsilon^{ijkl}  n_i\partial_z n_j \partial_t n_k \partial_u n_l\, ,
\end{split}
\end{equation}
where $\omega_i$, $i=x,y$, is the spin connection field on the $x$-$y$ plane. 
The curl of spin connection $(\partial_x \omega_y-\partial_y \omega_x)$ gives the $\pi/2$ disclination which is exactly the symmetry flux of $C_4$ rotation. The coupling between the disclination and the lower dimensional WZW term indicates the existence of an $O(3)/O(4)$ WZW theory with bound to the disclination point/line.

\subsection{Gauging the HOSPT phase}
For a conventional SPT states with internal unitary symmetry, gauging the theory results in a deconfined gauge theory containing nontrivial 3-loop braiding statistics between symmetry fluxes~\cite{jian2014layer,wang2014interacting}. We now discuss how a similar phenomenon arises in HOSPT phases protected by discrete rotations and and an internal symmetry $G$. Concretely, we consider the example of a 3D HOSPT with $C_4$ and $G$ symmetry, where we gauge the $G$ symmetry in the presence of lattice disclinations (which for our purposes can be viewed as gauging both symmetries, as the energy associated with the disclination will not play a role). We will show that the corresponding gauge theory, together with the geometry metric, has nontrivial 3-loop braiding statistics. 
Specifically, we will show that taking one gauge flux loop through another can lead to nontrivial 3- loop braiding statistics when both flux loops simultaneously enclose a disclination line. 

Before demonstrating this, we first review the three loop statistics in 3D $\mathbb{Z}_2 \times \mathbb{Z}_2$ SPT with a NL$\sigma$M description.
Consider the Lagrangian density
\begin{equation}
\mathcal{L}=\frac{1}{g}(\partial_i \vec{n})^2+\frac{2\pi}{\Omega^3} \epsilon^{ijkl}  n_i\partial_x n_j \partial_t n_k\partial_z n_l
\end{equation}
together with the two $\mathbb{Z}_2$ symmetries
\begin{equation}
\begin{split} 
&\mathbb{Z}_2^a: (n_1,n_2, n_3,n_4,n_5) \rightarrow (-n_1,-n_2,-n_3,-n_4,n_5), \\
&\mathbb{Z}_2^b: (n_1,n_2, n_3,n_4,n_5) \rightarrow (n_1,n_2,n_3,-n_4,-n_5).
\end{split}
\end{equation}
This SPT has a decorated domain wall structure. We can add an anisotropy term $n_2^2+n_3^2+n_4^2$ that enforces $n_2=n_3=n_4=0$. The domain wall membrane of $n_5$ contains an embedded 2D Levin-Gu-type SPT which consists of the proliferation of domain wall loops of $n_1$. 
Due to the Theta term, adding a domain wall loop of $n_1$ inside the domain wall membrane (of $n_5$) would introduce an additional minus sign to the wave function. 
Gauging the $\mathbb{Z}_2^b$ symmetry effectively allows these domain walls to end on $\mathbb{Z}_2^b$ vortex loops.  
The result is a domain plane decorated with a 2D Levin-Gu model, whose gapless boundary lies on a $\mathbb{Z}_2^b$ vortex loop. 
A vortex loop of 
 $\mathbb{Z}_2^a$ that pierces this Levin-Gu plane then creates a $\mathbb{Z}_2$ flux in this 2D Levin-Gu system.  The three-loop braiding process where two flux loops for $\mathbb{Z}_2^a$, penetrated by the flux loop for $\mathbb{Z}_2^b$, braid with each other is thus akin to braiding a pair of $\mathbb{Z}_2$ fluxes in the Levin-Gu model. 
 Since these fluxes have non-trivial statistics in 2D,  the 3-loop braiding results in a net phase of $\pi$~\cite{levin2012braiding,chen2011two,wang2014interacting,you2016decorated,you2016stripe}. (This implies that braiding a $\mathbb{Z}_2^a$ flux loop with a $\mathbb{Z}_2^b$ flux loop, with a $\mathbb{Z}_2^b$ base loop, gives a phase of $- \pi/2$, indicating a Berry phase that cannot be obtained by attaching charges to the relevant flux loops~\cite{wang2014interacting}.)

\begin{figure}[t]
  \centering
      \includegraphics[width=0.5\textwidth]{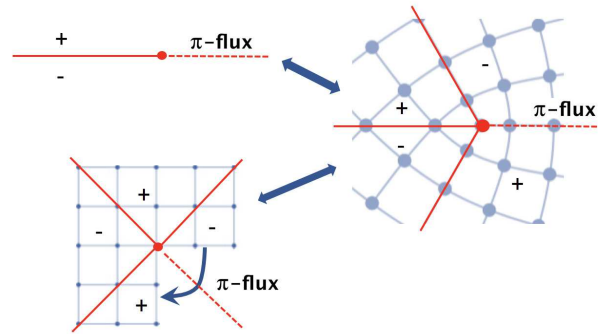}
  \caption{Disclination as a $\pi$-flux defect. Top Left: The $\mathbb{Z}_2$ gauge flux connection correspoding the open domain wall.
  Bottom Left:  After removing a quadrant of the cubic lattice, the two boundaries are connected with twisted boundary condition $n_5(x,y)=-n_5(y,-x)$.
  Right: The resulting disclination becomes the $\pi$ flux for the $C^z_4 \mathbb{Z}_2^b$ symmetry, connecting the open domain wall.}
  \label{figflux}
\end{figure}

Starting from this phase, we construct a higher-order topology by breaking the $\mathbb{Z}_2^b$ symmetry via polarizing $n_5$ with the pattern shown in Fig.~\ref{c4}. The corresponding theory breaks $\mathbb{Z}_2^b$ symmetry but preserves the combination of $C^z_4$ and $\mathbb{Z}_2^b$ symmetry, which becomes the new $\tilde{C}^z_4$ symmetry. Between the surfaces parallel to the $x$-$z$ and $y$-$z$ side planes, the sign of $n_5$ changes at the hinge to respect the new $\tilde{C}^z_4$ symmetry.  
The $\pi$ gauge flux for this $\tilde{C}^z_4$ involves the regular $\pi$ flux for $\mathbb{Z}_2^b$, which creates an open domain wall for $n_5$, along with a $\pi/2$ rotation flux, which is exactly the disclination line. In Fig.~\ref{figflux}, we depict the general procedure to create a $\pi$ gauge flux for the combined $C^z_4$ and $\mathbb{Z}_2^b$ symmetry~\cite{bi2014anyon,xu2013wave,cheng2016series,jian2014layer,agterberg2008dislocations}. One cuts out a $\pi/2$  corner and enforces $n_5(x,y)=-n_5(y,-x)$ as the boundary condition to connect the two side faces after the cut. The corresponding disclination becomes the $\pi$ gauge flux loop for the combined $C^z_4$ and $\mathbb{Z}_2^b$ symmetry, and serves as the open domain wall boundary where $n_5$ changes sign. 

\begin{figure}[t]
  \centering
      \includegraphics[width=0.2\textwidth]{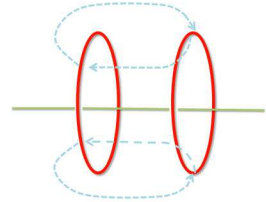}
  \caption{Three loop statistics process: The two $\mathbb{Z}_2$ flux loop (red) wind around each other with the penetration of a $\pi/2$ disclination loop(green). }
  \label{threeloop}
\end{figure}

If we further gauge the $\mathbb{Z}_2^a$ symmetry, the two $\pi$ flux loops of $\mathbb{Z}_2^a$ have semionic statistics when both of them are penetrated by the disclination loop (see Fig.~\ref{threeloop}). This $\pi$ statistical phase generated by the three-loop braiding between $\mathbb{Z}_2$ flux and rotation flux can be seen as a hallmark of the nontrivial topological structure of the parent HOSPT phase~\cite{huang2017building,cho2015topological}.

\section{Summary}\label{sec: summary}

In this work, we studied higher-order topology in interacting SPT phases, presenting both exactly soluble lattice models and an effective field theory. We focussed on $C_4$ and mirror as representative spatial symmetries to stabilize HOSPT phases. As a foundation for our analysis, we construct a higher-order bulk boundary correspondence similar to the one known from noninteracting fermionic HOTIs: corner states in 2D, as well as hinge and corner states in 3D. Beyond that, we studied aspects unique to the strongly interacting SPT setting: (i) bosonic phases, (ii) phases with a $\mathbb{Z}_m\times\mathbb{Z}_n$ symmetry, for general $m$ and $n$, (iii) fermionic HOSPTs with reduced classification compared to the noninteracting case, and (iv) gauged HOSPTs with nontrivial loop braiding statistics. Our key results include a topological field theory describing these HOSPT phases, wich reveals their relationship to conventional SPT phases.  
Based on our results, various directions for future studies present themselves, including a more systematic exploration of 
higher order topologically ordered phases, expressions for bulk topological invariants of HOSPTs, and an understanding for the topological response functions of such phases.

Our work also raises several interesting questions.  For example, for non-interacting HOTI phases, the classification of phases with $C_4$ rotation and reflection symmetries is known to be different. It would be interesting to explore examples where the difference between reflection and rotation symmetries is more manifest.

\textit{ Note added} ---
During completion of this work, we became aware of another paper by  O. Dubinkin and T. Hughes~\cite{2018arXiv180709781D} which has some overlap with our results.

\begin{acknowledgments}
YY is supported by PCTS Fellowship at Princeton University. FJB is grateful for the financial support of NSF-DMR 1352271 and the Sloan Foundation FG-2015-65927. This work(YY,FJB) was performed in part at Aspen Center for Physics, which is supported by National Science Foundation grant PHY-1607611.
TN acknowledges support from the Swiss National Science Foundation (grant number: 200021\_169061) and from the European Union’s Horizon 2020 research and innovation program (ERC-StG-Neupert-757867-PARATOP).

\end{acknowledgments}

\appendix

\section{Fermion version of 2nd order SPT with $\mathbb{Z}_2$ and $C_4$ symmetry}
Due to the growing interest on fermionic SPT phases, in this section we extend our 3D CZX model to interacting fermion systems~\cite{chen2011two}. 
The construction, together with the projection Hamiltonian is totally inherited from the spin model in Sec.~\ref{3dz}, while the spin state $| 0 \rangle, | 1 \rangle$ states are replaced by the spinless fermion occupancy. Each site contains eight fermion modes and therefore the system always respects the fermion parity symmetry. Meanwhile, we can define the $\mathbb{Z}_2$ symmetry as,
\begin{equation}
    \begin{split}
U_{czx}=&\,U_x U_{cz},\\
U_x=&\,\prod_{i=1}^8 (c_i^{\dagger}+c_i),\\
U_{cz}=&\,\text{CZ}_{13}~\text{CZ}_{24}~\text{CZ}_{57}~\text{CZ}_{68} ~\text{CZ}_{56}~\text{CZ}_{12}\\
&\times\,\text{CZ}_{34}~\text{CZ}_{78}~\text{CZ}_{15}~\text{CZ}_{26}~\text{CZ}_{37}~\text{CZ}_{48},\\
\text{CZ}_{ij}=&(1-2c_i^{\dagger}c_ic_j^{\dagger}c_j).
    \end{split}
\end{equation}

The Hamiltonian in the bulk involves the local projection operator on 8 fermions at the corner of the cube,
\begin{align} 
&H_{\text{cube}}=-\prod_{i \in \text{cube}} c_i^{\dagger}-\prod_{i \in \text{cube}} c_i.
\end{align}
Meanwhile, the surface could be gapped by adding a local projection operator on four fermions at the corner of the surface plaquette,
\begin{align} 
&H_{\text{plaquette}}=-\prod_{i \in \text{plaq}} c_i^{\dagger}-\prod_{i \in \text{plaq}} c_i.
\end{align}

When it comes to the hinge between two surfaces, the edge Hamiltonian involves the two-fermion interaction on each hinge bond
\begin{align} 
&H_{\text{bond}}=-c_i^{\dagger}c_ic_j^{\dagger}c_j-c_ic_i^{\dagger}c_jc_j^{\dagger}.
\end{align}
This interaction favors even fermion parity on each bond, and leaves the hinge with $2^N$ degeneracy, where $N$ is the number of hinge bonds.

\section{Bulk field theory for 3D HOSPT with $Z_2$ and $C_4$ symmetry} \label{Appendix:FieldTheory}

The bulk theory for 3D HOSPT with $Z_2$ and $C_4$ symmetry can be reduced from a conventional 3D SPT phase with $\mathbb{Z}_2\times \mathbb{Z}_2$ symmetry described by the $O(4)$ NL$\sigma$M,
\begin{equation} 
\mathcal{L}=\frac{1}{g}(\partial_{\mu} \vec{n})^2+\frac{\Theta}{\Omega^4} \epsilon^{ijklm}  n_i\partial_x n_j \partial_t n_k\partial_y n_l\nonumber \partial_z n_m,
\label{omg2}
\end{equation}
with $\Theta=2\pi$, and $\Omega^4 = p \pi^2/3$. The on-site symmetries act as
\begin{equation}
\begin{split}
&\mathbb{Z}_2^a: (n_1,n_2,n_3,n_4) \rightarrow (-n_1,-n_2,-n_3,-n_4), \\
&\mathbb{Z}_2^b: (n_5,n_4) \rightarrow (-n_5,-n_4) . \\
\end{split}
\end{equation}

We will now gap the edge by polarizing $n_5$, which breaks the local
$\mathbb{Z}_2^b$ symmetry.  
However, we chose to do this in a way that the product of $C_4$ and $\mathbb{Z}_2^b$ symmetry is preserved (while $C_4$ itself must then also be broken).

To do this we begin with Eq.~\eqref{omg2} in cylinder coordinates $(r,\phi,z)$, and take  $n_5 \equiv \langle n_5 \rangle=\cos(2\phi)$, which preserves both $C_4$ rotations.  We define a new $O(4)$ vector boson field $\vec{N}$ normalized as $\sum_i N_i^2 = 1$, via
\begin{equation} 
\begin{split}
n_i=&N_i \sin(2\phi),\quad i=1,2,3,4\\
n_5=&\cos(2\phi)
\end{split}
\end{equation}
We further let $\Theta$ be spatially depended as
\begin{equation}
\Theta(r)=\left[1-\mathrm{sgn}(r-R)\right]\pi,
\end{equation}
where $R$ is the radius of the system on the x-y plane.

\begin{widetext} 

The resulting topological term has the form
 \begin{equation} 
 \begin{split}
L_{\Theta}=& 
\int dz~ dr \int_0^{2\pi} d \phi~\frac{\Theta(r)}{\Omega^4} \epsilon^{ijkl} \left [ 2\sin^3(2\phi) N_i \partial_r N_j \partial_t N_k \partial_z N_l 
+\cos(2\phi)\sin^4(2\phi) \partial_{\phi} N_i \partial_r N_j \partial_t N_k \partial_z N_m \right].
 \end{split}
\end{equation}
Because $n_5$ is ordered, the bulk topological term is trivial, and we can integrate over $r$.  To do this, first note that
up to boundary terms the second term in parentheses is a total derivative in $r$. 
The first term is not a total derivative, but can be made to be one by introducing an extra dimension $u$, and exploiting the fact that 
\begin{align}
\partial_u \left( \epsilon^{ijkl}  N_i \partial_r N_j \partial_t N_k  \partial_z N_l \right ) =  \epsilon^{ijkl} \partial_u N_i \partial_r N_j \partial_t N_k \partial_z N_l
\end{align}
to write this term as an integral over $u$.
After doing so we can integrate both terms by parts in $r$, to obtain
\begin{equation}
\begin{split}
\label{LtopFinal}
L_{\Theta}=&\int  dz~dr \int_0^{2\pi} d \phi~ \epsilon^{ijkl}~ \frac{\delta(r-R)2\pi}{\Omega^4} \left [\int_0^1 du~  2\sin^3(2\phi) N_i \partial_u N_j \partial_t N_k \partial_z N_l+\cos(2\phi)\sin^4(2\phi) N_i \partial_{\phi} N_j \partial_t N_k \partial_z N_l \right]\\
=&\int dz \int_0^{2\pi} d \phi~ \left [\int_0^1 du~\frac{2\pi}{\Omega^4} \epsilon^{ijkl}~ 2\sin^3(2\phi) N_i \partial_u N_j \partial_t N_k \partial_z N_l+\frac{2\pi \cos(2\phi)\sin^4(2\phi)}{\Omega^4} \epsilon^{ijk}~ N_i \partial_{\phi} N_j \partial_t N_k \partial_z N_l \right ],
\end{split}
\end{equation}
with the boundary conditions $\vec{N}(\phi,t,z,u=0)=(1,0,0,0),\vec{N}(\phi,t,z,u=1)=\vec{N}(\phi,t)$.
\end{widetext}

The second term is precisely the O(4) theta term in $(2+1)d$ that we encountered in Eq.~(\ref{BdyQFT}). 
However, its coefficient $\Theta = \cos(2\phi)\sin^4(2\phi)2\pi$ is not quantized. In the infrared limit of the renormalization group, $\Theta$ will flow to one of the discrete stable fixed points $\Theta=2\pi K,\ K\in \mathbb{Z}$~\cite{xu2013nonperturbative}, depending on its microscopic magnitude. In our case this magnitude is small, and we expect $\Theta$ to flow to 0 in the infrared, corresponding to two topologically trivial boundaries.  However choosing a slightly different ordering configuration for $n_5$ (for example, with an abrupt sign change, as in Fig.~\ref{c4}), we could equally arrive at the conclusion that $\Theta$ flows to $2 \pi$ along each boundary.  The important thing is that reflection symmetry ensures that the magnitude of this term is the same on both sides of the ``hinge" (which on the disk corresponds to the lines across which $n_5$ changes sign), so that the net contribution of both edges is an integer spin. With only $C_4$ rotational symmetry, near a particular hinge we have the freedom to choose the coefficient associated with (say) the $x$ edge to be 0 and that of the $y$ edge to be $2 \pi$ -- but rotational invariance then forces a domain wall where $\Theta$ changes from $0$ to $2 \pi$ somewhere along each surface distributed in a $C_4$ symmetric way.

The first term in Eq.~(\ref{LtopFinal}) resembles a $(1+1)d$ WZW term, delocalized along the hinge.  To make this more precise, consider integrating along one quarter of the integration domain, from $\phi = 0$ to $\phi = \pi/2$, such that the coefficient is non-vanishing at both ends of the range of integration.  
Ignoring the $\phi$-dependence of the O(4) rotor $\vec{N}$, we obtain
\begin{equation} 
\begin{split}
    \mathcal{L}_{\Theta}=&\int_0^{\pi/2} d \phi \int_0^1  du~\frac{2\pi}{\Omega^4} \epsilon^{ijkl}~ 2\sin^3(2\phi) N_i \partial_u N_j \partial_t N_k \partial_z N_l \\
=& \int_0^1  du~\frac{2\pi}{\Omega^3} \epsilon^{ijkl} N_i \partial_u N_j \partial_t N_k \partial_z N_l.
\end{split}
\end{equation}
where $\Omega^3 = 2 \pi^2$.  Thus each quarter of the system on the side surface contains an $O(4)_1$ WZW term in $(1+1)d$.


\begin{thebibliography}{100}%
\makeatletter
\providecommand \@ifxundefined [1]{%
 \@ifx{#1\undefined}
}%
\providecommand \@ifnum [1]{%
 \ifnum #1\expandafter \@firstoftwo
 \else \expandafter \@secondoftwo
 \fi
}%
\providecommand \@ifx [1]{%
 \ifx #1\expandafter \@firstoftwo
 \else \expandafter \@secondoftwo
 \fi
}%
\providecommand \natexlab [1]{#1}%
\providecommand \enquote  [1]{``#1''}%
\providecommand \bibnamefont  [1]{#1}%
\providecommand \bibfnamefont [1]{#1}%
\providecommand \citenamefont [1]{#1}%
\providecommand \href@noop [0]{\@secondoftwo}%
\providecommand \href [0]{\begingroup \@sanitize@url \@href}%
\providecommand \@href[1]{\@@startlink{#1}\@@href}%
\providecommand \@@href[1]{\endgroup#1\@@endlink}%
\providecommand \@sanitize@url [0]{\catcode `\\12\catcode `\$12\catcode
  `\&12\catcode `\#12\catcode `\^12\catcode `\_12\catcode `\%12\relax}%
\providecommand \@@startlink[1]{}%
\providecommand \@@endlink[0]{}%
\providecommand \url  [0]{\begingroup\@sanitize@url \@url }%
\providecommand \@url [1]{\endgroup\@href {#1}{\urlprefix }}%
\providecommand \urlprefix  [0]{URL }%
\providecommand \Eprint [0]{\href }%
\providecommand \doibase [0]{http://dx.doi.org/}%
\providecommand \selectlanguage [0]{\@gobble}%
\providecommand \bibinfo  [0]{\@secondoftwo}%
\providecommand \bibfield  [0]{\@secondoftwo}%
\providecommand \translation [1]{[#1]}%
\providecommand \BibitemOpen [0]{}%
\providecommand \bibitemStop [0]{}%
\providecommand \bibitemNoStop [0]{.\EOS\space}%
\providecommand \EOS [0]{\spacefactor3000\relax}%
\providecommand \BibitemShut  [1]{\csname bibitem#1\endcsname}%
\let\auto@bib@innerbib\@empty
\bibitem [{\citenamefont {Kane}\ and\ \citenamefont
  {Mele}(2005)}]{kane2005quantum}%
  \BibitemOpen
  \bibfield  {author} {\bibinfo {author} {\bibfnamefont {C.~L.}\ \bibnamefont
  {Kane}}\ and\ \bibinfo {author} {\bibfnamefont {E.~J.}\ \bibnamefont
  {Mele}},\ }\href@noop {} {\bibfield  {journal} {\bibinfo  {journal} {Physical
  review letters}\ }\textbf {\bibinfo {volume} {95}},\ \bibinfo {pages}
  {226801} (\bibinfo {year} {2005})}\BibitemShut {NoStop}%
\bibitem [{\citenamefont {Ryu}\ \emph {et~al.}(2010)\citenamefont {Ryu},
  \citenamefont {Schnyder}, \citenamefont {Furusaki},\ and\ \citenamefont
  {Ludwig}}]{ryu2010topological}%
  \BibitemOpen
  \bibfield  {author} {\bibinfo {author} {\bibfnamefont {S.}~\bibnamefont
  {Ryu}}, \bibinfo {author} {\bibfnamefont {A.~P.}\ \bibnamefont {Schnyder}},
  \bibinfo {author} {\bibfnamefont {A.}~\bibnamefont {Furusaki}}, \ and\
  \bibinfo {author} {\bibfnamefont {A.~W.}\ \bibnamefont {Ludwig}},\
  }\href@noop {} {\bibfield  {journal} {\bibinfo  {journal} {New Journal of
  Physics}\ }\textbf {\bibinfo {volume} {12}},\ \bibinfo {pages} {065010}
  (\bibinfo {year} {2010})}\BibitemShut {NoStop}%
\bibitem [{\citenamefont {Fu}\ and\ \citenamefont
  {Kane}(2008)}]{fu2008superconducting}%
  \BibitemOpen
  \bibfield  {author} {\bibinfo {author} {\bibfnamefont {L.}~\bibnamefont
  {Fu}}\ and\ \bibinfo {author} {\bibfnamefont {C.~L.}\ \bibnamefont {Kane}},\
  }\href@noop {} {\bibfield  {journal} {\bibinfo  {journal} {Physical review
  letters}\ }\textbf {\bibinfo {volume} {100}},\ \bibinfo {pages} {096407}
  (\bibinfo {year} {2008})}\BibitemShut {NoStop}%
\bibitem [{\citenamefont {Bernevig}\ \emph {et~al.}(2006)\citenamefont
  {Bernevig}, \citenamefont {Hughes},\ and\ \citenamefont
  {Zhang}}]{bernevig2006quantum}%
  \BibitemOpen
  \bibfield  {author} {\bibinfo {author} {\bibfnamefont {B.~A.}\ \bibnamefont
  {Bernevig}}, \bibinfo {author} {\bibfnamefont {T.~L.}\ \bibnamefont
  {Hughes}}, \ and\ \bibinfo {author} {\bibfnamefont {S.-C.}\ \bibnamefont
  {Zhang}},\ }\href@noop {} {\bibfield  {journal} {\bibinfo  {journal}
  {Science}\ }\textbf {\bibinfo {volume} {314}},\ \bibinfo {pages} {1757}
  (\bibinfo {year} {2006})}\BibitemShut {NoStop}%
\bibitem [{\citenamefont {Kitaev}(2009)}]{kitaev2009periodic}%
  \BibitemOpen
  \bibfield  {author} {\bibinfo {author} {\bibfnamefont {A.}~\bibnamefont
  {Kitaev}},\ }in\ \href@noop {} {\emph {\bibinfo {booktitle} {AIP Conference
  Proceedings}}},\ Vol.\ \bibinfo {volume} {1134}\ (\bibinfo {organization}
  {AIP},\ \bibinfo {year} {2009})\ pp.\ \bibinfo {pages} {22--30}\BibitemShut
  {NoStop}%
\bibitem [{\citenamefont {Roy}(2009)}]{roy2009topological}%
  \BibitemOpen
  \bibfield  {author} {\bibinfo {author} {\bibfnamefont {R.}~\bibnamefont
  {Roy}},\ }\href@noop {} {\bibfield  {journal} {\bibinfo  {journal} {Physical
  Review B}\ }\textbf {\bibinfo {volume} {79}},\ \bibinfo {pages} {195322}
  (\bibinfo {year} {2009})}\BibitemShut {NoStop}%
\bibitem [{\citenamefont {Schnyder}\ \emph {et~al.}(2009)\citenamefont
  {Schnyder}, \citenamefont {Ryu}, \citenamefont {Furusaki},\ and\
  \citenamefont {Ludwig}}]{schnyder2009classification}%
  \BibitemOpen
  \bibfield  {author} {\bibinfo {author} {\bibfnamefont {A.~P.}\ \bibnamefont
  {Schnyder}}, \bibinfo {author} {\bibfnamefont {S.}~\bibnamefont {Ryu}},
  \bibinfo {author} {\bibfnamefont {A.}~\bibnamefont {Furusaki}}, \ and\
  \bibinfo {author} {\bibfnamefont {A.~W.}\ \bibnamefont {Ludwig}},\ }in\
  \href@noop {} {\emph {\bibinfo {booktitle} {AIP Conference Proceedings}}},\
  Vol.\ \bibinfo {volume} {1134}\ (\bibinfo {organization} {AIP},\ \bibinfo
  {year} {2009})\ pp.\ \bibinfo {pages} {10--21}\BibitemShut {NoStop}%
\bibitem [{\citenamefont {Chen}\ \emph
  {et~al.}(2011{\natexlab{a}})\citenamefont {Chen}, \citenamefont {Gu},\ and\
  \citenamefont {Wen}}]{Chen2011-et}%
  \BibitemOpen
  \bibfield  {author} {\bibinfo {author} {\bibfnamefont {X.}~\bibnamefont
  {Chen}}, \bibinfo {author} {\bibfnamefont {Z.-C.}\ \bibnamefont {Gu}}, \ and\
  \bibinfo {author} {\bibfnamefont {X.-G.}\ \bibnamefont {Wen}},\ }\href@noop
  {} {\bibfield  {journal} {\bibinfo  {journal} {Phys. Rev. B Condens. Matter}\
  }\textbf {\bibinfo {volume} {83}},\ \bibinfo {pages} {035107} (\bibinfo
  {year} {2011}{\natexlab{a}})}\BibitemShut {NoStop}%
\bibitem [{\citenamefont {Chen}\ \emph {et~al.}(2012)\citenamefont {Chen},
  \citenamefont {Gu}, \citenamefont {Liu},\ and\ \citenamefont
  {Wen}}]{chen2012symmetry}%
  \BibitemOpen
  \bibfield  {author} {\bibinfo {author} {\bibfnamefont {X.}~\bibnamefont
  {Chen}}, \bibinfo {author} {\bibfnamefont {Z.-C.}\ \bibnamefont {Gu}},
  \bibinfo {author} {\bibfnamefont {Z.-X.}\ \bibnamefont {Liu}}, \ and\
  \bibinfo {author} {\bibfnamefont {X.-G.}\ \bibnamefont {Wen}},\ }\href@noop
  {} {\bibfield  {journal} {\bibinfo  {journal} {Science}\ }\textbf {\bibinfo
  {volume} {338}},\ \bibinfo {pages} {1604} (\bibinfo {year}
  {2012})}\BibitemShut {NoStop}%
\bibitem [{\citenamefont {Pollmann}\ \emph {et~al.}(2012)\citenamefont
  {Pollmann}, \citenamefont {Berg}, \citenamefont {Turner},\ and\ \citenamefont
  {Oshikawa}}]{pollmann2012symmetry}%
  \BibitemOpen
  \bibfield  {author} {\bibinfo {author} {\bibfnamefont {F.}~\bibnamefont
  {Pollmann}}, \bibinfo {author} {\bibfnamefont {E.}~\bibnamefont {Berg}},
  \bibinfo {author} {\bibfnamefont {A.~M.}\ \bibnamefont {Turner}}, \ and\
  \bibinfo {author} {\bibfnamefont {M.}~\bibnamefont {Oshikawa}},\ }\href@noop
  {} {\bibfield  {journal} {\bibinfo  {journal} {Physical review b}\ }\textbf
  {\bibinfo {volume} {85}},\ \bibinfo {pages} {075125} (\bibinfo {year}
  {2012})}\BibitemShut {NoStop}%
\bibitem [{\citenamefont {Lu}\ and\ \citenamefont
  {Vishwanath}(2012)}]{lu2012theory}%
  \BibitemOpen
  \bibfield  {author} {\bibinfo {author} {\bibfnamefont {Y.-M.}\ \bibnamefont
  {Lu}}\ and\ \bibinfo {author} {\bibfnamefont {A.}~\bibnamefont
  {Vishwanath}},\ }\href@noop {} {\bibfield  {journal} {\bibinfo  {journal}
  {Physical Review B}\ }\textbf {\bibinfo {volume} {86}},\ \bibinfo {pages}
  {125119} (\bibinfo {year} {2012})}\BibitemShut {NoStop}%
\bibitem [{\citenamefont {Xu}\ and\ \citenamefont
  {Senthil}(2013)}]{xu2013wave}%
  \BibitemOpen
  \bibfield  {author} {\bibinfo {author} {\bibfnamefont {C.}~\bibnamefont
  {Xu}}\ and\ \bibinfo {author} {\bibfnamefont {T.}~\bibnamefont {Senthil}},\
  }\href@noop {} {\bibfield  {journal} {\bibinfo  {journal} {Physical Review
  B}\ }\textbf {\bibinfo {volume} {87}},\ \bibinfo {pages} {174412} (\bibinfo
  {year} {2013})}\BibitemShut {NoStop}%
\bibitem [{\citenamefont {Bi}\ \emph {et~al.}(2015)\citenamefont {Bi},
  \citenamefont {Rasmussen}, \citenamefont {Slagle},\ and\ \citenamefont
  {Xu}}]{bi2015classification}%
  \BibitemOpen
  \bibfield  {author} {\bibinfo {author} {\bibfnamefont {Z.}~\bibnamefont
  {Bi}}, \bibinfo {author} {\bibfnamefont {A.}~\bibnamefont {Rasmussen}},
  \bibinfo {author} {\bibfnamefont {K.}~\bibnamefont {Slagle}}, \ and\ \bibinfo
  {author} {\bibfnamefont {C.}~\bibnamefont {Xu}},\ }\href@noop {} {\bibfield
  {journal} {\bibinfo  {journal} {Physical Review B}\ }\textbf {\bibinfo
  {volume} {91}},\ \bibinfo {pages} {134404} (\bibinfo {year}
  {2015})}\BibitemShut {NoStop}%
\bibitem [{\citenamefont {Vishwanath}\ and\ \citenamefont
  {Senthil}(2013)}]{vishwanath2013physics}%
  \BibitemOpen
  \bibfield  {author} {\bibinfo {author} {\bibfnamefont {A.}~\bibnamefont
  {Vishwanath}}\ and\ \bibinfo {author} {\bibfnamefont {T.}~\bibnamefont
  {Senthil}},\ }\href@noop {} {\bibfield  {journal} {\bibinfo  {journal}
  {Physical Review X}\ }\textbf {\bibinfo {volume} {3}},\ \bibinfo {pages}
  {011016} (\bibinfo {year} {2013})}\BibitemShut {NoStop}%
\bibitem [{\citenamefont {Wen}(2015)}]{wen2015construction}%
  \BibitemOpen
  \bibfield  {author} {\bibinfo {author} {\bibfnamefont {X.-G.}\ \bibnamefont
  {Wen}},\ }\href@noop {} {\bibfield  {journal} {\bibinfo  {journal} {Physical
  Review B}\ }\textbf {\bibinfo {volume} {91}},\ \bibinfo {pages} {205101}
  (\bibinfo {year} {2015})}\BibitemShut {NoStop}%
\bibitem [{\citenamefont {Cheng}\ \emph
  {et~al.}(2016{\natexlab{a}})\citenamefont {Cheng}, \citenamefont {Zaletel},
  \citenamefont {Barkeshli}, \citenamefont {Vishwanath},\ and\ \citenamefont
  {Bonderson}}]{PhysRevX.6.041068}%
  \BibitemOpen
  \bibfield  {author} {\bibinfo {author} {\bibfnamefont {M.}~\bibnamefont
  {Cheng}}, \bibinfo {author} {\bibfnamefont {M.}~\bibnamefont {Zaletel}},
  \bibinfo {author} {\bibfnamefont {M.}~\bibnamefont {Barkeshli}}, \bibinfo
  {author} {\bibfnamefont {A.}~\bibnamefont {Vishwanath}}, \ and\ \bibinfo
  {author} {\bibfnamefont {P.}~\bibnamefont {Bonderson}},\ }\href {\doibase
  10.1103/PhysRevX.6.041068} {\bibfield  {journal} {\bibinfo  {journal} {Phys.
  Rev. X}\ }\textbf {\bibinfo {volume} {6}},\ \bibinfo {pages} {041068}
  (\bibinfo {year} {2016}{\natexlab{a}})}\BibitemShut {NoStop}%
\bibitem [{\citenamefont {Kapustin}\ and\ \citenamefont
  {Thorngren}(2017)}]{kapustin2017fermionic}%
  \BibitemOpen
  \bibfield  {author} {\bibinfo {author} {\bibfnamefont {A.}~\bibnamefont
  {Kapustin}}\ and\ \bibinfo {author} {\bibfnamefont {R.}~\bibnamefont
  {Thorngren}},\ }\href@noop {} {\bibfield  {journal} {\bibinfo  {journal}
  {Journal of High Energy Physics}\ }\textbf {\bibinfo {volume} {2017}},\
  \bibinfo {pages} {80} (\bibinfo {year} {2017})}\BibitemShut {NoStop}%
\bibitem [{\citenamefont {{Metlitski}}\ \emph {et~al.}(2014)\citenamefont
  {{Metlitski}}, \citenamefont {{Fidkowski}}, \citenamefont {{Chen}},\ and\
  \citenamefont {{Vishwanath}}}]{2014arXiv1406.3032M}%
  \BibitemOpen
  \bibfield  {author} {\bibinfo {author} {\bibfnamefont {M.~A.}\ \bibnamefont
  {{Metlitski}}}, \bibinfo {author} {\bibfnamefont {L.}~\bibnamefont
  {{Fidkowski}}}, \bibinfo {author} {\bibfnamefont {X.}~\bibnamefont {{Chen}}},
  \ and\ \bibinfo {author} {\bibfnamefont {A.}~\bibnamefont {{Vishwanath}}},\
  }\href@noop {} {\bibfield  {journal} {\bibinfo  {journal} {ArXiv e-prints}\ }
  (\bibinfo {year} {2014})},\ \Eprint {http://arxiv.org/abs/1406.3032}
  {arXiv:1406.3032 [cond-mat.str-el]} \BibitemShut {NoStop}%
\bibitem [{\citenamefont {{Fidkowski}}\ \emph {et~al.}(2018)\citenamefont
  {{Fidkowski}}, \citenamefont {{Vishwanath}},\ and\ \citenamefont
  {{Metlitski}}}]{2018arXiv180408628F}%
  \BibitemOpen
  \bibfield  {author} {\bibinfo {author} {\bibfnamefont {L.}~\bibnamefont
  {{Fidkowski}}}, \bibinfo {author} {\bibfnamefont {A.}~\bibnamefont
  {{Vishwanath}}}, \ and\ \bibinfo {author} {\bibfnamefont {M.~A.}\
  \bibnamefont {{Metlitski}}},\ }\href@noop {} {\bibfield  {journal} {\bibinfo
  {journal} {ArXiv e-prints}\ } (\bibinfo {year} {2018})},\ \Eprint
  {http://arxiv.org/abs/1804.08628} {arXiv:1804.08628 [cond-mat.str-el]}
  \BibitemShut {NoStop}%
\bibitem [{\citenamefont {Fu}(2011)}]{fu2011topological}%
  \BibitemOpen
  \bibfield  {author} {\bibinfo {author} {\bibfnamefont {L.}~\bibnamefont
  {Fu}},\ }\href@noop {} {\bibfield  {journal} {\bibinfo  {journal} {Physical
  Review Letters}\ }\textbf {\bibinfo {volume} {106}},\ \bibinfo {pages}
  {106802} (\bibinfo {year} {2011})}\BibitemShut {NoStop}%
\bibitem [{\citenamefont {Hsieh}\ \emph {et~al.}(2012)\citenamefont {Hsieh},
  \citenamefont {Lin}, \citenamefont {Liu}, \citenamefont {Duan}, \citenamefont
  {Bansil},\ and\ \citenamefont {Fu}}]{hsieh2012topological}%
  \BibitemOpen
  \bibfield  {author} {\bibinfo {author} {\bibfnamefont {T.~H.}\ \bibnamefont
  {Hsieh}}, \bibinfo {author} {\bibfnamefont {H.}~\bibnamefont {Lin}}, \bibinfo
  {author} {\bibfnamefont {J.}~\bibnamefont {Liu}}, \bibinfo {author}
  {\bibfnamefont {W.}~\bibnamefont {Duan}}, \bibinfo {author} {\bibfnamefont
  {A.}~\bibnamefont {Bansil}}, \ and\ \bibinfo {author} {\bibfnamefont
  {L.}~\bibnamefont {Fu}},\ }\href@noop {} {\bibfield  {journal} {\bibinfo
  {journal} {Nature communications}\ }\textbf {\bibinfo {volume} {3}},\
  \bibinfo {pages} {982} (\bibinfo {year} {2012})}\BibitemShut {NoStop}%
\bibitem [{\citenamefont {Cheng}\ \emph
  {et~al.}(2016{\natexlab{b}})\citenamefont {Cheng}, \citenamefont {Zaletel},
  \citenamefont {Barkeshli}, \citenamefont {Vishwanath},\ and\ \citenamefont
  {Bonderson}}]{cheng2016translational}%
  \BibitemOpen
  \bibfield  {author} {\bibinfo {author} {\bibfnamefont {M.}~\bibnamefont
  {Cheng}}, \bibinfo {author} {\bibfnamefont {M.}~\bibnamefont {Zaletel}},
  \bibinfo {author} {\bibfnamefont {M.}~\bibnamefont {Barkeshli}}, \bibinfo
  {author} {\bibfnamefont {A.}~\bibnamefont {Vishwanath}}, \ and\ \bibinfo
  {author} {\bibfnamefont {P.}~\bibnamefont {Bonderson}},\ }\href@noop {}
  {\bibfield  {journal} {\bibinfo  {journal} {Physical Review X}\ }\textbf
  {\bibinfo {volume} {6}},\ \bibinfo {pages} {041068} (\bibinfo {year}
  {2016}{\natexlab{b}})}\BibitemShut {NoStop}%
\bibitem [{\citenamefont {Shiozaki}\ \emph {et~al.}(2017)\citenamefont
  {Shiozaki}, \citenamefont {Shapourian},\ and\ \citenamefont
  {Ryu}}]{shiozaki2017many}%
  \BibitemOpen
  \bibfield  {author} {\bibinfo {author} {\bibfnamefont {K.}~\bibnamefont
  {Shiozaki}}, \bibinfo {author} {\bibfnamefont {H.}~\bibnamefont
  {Shapourian}}, \ and\ \bibinfo {author} {\bibfnamefont {S.}~\bibnamefont
  {Ryu}},\ }\href@noop {} {\bibfield  {journal} {\bibinfo  {journal} {Physical
  Review B}\ }\textbf {\bibinfo {volume} {95}},\ \bibinfo {pages} {205139}
  (\bibinfo {year} {2017})}\BibitemShut {NoStop}%
\bibitem [{\citenamefont {Ando}\ and\ \citenamefont
  {Fu}(2015)}]{ando2015topological}%
  \BibitemOpen
  \bibfield  {author} {\bibinfo {author} {\bibfnamefont {Y.}~\bibnamefont
  {Ando}}\ and\ \bibinfo {author} {\bibfnamefont {L.}~\bibnamefont {Fu}},\
  }\href@noop {} {\bibfield  {journal} {\bibinfo  {journal} {Annu. Rev.
  Condens. Matter Phys.}\ }\textbf {\bibinfo {volume} {6}},\ \bibinfo {pages}
  {361} (\bibinfo {year} {2015})}\BibitemShut {NoStop}%
\bibitem [{\citenamefont {{Shapourian}}\ \emph {et~al.}(2018)\citenamefont
  {{Shapourian}}, \citenamefont {{Wang}},\ and\ \citenamefont
  {{Ryu}}}]{2018PhRvB..97i4508S}%
  \BibitemOpen
  \bibfield  {author} {\bibinfo {author} {\bibfnamefont {H.}~\bibnamefont
  {{Shapourian}}}, \bibinfo {author} {\bibfnamefont {Y.}~\bibnamefont
  {{Wang}}}, \ and\ \bibinfo {author} {\bibfnamefont {S.}~\bibnamefont
  {{Ryu}}},\ }\href {\doibase 10.1103/PhysRevB.97.094508} {\bibfield  {journal}
  {\bibinfo  {journal} {\prb}\ }\textbf {\bibinfo {volume} {97}},\ \bibinfo
  {eid} {094508} (\bibinfo {year} {2018})},\ \Eprint
  {http://arxiv.org/abs/1711.02122} {arXiv:1711.02122 [cond-mat.str-el]}
  \BibitemShut {NoStop}%
\bibitem [{\citenamefont {{Shiozaki}}\ \emph {et~al.}(2017)\citenamefont
  {{Shiozaki}}, \citenamefont {{Shapourian}},\ and\ \citenamefont
  {{Ryu}}}]{2017PhRvB..95t5139S}%
  \BibitemOpen
  \bibfield  {author} {\bibinfo {author} {\bibfnamefont {K.}~\bibnamefont
  {{Shiozaki}}}, \bibinfo {author} {\bibfnamefont {H.}~\bibnamefont
  {{Shapourian}}}, \ and\ \bibinfo {author} {\bibfnamefont {S.}~\bibnamefont
  {{Ryu}}},\ }\href {\doibase 10.1103/PhysRevB.95.205139} {\bibfield  {journal}
  {\bibinfo  {journal} {\prb}\ }\textbf {\bibinfo {volume} {95}},\ \bibinfo
  {eid} {205139} (\bibinfo {year} {2017})},\ \Eprint
  {http://arxiv.org/abs/1609.05970} {arXiv:1609.05970 [cond-mat.str-el]}
  \BibitemShut {NoStop}%
\bibitem [{\citenamefont {Hong}\ and\ \citenamefont
  {Fu}(2017)}]{hong2017topological}%
  \BibitemOpen
  \bibfield  {author} {\bibinfo {author} {\bibfnamefont {S.}~\bibnamefont
  {Hong}}\ and\ \bibinfo {author} {\bibfnamefont {L.}~\bibnamefont {Fu}},\
  }\href@noop {} {\bibfield  {journal} {\bibinfo  {journal} {arXiv preprint
  arXiv:1707.02594}\ } (\bibinfo {year} {2017})}\BibitemShut {NoStop}%
\bibitem [{\citenamefont {Qi}\ and\ \citenamefont
  {Fu}(2015)}]{qi2015anomalous}%
  \BibitemOpen
  \bibfield  {author} {\bibinfo {author} {\bibfnamefont {Y.}~\bibnamefont
  {Qi}}\ and\ \bibinfo {author} {\bibfnamefont {L.}~\bibnamefont {Fu}},\
  }\href@noop {} {\bibfield  {journal} {\bibinfo  {journal} {Physical review
  letters}\ }\textbf {\bibinfo {volume} {115}},\ \bibinfo {pages} {236801}
  (\bibinfo {year} {2015})}\BibitemShut {NoStop}%
\bibitem [{\citenamefont {Huang}\ \emph {et~al.}(2017)\citenamefont {Huang},
  \citenamefont {Song}, \citenamefont {Huang},\ and\ \citenamefont
  {Hermele}}]{huang2017building}%
  \BibitemOpen
  \bibfield  {author} {\bibinfo {author} {\bibfnamefont {S.-J.}\ \bibnamefont
  {Huang}}, \bibinfo {author} {\bibfnamefont {H.}~\bibnamefont {Song}},
  \bibinfo {author} {\bibfnamefont {Y.-P.}\ \bibnamefont {Huang}}, \ and\
  \bibinfo {author} {\bibfnamefont {M.}~\bibnamefont {Hermele}},\ }\href@noop
  {} {\bibfield  {journal} {\bibinfo  {journal} {Physical Review B}\ }\textbf
  {\bibinfo {volume} {96}},\ \bibinfo {pages} {205106} (\bibinfo {year}
  {2017})}\BibitemShut {NoStop}%
\bibitem [{\citenamefont {Teo}\ and\ \citenamefont
  {Hughes}(2013)}]{teo2013existence}%
  \BibitemOpen
  \bibfield  {author} {\bibinfo {author} {\bibfnamefont {J.~C.}\ \bibnamefont
  {Teo}}\ and\ \bibinfo {author} {\bibfnamefont {T.~L.}\ \bibnamefont
  {Hughes}},\ }\href@noop {} {\bibfield  {journal} {\bibinfo  {journal}
  {Physical review letters}\ }\textbf {\bibinfo {volume} {111}},\ \bibinfo
  {pages} {047006} (\bibinfo {year} {2013})}\BibitemShut {NoStop}%
\bibitem [{\citenamefont {Song}\ \emph
  {et~al.}(2017{\natexlab{a}})\citenamefont {Song}, \citenamefont {Huang},
  \citenamefont {Fu},\ and\ \citenamefont {Hermele}}]{song2017topological}%
  \BibitemOpen
  \bibfield  {author} {\bibinfo {author} {\bibfnamefont {H.}~\bibnamefont
  {Song}}, \bibinfo {author} {\bibfnamefont {S.-J.}\ \bibnamefont {Huang}},
  \bibinfo {author} {\bibfnamefont {L.}~\bibnamefont {Fu}}, \ and\ \bibinfo
  {author} {\bibfnamefont {M.}~\bibnamefont {Hermele}},\ }\href@noop {}
  {\bibfield  {journal} {\bibinfo  {journal} {Physical Review X}\ }\textbf
  {\bibinfo {volume} {7}},\ \bibinfo {pages} {011020} (\bibinfo {year}
  {2017}{\natexlab{a}})}\BibitemShut {NoStop}%
\bibitem [{\citenamefont {Cheng}(2018)}]{cheng2018microscopic}%
  \BibitemOpen
  \bibfield  {author} {\bibinfo {author} {\bibfnamefont {M.}~\bibnamefont
  {Cheng}},\ }\href@noop {} {\bibfield  {journal} {\bibinfo  {journal}
  {Physical Review Letters}\ }\textbf {\bibinfo {volume} {120}},\ \bibinfo
  {pages} {036801} (\bibinfo {year} {2018})}\BibitemShut {NoStop}%
\bibitem [{\citenamefont {Huang}\ and\ \citenamefont
  {Hermele}(2018)}]{huang2018surface}%
  \BibitemOpen
  \bibfield  {author} {\bibinfo {author} {\bibfnamefont {S.-J.}\ \bibnamefont
  {Huang}}\ and\ \bibinfo {author} {\bibfnamefont {M.}~\bibnamefont
  {Hermele}},\ }\href@noop {} {\bibfield  {journal} {\bibinfo  {journal}
  {Physical Review B}\ }\textbf {\bibinfo {volume} {97}},\ \bibinfo {pages}
  {075145} (\bibinfo {year} {2018})}\BibitemShut {NoStop}%
\bibitem [{\citenamefont {Watanabe}\ \emph {et~al.}(2017)\citenamefont
  {Watanabe}, \citenamefont {Po},\ and\ \citenamefont
  {Vishwanath}}]{watanabe2017structure}%
  \BibitemOpen
  \bibfield  {author} {\bibinfo {author} {\bibfnamefont {H.}~\bibnamefont
  {Watanabe}}, \bibinfo {author} {\bibfnamefont {H.~C.}\ \bibnamefont {Po}}, \
  and\ \bibinfo {author} {\bibfnamefont {A.}~\bibnamefont {Vishwanath}},\
  }\href@noop {} {\bibfield  {journal} {\bibinfo  {journal} {arXiv preprint
  arXiv:1707.01903}\ } (\bibinfo {year} {2017})}\BibitemShut {NoStop}%
\bibitem [{\citenamefont {Po}\ \emph {et~al.}(2017)\citenamefont {Po},
  \citenamefont {Vishwanath},\ and\ \citenamefont {Watanabe}}]{po2017symmetry}%
  \BibitemOpen
  \bibfield  {author} {\bibinfo {author} {\bibfnamefont {H.~C.}\ \bibnamefont
  {Po}}, \bibinfo {author} {\bibfnamefont {A.}~\bibnamefont {Vishwanath}}, \
  and\ \bibinfo {author} {\bibfnamefont {H.}~\bibnamefont {Watanabe}},\
  }\href@noop {} {\bibfield  {journal} {\bibinfo  {journal} {Nature
  Communications}\ }\textbf {\bibinfo {volume} {8}},\ \bibinfo {pages} {50}
  (\bibinfo {year} {2017})}\BibitemShut {NoStop}%
\bibitem [{\citenamefont {Schindler}\ \emph {et~al.}(2017)\citenamefont
  {Schindler}, \citenamefont {Cook}, \citenamefont {Vergniory}, \citenamefont
  {Wang}, \citenamefont {Parkin}, \citenamefont {Bernevig},\ and\ \citenamefont
  {Neupert}}]{schindler2017higher}%
  \BibitemOpen
  \bibfield  {author} {\bibinfo {author} {\bibfnamefont {F.}~\bibnamefont
  {Schindler}}, \bibinfo {author} {\bibfnamefont {A.~M.}\ \bibnamefont {Cook}},
  \bibinfo {author} {\bibfnamefont {M.~G.}\ \bibnamefont {Vergniory}}, \bibinfo
  {author} {\bibfnamefont {Z.}~\bibnamefont {Wang}}, \bibinfo {author}
  {\bibfnamefont {S.~S.}\ \bibnamefont {Parkin}}, \bibinfo {author}
  {\bibfnamefont {B.~A.}\ \bibnamefont {Bernevig}}, \ and\ \bibinfo {author}
  {\bibfnamefont {T.}~\bibnamefont {Neupert}},\ }\href@noop {} {\bibfield
  {journal} {\bibinfo  {journal} {arXiv preprint arXiv:1708.03636}\ } (\bibinfo
  {year} {2017})}\BibitemShut {NoStop}%
\bibitem [{\citenamefont {Kunst}\ \emph {et~al.}(2017)\citenamefont {Kunst},
  \citenamefont {van Miert},\ and\ \citenamefont
  {Bergholtz}}]{kunst2017lattice}%
  \BibitemOpen
  \bibfield  {author} {\bibinfo {author} {\bibfnamefont {F.~K.}\ \bibnamefont
  {Kunst}}, \bibinfo {author} {\bibfnamefont {G.}~\bibnamefont {van Miert}}, \
  and\ \bibinfo {author} {\bibfnamefont {E.~J.}\ \bibnamefont {Bergholtz}},\
  }\href@noop {} {\bibfield  {journal} {\bibinfo  {journal} {arXiv preprint
  arXiv:1712.07911}\ } (\bibinfo {year} {2017})}\BibitemShut {NoStop}%
\bibitem [{\citenamefont {Song}\ \emph
  {et~al.}(2017{\natexlab{b}})\citenamefont {Song}, \citenamefont {Fang},\ and\
  \citenamefont {Fang}}]{song2017d}%
  \BibitemOpen
  \bibfield  {author} {\bibinfo {author} {\bibfnamefont {Z.}~\bibnamefont
  {Song}}, \bibinfo {author} {\bibfnamefont {Z.}~\bibnamefont {Fang}}, \ and\
  \bibinfo {author} {\bibfnamefont {C.}~\bibnamefont {Fang}},\ }\href@noop {}
  {\bibfield  {journal} {\bibinfo  {journal} {Physical review letters}\
  }\textbf {\bibinfo {volume} {119}},\ \bibinfo {pages} {246402} (\bibinfo
  {year} {2017}{\natexlab{b}})}\BibitemShut {NoStop}%
\bibitem [{\citenamefont {Wang}\ \emph {et~al.}(2018)\citenamefont {Wang},
  \citenamefont {Lin},\ and\ \citenamefont {Hughes}}]{wang2018weak}%
  \BibitemOpen
  \bibfield  {author} {\bibinfo {author} {\bibfnamefont {Y.}~\bibnamefont
  {Wang}}, \bibinfo {author} {\bibfnamefont {M.}~\bibnamefont {Lin}}, \ and\
  \bibinfo {author} {\bibfnamefont {T.~L.}\ \bibnamefont {Hughes}},\
  }\href@noop {} {\bibfield  {journal} {\bibinfo  {journal} {arXiv preprint
  arXiv:1804.01531}\ } (\bibinfo {year} {2018})}\BibitemShut {NoStop}%
\bibitem [{\citenamefont {Ezawa}(2018)}]{ezawa2018minimal}%
  \BibitemOpen
  \bibfield  {author} {\bibinfo {author} {\bibfnamefont {M.}~\bibnamefont
  {Ezawa}},\ }\href@noop {} {\bibfield  {journal} {\bibinfo  {journal} {arXiv
  preprint arXiv:1801.00437}\ } (\bibinfo {year} {2018})}\BibitemShut {NoStop}%
\bibitem [{\citenamefont {Benalcazar}\ \emph
  {et~al.}(2017{\natexlab{a}})\citenamefont {Benalcazar}, \citenamefont
  {Bernevig},\ and\ \citenamefont {Hughes}}]{benalcazar2017electric}%
  \BibitemOpen
  \bibfield  {author} {\bibinfo {author} {\bibfnamefont {W.~A.}\ \bibnamefont
  {Benalcazar}}, \bibinfo {author} {\bibfnamefont {B.~A.}\ \bibnamefont
  {Bernevig}}, \ and\ \bibinfo {author} {\bibfnamefont {T.~L.}\ \bibnamefont
  {Hughes}},\ }\href@noop {} {\bibfield  {journal} {\bibinfo  {journal}
  {Physical Review B}\ }\textbf {\bibinfo {volume} {96}},\ \bibinfo {pages}
  {245115} (\bibinfo {year} {2017}{\natexlab{a}})}\BibitemShut {NoStop}%
\bibitem [{\citenamefont {Benalcazar}\ \emph
  {et~al.}(2017{\natexlab{b}})\citenamefont {Benalcazar}, \citenamefont
  {Bernevig},\ and\ \citenamefont {Hughes}}]{benalcazar2017quantized}%
  \BibitemOpen
  \bibfield  {author} {\bibinfo {author} {\bibfnamefont {W.~A.}\ \bibnamefont
  {Benalcazar}}, \bibinfo {author} {\bibfnamefont {B.~A.}\ \bibnamefont
  {Bernevig}}, \ and\ \bibinfo {author} {\bibfnamefont {T.~L.}\ \bibnamefont
  {Hughes}},\ }\href@noop {} {\bibfield  {journal} {\bibinfo  {journal}
  {Science}\ }\textbf {\bibinfo {volume} {357}},\ \bibinfo {pages} {61}
  (\bibinfo {year} {2017}{\natexlab{b}})}\BibitemShut {NoStop}%
\bibitem [{\citenamefont {Khalaf}(2018)}]{khalaf2018higher}%
  \BibitemOpen
  \bibfield  {author} {\bibinfo {author} {\bibfnamefont {E.}~\bibnamefont
  {Khalaf}},\ }\href@noop {} {\bibfield  {journal} {\bibinfo  {journal} {arXiv
  preprint arXiv:1801.10050}\ } (\bibinfo {year} {2018})}\BibitemShut {NoStop}%
\bibitem [{\citenamefont {Matsugatani}\ and\ \citenamefont
  {Watanabe}(2018)}]{matsugatani2018connecting}%
  \BibitemOpen
  \bibfield  {author} {\bibinfo {author} {\bibfnamefont {A.}~\bibnamefont
  {Matsugatani}}\ and\ \bibinfo {author} {\bibfnamefont {H.}~\bibnamefont
  {Watanabe}},\ }\href@noop {} {\bibfield  {journal} {\bibinfo  {journal}
  {arXiv preprint arXiv:1804.02794}\ } (\bibinfo {year} {2018})}\BibitemShut
  {NoStop}%
\bibitem [{\citenamefont {Lin}\ and\ \citenamefont
  {Hughes}(2017)}]{lin2017topological}%
  \BibitemOpen
  \bibfield  {author} {\bibinfo {author} {\bibfnamefont {M.}~\bibnamefont
  {Lin}}\ and\ \bibinfo {author} {\bibfnamefont {T.~L.}\ \bibnamefont
  {Hughes}},\ }\href@noop {} {\bibfield  {journal} {\bibinfo  {journal} {arXiv
  preprint arXiv:1708.08457}\ } (\bibinfo {year} {2017})}\BibitemShut {NoStop}%
\bibitem [{\citenamefont {Dwivedi}\ \emph {et~al.}(2018)\citenamefont
  {Dwivedi}, \citenamefont {Hickey}, \citenamefont {Eschmann},\ and\
  \citenamefont {Trebst}}]{dwivedi2018majorana}%
  \BibitemOpen
  \bibfield  {author} {\bibinfo {author} {\bibfnamefont {V.}~\bibnamefont
  {Dwivedi}}, \bibinfo {author} {\bibfnamefont {C.}~\bibnamefont {Hickey}},
  \bibinfo {author} {\bibfnamefont {T.}~\bibnamefont {Eschmann}}, \ and\
  \bibinfo {author} {\bibfnamefont {S.}~\bibnamefont {Trebst}},\ }\href@noop {}
  {\bibfield  {journal} {\bibinfo  {journal} {arXiv preprint arXiv:1803.08922}\
  } (\bibinfo {year} {2018})}\BibitemShut {NoStop}%
\bibitem [{\citenamefont {Langbehn}\ \emph {et~al.}(2017)\citenamefont
  {Langbehn}, \citenamefont {Peng}, \citenamefont {Trifunovic}, \citenamefont
  {von Oppen},\ and\ \citenamefont {Brouwer}}]{langbehn2017reflection}%
  \BibitemOpen
  \bibfield  {author} {\bibinfo {author} {\bibfnamefont {J.}~\bibnamefont
  {Langbehn}}, \bibinfo {author} {\bibfnamefont {Y.}~\bibnamefont {Peng}},
  \bibinfo {author} {\bibfnamefont {L.}~\bibnamefont {Trifunovic}}, \bibinfo
  {author} {\bibfnamefont {F.}~\bibnamefont {von Oppen}}, \ and\ \bibinfo
  {author} {\bibfnamefont {P.~W.}\ \bibnamefont {Brouwer}},\ }\href@noop {}
  {\bibfield  {journal} {\bibinfo  {journal} {Physical review letters}\
  }\textbf {\bibinfo {volume} {119}},\ \bibinfo {pages} {246401} (\bibinfo
  {year} {2017})}\BibitemShut {NoStop}%
\bibitem [{\citenamefont {{Queiroz}}\ and\ \citenamefont
  {{Stern}}(2018)}]{2018arXiv180704141Q}%
  \BibitemOpen
  \bibfield  {author} {\bibinfo {author} {\bibfnamefont {R.}~\bibnamefont
  {{Queiroz}}}\ and\ \bibinfo {author} {\bibfnamefont {A.}~\bibnamefont
  {{Stern}}},\ }\href@noop {} {\bibfield  {journal} {\bibinfo  {journal} {ArXiv
  e-prints}\ } (\bibinfo {year} {2018})},\ \Eprint
  {http://arxiv.org/abs/1807.04141} {arXiv:1807.04141 [cond-mat.mes-hall]}
  \BibitemShut {NoStop}%
\bibitem [{\citenamefont {{Ezawa}}(2018)}]{2018arXiv180603007E}%
  \BibitemOpen
  \bibfield  {author} {\bibinfo {author} {\bibfnamefont {M.}~\bibnamefont
  {{Ezawa}}},\ }\href@noop {} {\bibfield  {journal} {\bibinfo  {journal} {ArXiv
  e-prints}\ ,\ \bibinfo {eid} {arXiv:1806.03007}} (\bibinfo {year} {2018})},\
  \Eprint {http://arxiv.org/abs/1806.03007} {arXiv:1806.03007} \BibitemShut
  {NoStop}%
\bibitem [{\citenamefont {{Wang}}\ \emph {et~al.}(2018)\citenamefont {{Wang}},
  \citenamefont {{Wieder}}, \citenamefont {{Li}}, \citenamefont {{Yan}},\ and\
  \citenamefont {{Bernevig}}}]{2018arXiv180611116W}%
  \BibitemOpen
  \bibfield  {author} {\bibinfo {author} {\bibfnamefont {Z.}~\bibnamefont
  {{Wang}}}, \bibinfo {author} {\bibfnamefont {B.~J.}\ \bibnamefont
  {{Wieder}}}, \bibinfo {author} {\bibfnamefont {J.}~\bibnamefont {{Li}}},
  \bibinfo {author} {\bibfnamefont {B.}~\bibnamefont {{Yan}}}, \ and\ \bibinfo
  {author} {\bibfnamefont {B.~A.}\ \bibnamefont {{Bernevig}}},\ }\href@noop {}
  {\bibfield  {journal} {\bibinfo  {journal} {ArXiv e-prints}\ } (\bibinfo
  {year} {2018})},\ \Eprint {http://arxiv.org/abs/1806.11116} {arXiv:1806.11116
  [cond-mat.mtrl-sci]} \BibitemShut {NoStop}%
\bibitem [{\citenamefont {Parameswaran}\ and\ \citenamefont
  {Wan}(2017)}]{parameswaran2017topological}%
  \BibitemOpen
  \bibfield  {author} {\bibinfo {author} {\bibfnamefont {S.~A.}\ \bibnamefont
  {Parameswaran}}\ and\ \bibinfo {author} {\bibfnamefont {Y.}~\bibnamefont
  {Wan}},\ }\href@noop {} {\bibfield  {journal} {\bibinfo  {journal} {Physics}\
  }\textbf {\bibinfo {volume} {10}},\ \bibinfo {pages} {132} (\bibinfo {year}
  {2017})}\BibitemShut {NoStop}%
\bibitem [{\citenamefont {Yan}\ \emph {et~al.}(2018)\citenamefont {Yan},
  \citenamefont {Song},\ and\ \citenamefont {Wang}}]{yan2018majorana}%
  \BibitemOpen
  \bibfield  {author} {\bibinfo {author} {\bibfnamefont {Z.}~\bibnamefont
  {Yan}}, \bibinfo {author} {\bibfnamefont {F.}~\bibnamefont {Song}}, \ and\
  \bibinfo {author} {\bibfnamefont {Z.}~\bibnamefont {Wang}},\ }\href@noop {}
  {\bibfield  {journal} {\bibinfo  {journal} {arXiv preprint arXiv:1803.08545}\
  } (\bibinfo {year} {2018})}\BibitemShut {NoStop}%
\bibitem [{\citenamefont {Zhu}(2018)}]{zhu2018tunable}%
  \BibitemOpen
  \bibfield  {author} {\bibinfo {author} {\bibfnamefont {X.}~\bibnamefont
  {Zhu}},\ }\href@noop {} {\bibfield  {journal} {\bibinfo  {journal} {arXiv
  preprint arXiv:1802.00270}\ } (\bibinfo {year} {2018})}\BibitemShut {NoStop}%
\bibitem [{\citenamefont {Ezawa}(2018)}]{ezawa2018higher}%
  \BibitemOpen
  \bibfield  {author} {\bibinfo {author} {\bibfnamefont {M.}~\bibnamefont
  {Ezawa}},\ }\href@noop {} {\bibfield  {journal} {\bibinfo  {journal}
  {Physical Review Letters}\ }\textbf {\bibinfo {volume} {120}},\ \bibinfo
  {pages} {026801} (\bibinfo {year} {2018})}\BibitemShut {NoStop}%
\bibitem [{\citenamefont {Qi}\ \emph {et~al.}(2017)\citenamefont {Qi},
  \citenamefont {Jian},\ and\ \citenamefont {Wang}}]{qi2017folding}%
  \BibitemOpen
  \bibfield  {author} {\bibinfo {author} {\bibfnamefont {Y.}~\bibnamefont
  {Qi}}, \bibinfo {author} {\bibfnamefont {C.-M.}\ \bibnamefont {Jian}}, \ and\
  \bibinfo {author} {\bibfnamefont {C.}~\bibnamefont {Wang}},\ }\href@noop {}
  {\bibfield  {journal} {\bibinfo  {journal} {arXiv preprint arXiv:1710.09391}\
  } (\bibinfo {year} {2017})}\BibitemShut {NoStop}%
\bibitem [{\citenamefont {Peterson}\ \emph {et~al.}(2017)\citenamefont
  {Peterson}, \citenamefont {Benalcazar}, \citenamefont {Hughes},\ and\
  \citenamefont {Bahl}}]{peterson2017demonstration}%
  \BibitemOpen
  \bibfield  {author} {\bibinfo {author} {\bibfnamefont {C.~W.}\ \bibnamefont
  {Peterson}}, \bibinfo {author} {\bibfnamefont {W.~A.}\ \bibnamefont
  {Benalcazar}}, \bibinfo {author} {\bibfnamefont {T.~L.}\ \bibnamefont
  {Hughes}}, \ and\ \bibinfo {author} {\bibfnamefont {G.}~\bibnamefont
  {Bahl}},\ }\href@noop {} {\bibfield  {journal} {\bibinfo  {journal} {arXiv
  preprint arXiv:1710.03231}\ } (\bibinfo {year} {2017})}\BibitemShut {NoStop}%
\bibitem [{\citenamefont {Isobe}\ and\ \citenamefont
  {Fu}(2015)}]{isobe2015theory}%
  \BibitemOpen
  \bibfield  {author} {\bibinfo {author} {\bibfnamefont {H.}~\bibnamefont
  {Isobe}}\ and\ \bibinfo {author} {\bibfnamefont {L.}~\bibnamefont {Fu}},\
  }\href@noop {} {\bibfield  {journal} {\bibinfo  {journal} {Physical Review
  B}\ }\textbf {\bibinfo {volume} {92}},\ \bibinfo {pages} {081304} (\bibinfo
  {year} {2015})}\BibitemShut {NoStop}%
\bibitem [{\citenamefont {Song}\ and\ \citenamefont
  {Schnyder}(2017)}]{song2017interaction}%
  \BibitemOpen
  \bibfield  {author} {\bibinfo {author} {\bibfnamefont {X.-Y.}\ \bibnamefont
  {Song}}\ and\ \bibinfo {author} {\bibfnamefont {A.~P.}\ \bibnamefont
  {Schnyder}},\ }\href@noop {} {\bibfield  {journal} {\bibinfo  {journal}
  {Physical Review B}\ }\textbf {\bibinfo {volume} {95}},\ \bibinfo {pages}
  {195108} (\bibinfo {year} {2017})}\BibitemShut {NoStop}%
\bibitem [{\citenamefont {Benalcazar}\ \emph {et~al.}(2014)\citenamefont
  {Benalcazar}, \citenamefont {Teo},\ and\ \citenamefont
  {Hughes}}]{benalcazar2014classification}%
  \BibitemOpen
  \bibfield  {author} {\bibinfo {author} {\bibfnamefont {W.~A.}\ \bibnamefont
  {Benalcazar}}, \bibinfo {author} {\bibfnamefont {J.~C.}\ \bibnamefont {Teo}},
  \ and\ \bibinfo {author} {\bibfnamefont {T.~L.}\ \bibnamefont {Hughes}},\
  }\href@noop {} {\bibfield  {journal} {\bibinfo  {journal} {Physical Review
  B}\ }\textbf {\bibinfo {volume} {89}},\ \bibinfo {pages} {224503} (\bibinfo
  {year} {2014})}\BibitemShut {NoStop}%
\bibitem [{\citenamefont {Lapa}\ \emph {et~al.}(2016)\citenamefont {Lapa},
  \citenamefont {Teo},\ and\ \citenamefont {Hughes}}]{lapa2016interaction}%
  \BibitemOpen
  \bibfield  {author} {\bibinfo {author} {\bibfnamefont {M.~F.}\ \bibnamefont
  {Lapa}}, \bibinfo {author} {\bibfnamefont {J.~C.}\ \bibnamefont {Teo}}, \
  and\ \bibinfo {author} {\bibfnamefont {T.~L.}\ \bibnamefont {Hughes}},\
  }\href@noop {} {\bibfield  {journal} {\bibinfo  {journal} {Physical Review
  B}\ }\textbf {\bibinfo {volume} {93}},\ \bibinfo {pages} {115131} (\bibinfo
  {year} {2016})}\BibitemShut {NoStop}%
\bibitem [{\citenamefont {Gopalakrishnan}\ \emph {et~al.}(2013)\citenamefont
  {Gopalakrishnan}, \citenamefont {Teo},\ and\ \citenamefont
  {Hughes}}]{gopalakrishnan2013disclination}%
  \BibitemOpen
  \bibfield  {author} {\bibinfo {author} {\bibfnamefont {S.}~\bibnamefont
  {Gopalakrishnan}}, \bibinfo {author} {\bibfnamefont {J.~C.}\ \bibnamefont
  {Teo}}, \ and\ \bibinfo {author} {\bibfnamefont {T.~L.}\ \bibnamefont
  {Hughes}},\ }\href@noop {} {\bibfield  {journal} {\bibinfo  {journal}
  {Physical review letters}\ }\textbf {\bibinfo {volume} {111}},\ \bibinfo
  {pages} {025304} (\bibinfo {year} {2013})}\BibitemShut {NoStop}%
\bibitem [{\citenamefont {You}\ \emph {et~al.}(2016)\citenamefont {You},
  \citenamefont {Cho},\ and\ \citenamefont {Hughes}}]{you2016response}%
  \BibitemOpen
  \bibfield  {author} {\bibinfo {author} {\bibfnamefont {Y.}~\bibnamefont
  {You}}, \bibinfo {author} {\bibfnamefont {G.~Y.}\ \bibnamefont {Cho}}, \ and\
  \bibinfo {author} {\bibfnamefont {T.~L.}\ \bibnamefont {Hughes}},\
  }\href@noop {} {\bibfield  {journal} {\bibinfo  {journal} {Physical Review
  B}\ }\textbf {\bibinfo {volume} {94}},\ \bibinfo {pages} {085102} (\bibinfo
  {year} {2016})}\BibitemShut {NoStop}%
\bibitem [{\citenamefont {Teo}\ and\ \citenamefont
  {Kane}(2010)}]{teo2010topological}%
  \BibitemOpen
  \bibfield  {author} {\bibinfo {author} {\bibfnamefont {J.~C.}\ \bibnamefont
  {Teo}}\ and\ \bibinfo {author} {\bibfnamefont {C.~L.}\ \bibnamefont {Kane}},\
  }\href@noop {} {\bibfield  {journal} {\bibinfo  {journal} {Physical Review
  B}\ }\textbf {\bibinfo {volume} {82}},\ \bibinfo {pages} {115120} (\bibinfo
  {year} {2010})}\BibitemShut {NoStop}%
\bibitem [{\citenamefont {Thorngren}\ and\ \citenamefont
  {Else}(2018)}]{thorngren2018gauging}%
  \BibitemOpen
  \bibfield  {author} {\bibinfo {author} {\bibfnamefont {R.}~\bibnamefont
  {Thorngren}}\ and\ \bibinfo {author} {\bibfnamefont {D.~V.}\ \bibnamefont
  {Else}},\ }\href@noop {} {\bibfield  {journal} {\bibinfo  {journal} {Physical
  Review X}\ }\textbf {\bibinfo {volume} {8}},\ \bibinfo {pages} {011040}
  (\bibinfo {year} {2018})}\BibitemShut {NoStop}%
\bibitem [{\citenamefont {Wang}\ and\ \citenamefont
  {Levin}(2014)}]{wang2014braiding}%
  \BibitemOpen
  \bibfield  {author} {\bibinfo {author} {\bibfnamefont {C.}~\bibnamefont
  {Wang}}\ and\ \bibinfo {author} {\bibfnamefont {M.}~\bibnamefont {Levin}},\
  }\href@noop {} {\bibfield  {journal} {\bibinfo  {journal} {Physical review
  letters}\ }\textbf {\bibinfo {volume} {113}},\ \bibinfo {pages} {080403}
  (\bibinfo {year} {2014})}\BibitemShut {NoStop}%
\bibitem [{\citenamefont {Levin}\ and\ \citenamefont
  {Gu}(2012)}]{levin2012braiding}%
  \BibitemOpen
  \bibfield  {author} {\bibinfo {author} {\bibfnamefont {M.}~\bibnamefont
  {Levin}}\ and\ \bibinfo {author} {\bibfnamefont {Z.-C.}\ \bibnamefont {Gu}},\
  }\href@noop {} {\bibfield  {journal} {\bibinfo  {journal} {Physical Review
  B}\ }\textbf {\bibinfo {volume} {86}},\ \bibinfo {pages} {115109} (\bibinfo
  {year} {2012})}\BibitemShut {NoStop}%
\bibitem [{\citenamefont {Jian}\ and\ \citenamefont
  {Qi}(2014)}]{jian2014layer}%
  \BibitemOpen
  \bibfield  {author} {\bibinfo {author} {\bibfnamefont {C.-M.}\ \bibnamefont
  {Jian}}\ and\ \bibinfo {author} {\bibfnamefont {X.-L.}\ \bibnamefont {Qi}},\
  }\href@noop {} {\bibfield  {journal} {\bibinfo  {journal} {Physical Review
  X}\ }\textbf {\bibinfo {volume} {4}},\ \bibinfo {pages} {041043} (\bibinfo
  {year} {2014})}\BibitemShut {NoStop}%
\bibitem [{\citenamefont {Bi}\ \emph {et~al.}(2014)\citenamefont {Bi},
  \citenamefont {You},\ and\ \citenamefont {Xu}}]{bi2014anyon}%
  \BibitemOpen
  \bibfield  {author} {\bibinfo {author} {\bibfnamefont {Z.}~\bibnamefont
  {Bi}}, \bibinfo {author} {\bibfnamefont {Y.-Z.}\ \bibnamefont {You}}, \ and\
  \bibinfo {author} {\bibfnamefont {C.}~\bibnamefont {Xu}},\ }\href@noop {}
  {\bibfield  {journal} {\bibinfo  {journal} {Physical Review B}\ }\textbf
  {\bibinfo {volume} {90}},\ \bibinfo {pages} {081110} (\bibinfo {year}
  {2014})}\BibitemShut {NoStop}%
\bibitem [{\citenamefont {Jiang}\ \emph {et~al.}(2014)\citenamefont {Jiang},
  \citenamefont {Mesaros},\ and\ \citenamefont {Ran}}]{jiang2014generalized}%
  \BibitemOpen
  \bibfield  {author} {\bibinfo {author} {\bibfnamefont {S.}~\bibnamefont
  {Jiang}}, \bibinfo {author} {\bibfnamefont {A.}~\bibnamefont {Mesaros}}, \
  and\ \bibinfo {author} {\bibfnamefont {Y.}~\bibnamefont {Ran}},\ }\href@noop
  {} {\bibfield  {journal} {\bibinfo  {journal} {Physical Review X}\ }\textbf
  {\bibinfo {volume} {4}},\ \bibinfo {pages} {031048} (\bibinfo {year}
  {2014})}\BibitemShut {NoStop}%
\bibitem [{\citenamefont {Chen}\ \emph {et~al.}(2015)\citenamefont {Chen},
  \citenamefont {Burnell}, \citenamefont {Vishwanath},\ and\ \citenamefont
  {Fidkowski}}]{chen2015anomalous}%
  \BibitemOpen
  \bibfield  {author} {\bibinfo {author} {\bibfnamefont {X.}~\bibnamefont
  {Chen}}, \bibinfo {author} {\bibfnamefont {F.~J.}\ \bibnamefont {Burnell}},
  \bibinfo {author} {\bibfnamefont {A.}~\bibnamefont {Vishwanath}}, \ and\
  \bibinfo {author} {\bibfnamefont {L.}~\bibnamefont {Fidkowski}},\ }\href@noop
  {} {\bibfield  {journal} {\bibinfo  {journal} {Physical Review X}\ }\textbf
  {\bibinfo {volume} {5}},\ \bibinfo {pages} {041013} (\bibinfo {year}
  {2015})}\BibitemShut {NoStop}%
\bibitem [{\citenamefont {Abanov}\ and\ \citenamefont
  {Wiegmann}(2000)}]{abanov2000theta}%
  \BibitemOpen
  \bibfield  {author} {\bibinfo {author} {\bibfnamefont {A.}~\bibnamefont
  {Abanov}}\ and\ \bibinfo {author} {\bibfnamefont {P.~B.}\ \bibnamefont
  {Wiegmann}},\ }\href@noop {} {\bibfield  {journal} {\bibinfo  {journal}
  {Nuclear Physics B}\ }\textbf {\bibinfo {volume} {570}},\ \bibinfo {pages}
  {685} (\bibinfo {year} {2000})}\BibitemShut {NoStop}%
\bibitem [{\citenamefont {Senthil}(2015)}]{Senthil2015-tp}%
  \BibitemOpen
  \bibfield  {author} {\bibinfo {author} {\bibfnamefont {T.}~\bibnamefont
  {Senthil}},\ }\href@noop {} {\bibfield  {journal} {\bibinfo  {journal}
  {Annual Review of Condensed Matter Physics}\ }\textbf {\bibinfo {volume}
  {6}},\ \bibinfo {pages} {299} (\bibinfo {year} {2015})}\BibitemShut {NoStop}%
\bibitem [{\citenamefont {Chen}\ \emph {et~al.}(2014)\citenamefont {Chen},
  \citenamefont {Lu},\ and\ \citenamefont {Vishwanath}}]{chen2014symmetry}%
  \BibitemOpen
  \bibfield  {author} {\bibinfo {author} {\bibfnamefont {X.}~\bibnamefont
  {Chen}}, \bibinfo {author} {\bibfnamefont {Y.-M.}\ \bibnamefont {Lu}}, \ and\
  \bibinfo {author} {\bibfnamefont {A.}~\bibnamefont {Vishwanath}},\
  }\href@noop {} {\bibfield  {journal} {\bibinfo  {journal} {Nature
  communications}\ }\textbf {\bibinfo {volume} {5}},\ \bibinfo {pages} {3507}
  (\bibinfo {year} {2014})}\BibitemShut {NoStop}%
\bibitem [{\citenamefont {{You}}\ \emph {et~al.}(2018)\citenamefont {{You}},
  \citenamefont {{Devakul}}, \citenamefont {{Burnell}},\ and\ \citenamefont
  {{Sondhi}}}]{2018arXiv180302369Y}%
  \BibitemOpen
  \bibfield  {author} {\bibinfo {author} {\bibfnamefont {Y.}~\bibnamefont
  {{You}}}, \bibinfo {author} {\bibfnamefont {T.}~\bibnamefont {{Devakul}}},
  \bibinfo {author} {\bibfnamefont {F.~J.}\ \bibnamefont {{Burnell}}}, \ and\
  \bibinfo {author} {\bibfnamefont {S.~L.}\ \bibnamefont {{Sondhi}}},\
  }\href@noop {} {\bibfield  {journal} {\bibinfo  {journal} {ArXiv e-prints}\ }
  (\bibinfo {year} {2018})},\ \Eprint {http://arxiv.org/abs/1803.02369}
  {arXiv:1803.02369 [cond-mat.str-el]} \BibitemShut {NoStop}%
\bibitem [{\citenamefont {Chen}\ \emph
  {et~al.}(2011{\natexlab{b}})\citenamefont {Chen}, \citenamefont {Liu},\ and\
  \citenamefont {Wen}}]{chen2011two}%
  \BibitemOpen
  \bibfield  {author} {\bibinfo {author} {\bibfnamefont {X.}~\bibnamefont
  {Chen}}, \bibinfo {author} {\bibfnamefont {Z.-X.}\ \bibnamefont {Liu}}, \
  and\ \bibinfo {author} {\bibfnamefont {X.-G.}\ \bibnamefont {Wen}},\
  }\href@noop {} {\bibfield  {journal} {\bibinfo  {journal} {Physical Review
  B}\ }\textbf {\bibinfo {volume} {84}},\ \bibinfo {pages} {235141} (\bibinfo
  {year} {2011}{\natexlab{b}})}\BibitemShut {NoStop}%
\bibitem [{\citenamefont {Chen}\ \emph {et~al.}(2013)\citenamefont {Chen},
  \citenamefont {Wang}, \citenamefont {Lu},\ and\ \citenamefont
  {Lee}}]{chen2013critical}%
  \BibitemOpen
  \bibfield  {author} {\bibinfo {author} {\bibfnamefont {X.}~\bibnamefont
  {Chen}}, \bibinfo {author} {\bibfnamefont {F.}~\bibnamefont {Wang}}, \bibinfo
  {author} {\bibfnamefont {Y.-M.}\ \bibnamefont {Lu}}, \ and\ \bibinfo {author}
  {\bibfnamefont {D.-H.}\ \bibnamefont {Lee}},\ }\href@noop {} {\bibfield
  {journal} {\bibinfo  {journal} {Nuclear Physics B}\ }\textbf {\bibinfo
  {volume} {873}},\ \bibinfo {pages} {248} (\bibinfo {year}
  {2013})}\BibitemShut {NoStop}%
\bibitem [{\citenamefont {Xu}\ and\ \citenamefont
  {Ludwig}(2013)}]{xu2013nonperturbative}%
  \BibitemOpen
  \bibfield  {author} {\bibinfo {author} {\bibfnamefont {C.}~\bibnamefont
  {Xu}}\ and\ \bibinfo {author} {\bibfnamefont {A.~W.}\ \bibnamefont
  {Ludwig}},\ }\href@noop {} {\bibfield  {journal} {\bibinfo  {journal}
  {Physical review letters}\ }\textbf {\bibinfo {volume} {110}},\ \bibinfo
  {pages} {200405} (\bibinfo {year} {2013})}\BibitemShut {NoStop}%
\bibitem [{\citenamefont {You}\ and\ \citenamefont
  {Xu}(2014)}]{you2014symmetry}%
  \BibitemOpen
  \bibfield  {author} {\bibinfo {author} {\bibfnamefont {Y.-Z.}\ \bibnamefont
  {You}}\ and\ \bibinfo {author} {\bibfnamefont {C.}~\bibnamefont {Xu}},\
  }\href@noop {} {\bibfield  {journal} {\bibinfo  {journal} {Physical Review
  B}\ }\textbf {\bibinfo {volume} {90}},\ \bibinfo {pages} {245120} (\bibinfo
  {year} {2014})}\BibitemShut {NoStop}%
\bibitem [{Note1()}]{Note1}%
  \BibitemOpen
  \bibinfo {note} {Here we have ignored the $\phi $ dependence of the fields
  $N_i$. More generally, we could expand the topological term into modes of
  different angular momenta. For the $l=0$ mode the calculation described here
  applies. For modes with $l>0$ we should consider the integral over the entire
  boundary rather than over a single quadrant; doing so reveals that the net
  topological term for these modes is $0$.}\BibitemShut {Stop}%
\bibitem [{\citenamefont {You}(2016)}]{you2016decorated}%
  \BibitemOpen
  \bibfield  {author} {\bibinfo {author} {\bibfnamefont {Y.}~\bibnamefont
  {You}},\ }\href@noop {} {\bibfield  {journal} {\bibinfo  {journal} {Physical
  Review B}\ }\textbf {\bibinfo {volume} {94}},\ \bibinfo {pages} {195112}
  (\bibinfo {year} {2016})}\BibitemShut {NoStop}%
\bibitem [{\citenamefont {Peng}\ \emph
  {et~al.}(2017{\natexlab{a}})\citenamefont {Peng}, \citenamefont {Bao},\ and\
  \citenamefont {von Oppen}}]{PhysRevB.95.235143}%
  \BibitemOpen
  \bibfield  {author} {\bibinfo {author} {\bibfnamefont {Y.}~\bibnamefont
  {Peng}}, \bibinfo {author} {\bibfnamefont {Y.}~\bibnamefont {Bao}}, \ and\
  \bibinfo {author} {\bibfnamefont {F.}~\bibnamefont {von Oppen}},\ }\href
  {\doibase 10.1103/PhysRevB.95.235143} {\bibfield  {journal} {\bibinfo
  {journal} {Phys. Rev. B}\ }\textbf {\bibinfo {volume} {95}},\ \bibinfo
  {pages} {235143} (\bibinfo {year} {2017}{\natexlab{a}})}\BibitemShut
  {NoStop}%
\bibitem [{\citenamefont {Watanabe}\ and\ \citenamefont
  {Fu}(2017)}]{watanabe2017topological}%
  \BibitemOpen
  \bibfield  {author} {\bibinfo {author} {\bibfnamefont {H.}~\bibnamefont
  {Watanabe}}\ and\ \bibinfo {author} {\bibfnamefont {L.}~\bibnamefont {Fu}},\
  }\href@noop {} {\bibfield  {journal} {\bibinfo  {journal} {Physical Review
  B}\ }\textbf {\bibinfo {volume} {95}},\ \bibinfo {pages} {081107} (\bibinfo
  {year} {2017})}\BibitemShut {NoStop}%
\bibitem [{\citenamefont {Peng}\ \emph
  {et~al.}(2017{\natexlab{b}})\citenamefont {Peng}, \citenamefont {Bao},\ and\
  \citenamefont {von Oppen}}]{peng2017boundary}%
  \BibitemOpen
  \bibfield  {author} {\bibinfo {author} {\bibfnamefont {Y.}~\bibnamefont
  {Peng}}, \bibinfo {author} {\bibfnamefont {Y.}~\bibnamefont {Bao}}, \ and\
  \bibinfo {author} {\bibfnamefont {F.}~\bibnamefont {von Oppen}},\ }\href@noop
  {} {\bibfield  {journal} {\bibinfo  {journal} {Physical Review B}\ }\textbf
  {\bibinfo {volume} {95}},\ \bibinfo {pages} {235143} (\bibinfo {year}
  {2017}{\natexlab{b}})}\BibitemShut {NoStop}%
\bibitem [{\citenamefont {Schindler}\ \emph {et~al.}(2018)\citenamefont
  {Schindler}, \citenamefont {Cook}, \citenamefont {Vergniory}, \citenamefont
  {Wang}, \citenamefont {Parkin}, \citenamefont {Bernevig},\ and\ \citenamefont
  {Neupert}}]{schindler2018higher}%
  \BibitemOpen
  \bibfield  {author} {\bibinfo {author} {\bibfnamefont {F.}~\bibnamefont
  {Schindler}}, \bibinfo {author} {\bibfnamefont {A.~M.}\ \bibnamefont {Cook}},
  \bibinfo {author} {\bibfnamefont {M.~G.}\ \bibnamefont {Vergniory}}, \bibinfo
  {author} {\bibfnamefont {Z.}~\bibnamefont {Wang}}, \bibinfo {author}
  {\bibfnamefont {S.~S.}\ \bibnamefont {Parkin}}, \bibinfo {author}
  {\bibfnamefont {B.~A.}\ \bibnamefont {Bernevig}}, \ and\ \bibinfo {author}
  {\bibfnamefont {T.}~\bibnamefont {Neupert}},\ }\href@noop {} {\bibfield
  {journal} {\bibinfo  {journal} {Science advances}\ }\textbf {\bibinfo
  {volume} {4}},\ \bibinfo {pages} {eaat0346} (\bibinfo {year}
  {2018})}\BibitemShut {NoStop}%
\bibitem [{\citenamefont {Bradlyn}\ \emph {et~al.}(2018)\citenamefont
  {Bradlyn}, \citenamefont {Elcoro}, \citenamefont {Vergniory}, \citenamefont
  {Cano}, \citenamefont {Wang}, \citenamefont {Felser}, \citenamefont {Aroyo},\
  and\ \citenamefont {Bernevig}}]{PhysRevB.97.035138}%
  \BibitemOpen
  \bibfield  {author} {\bibinfo {author} {\bibfnamefont {B.}~\bibnamefont
  {Bradlyn}}, \bibinfo {author} {\bibfnamefont {L.}~\bibnamefont {Elcoro}},
  \bibinfo {author} {\bibfnamefont {M.~G.}\ \bibnamefont {Vergniory}}, \bibinfo
  {author} {\bibfnamefont {J.}~\bibnamefont {Cano}}, \bibinfo {author}
  {\bibfnamefont {Z.}~\bibnamefont {Wang}}, \bibinfo {author} {\bibfnamefont
  {C.}~\bibnamefont {Felser}}, \bibinfo {author} {\bibfnamefont {M.~I.}\
  \bibnamefont {Aroyo}}, \ and\ \bibinfo {author} {\bibfnamefont {B.~A.}\
  \bibnamefont {Bernevig}},\ }\href {\doibase 10.1103/PhysRevB.97.035138}
  {\bibfield  {journal} {\bibinfo  {journal} {Phys. Rev. B}\ }\textbf {\bibinfo
  {volume} {97}},\ \bibinfo {pages} {035138} (\bibinfo {year}
  {2018})}\BibitemShut {NoStop}%
\bibitem [{\citenamefont {{Trifunovic}}\ and\ \citenamefont
  {{Brouwer}}(2018)}]{2018arXiv180502598T}%
  \BibitemOpen
  \bibfield  {author} {\bibinfo {author} {\bibfnamefont {L.}~\bibnamefont
  {{Trifunovic}}}\ and\ \bibinfo {author} {\bibfnamefont {P.}~\bibnamefont
  {{Brouwer}}},\ }\href@noop {} {\bibfield  {journal} {\bibinfo  {journal}
  {ArXiv e-prints}\ } (\bibinfo {year} {2018})},\ \Eprint
  {http://arxiv.org/abs/1805.02598} {arXiv:1805.02598 [cond-mat.mes-hall]}
  \BibitemShut {NoStop}%
\bibitem [{\citenamefont {Kruthoff}\ \emph {et~al.}(2017)\citenamefont
  {Kruthoff}, \citenamefont {de~Boer}, \citenamefont {van Wezel}, \citenamefont
  {Kane},\ and\ \citenamefont {Slager}}]{PhysRevX.7.041069}%
  \BibitemOpen
  \bibfield  {author} {\bibinfo {author} {\bibfnamefont {J.}~\bibnamefont
  {Kruthoff}}, \bibinfo {author} {\bibfnamefont {J.}~\bibnamefont {de~Boer}},
  \bibinfo {author} {\bibfnamefont {J.}~\bibnamefont {van Wezel}}, \bibinfo
  {author} {\bibfnamefont {C.~L.}\ \bibnamefont {Kane}}, \ and\ \bibinfo
  {author} {\bibfnamefont {R.-J.}\ \bibnamefont {Slager}},\ }\href {\doibase
  10.1103/PhysRevX.7.041069} {\bibfield  {journal} {\bibinfo  {journal} {Phys.
  Rev. X}\ }\textbf {\bibinfo {volume} {7}},\ \bibinfo {pages} {041069}
  (\bibinfo {year} {2017})}\BibitemShut {NoStop}%
\bibitem [{\citenamefont {Slager}\ \emph {et~al.}(2013)\citenamefont {Slager},
  \citenamefont {Mesaros}, \citenamefont {Juri{\v{c}}i{\'c}},\ and\
  \citenamefont {Zaanen}}]{slager2013space}%
  \BibitemOpen
  \bibfield  {author} {\bibinfo {author} {\bibfnamefont {R.-J.}\ \bibnamefont
  {Slager}}, \bibinfo {author} {\bibfnamefont {A.}~\bibnamefont {Mesaros}},
  \bibinfo {author} {\bibfnamefont {V.}~\bibnamefont {Juri{\v{c}}i{\'c}}}, \
  and\ \bibinfo {author} {\bibfnamefont {J.}~\bibnamefont {Zaanen}},\
  }\href@noop {} {\bibfield  {journal} {\bibinfo  {journal} {Nature Physics}\
  }\textbf {\bibinfo {volume} {9}},\ \bibinfo {pages} {98} (\bibinfo {year}
  {2013})}\BibitemShut {NoStop}%
\bibitem [{\citenamefont {You}\ \emph {et~al.}(2015)\citenamefont {You},
  \citenamefont {Bi}, \citenamefont {Rasmussen}, \citenamefont {Cheng},\ and\
  \citenamefont {Xu}}]{you2015bridging}%
  \BibitemOpen
  \bibfield  {author} {\bibinfo {author} {\bibfnamefont {Y.-Z.}\ \bibnamefont
  {You}}, \bibinfo {author} {\bibfnamefont {Z.}~\bibnamefont {Bi}}, \bibinfo
  {author} {\bibfnamefont {A.}~\bibnamefont {Rasmussen}}, \bibinfo {author}
  {\bibfnamefont {M.}~\bibnamefont {Cheng}}, \ and\ \bibinfo {author}
  {\bibfnamefont {C.}~\bibnamefont {Xu}},\ }\href@noop {} {\bibfield  {journal}
  {\bibinfo  {journal} {New Journal of Physics}\ }\textbf {\bibinfo {volume}
  {17}},\ \bibinfo {pages} {075010} (\bibinfo {year} {2015})}\BibitemShut
  {NoStop}%
\bibitem [{\citenamefont {Morimoto}\ \emph {et~al.}(2015)\citenamefont
  {Morimoto}, \citenamefont {Furusaki},\ and\ \citenamefont
  {Mudry}}]{morimoto2015breakdown}%
  \BibitemOpen
  \bibfield  {author} {\bibinfo {author} {\bibfnamefont {T.}~\bibnamefont
  {Morimoto}}, \bibinfo {author} {\bibfnamefont {A.}~\bibnamefont {Furusaki}},
  \ and\ \bibinfo {author} {\bibfnamefont {C.}~\bibnamefont {Mudry}},\
  }\href@noop {} {\bibfield  {journal} {\bibinfo  {journal} {Physical Review
  B}\ }\textbf {\bibinfo {volume} {92}},\ \bibinfo {pages} {125104} (\bibinfo
  {year} {2015})}\BibitemShut {NoStop}%
\bibitem [{\citenamefont {Yoshida}\ \emph {et~al.}(2015)\citenamefont
  {Yoshida}, \citenamefont {Morimoto},\ and\ \citenamefont
  {Furusaki}}]{yoshida2015bosonic}%
  \BibitemOpen
  \bibfield  {author} {\bibinfo {author} {\bibfnamefont {T.}~\bibnamefont
  {Yoshida}}, \bibinfo {author} {\bibfnamefont {T.}~\bibnamefont {Morimoto}}, \
  and\ \bibinfo {author} {\bibfnamefont {A.}~\bibnamefont {Furusaki}},\
  }\href@noop {} {\bibfield  {journal} {\bibinfo  {journal} {Physical Review
  B}\ }\textbf {\bibinfo {volume} {92}},\ \bibinfo {pages} {245122} (\bibinfo
  {year} {2015})}\BibitemShut {NoStop}%
\bibitem [{\citenamefont {Wang}\ and\ \citenamefont
  {Senthil}(2014)}]{wang2014interacting}%
  \BibitemOpen
  \bibfield  {author} {\bibinfo {author} {\bibfnamefont {C.}~\bibnamefont
  {Wang}}\ and\ \bibinfo {author} {\bibfnamefont {T.}~\bibnamefont {Senthil}},\
  }\href@noop {} {\bibfield  {journal} {\bibinfo  {journal} {Physical Review
  B}\ }\textbf {\bibinfo {volume} {89}},\ \bibinfo {pages} {195124} (\bibinfo
  {year} {2014})}\BibitemShut {NoStop}%
\bibitem [{\citenamefont {Khalaf}\ \emph {et~al.}(2017)\citenamefont {Khalaf},
  \citenamefont {Po}, \citenamefont {Vishwanath},\ and\ \citenamefont
  {Watanabe}}]{khalaf2017symmetry}%
  \BibitemOpen
  \bibfield  {author} {\bibinfo {author} {\bibfnamefont {E.}~\bibnamefont
  {Khalaf}}, \bibinfo {author} {\bibfnamefont {H.~C.}\ \bibnamefont {Po}},
  \bibinfo {author} {\bibfnamefont {A.}~\bibnamefont {Vishwanath}}, \ and\
  \bibinfo {author} {\bibfnamefont {H.}~\bibnamefont {Watanabe}},\ }\href@noop
  {} {\bibfield  {journal} {\bibinfo  {journal} {arXiv preprint
  arXiv:1711.11589}\ } (\bibinfo {year} {2017})}\BibitemShut {NoStop}%
\bibitem [{\citenamefont {Fidkowski}\ and\ \citenamefont
  {Kitaev}(2010)}]{fidkowski2010effects}%
  \BibitemOpen
  \bibfield  {author} {\bibinfo {author} {\bibfnamefont {L.}~\bibnamefont
  {Fidkowski}}\ and\ \bibinfo {author} {\bibfnamefont {A.}~\bibnamefont
  {Kitaev}},\ }\href@noop {} {\bibfield  {journal} {\bibinfo  {journal}
  {Physical Review B}\ }\textbf {\bibinfo {volume} {81}},\ \bibinfo {pages}
  {134509} (\bibinfo {year} {2010})}\BibitemShut {NoStop}%
\bibitem [{\citenamefont {You}\ and\ \citenamefont
  {You}(2016{\natexlab{a}})}]{you2016geometry}%
  \BibitemOpen
  \bibfield  {author} {\bibinfo {author} {\bibfnamefont {Y.}~\bibnamefont
  {You}}\ and\ \bibinfo {author} {\bibfnamefont {Y.-Z.}\ \bibnamefont {You}},\
  }\href@noop {} {\bibfield  {journal} {\bibinfo  {journal} {Physical Review
  B}\ }\textbf {\bibinfo {volume} {93}},\ \bibinfo {pages} {245135} (\bibinfo
  {year} {2016}{\natexlab{a}})}\BibitemShut {NoStop}%
\bibitem [{\citenamefont {You}\ and\ \citenamefont
  {You}(2016{\natexlab{b}})}]{you2016stripe}%
  \BibitemOpen
  \bibfield  {author} {\bibinfo {author} {\bibfnamefont {Y.}~\bibnamefont
  {You}}\ and\ \bibinfo {author} {\bibfnamefont {Y.-Z.}\ \bibnamefont {You}},\
  }\href@noop {} {\bibfield  {journal} {\bibinfo  {journal} {Physical Review
  B}\ }\textbf {\bibinfo {volume} {93}},\ \bibinfo {pages} {195141} (\bibinfo
  {year} {2016}{\natexlab{b}})}\BibitemShut {NoStop}%
\bibitem [{\citenamefont {Cheng}\ and\ \citenamefont
  {Xu}(2016)}]{cheng2016series}%
  \BibitemOpen
  \bibfield  {author} {\bibinfo {author} {\bibfnamefont {M.}~\bibnamefont
  {Cheng}}\ and\ \bibinfo {author} {\bibfnamefont {C.}~\bibnamefont {Xu}},\
  }\href@noop {} {\bibfield  {journal} {\bibinfo  {journal} {Physical Review
  B}\ }\textbf {\bibinfo {volume} {94}},\ \bibinfo {pages} {214415} (\bibinfo
  {year} {2016})}\BibitemShut {NoStop}%
\bibitem [{\citenamefont {Agterberg}\ and\ \citenamefont
  {Tsunetsugu}(2008)}]{agterberg2008dislocations}%
  \BibitemOpen
  \bibfield  {author} {\bibinfo {author} {\bibfnamefont {D.}~\bibnamefont
  {Agterberg}}\ and\ \bibinfo {author} {\bibfnamefont {H.}~\bibnamefont
  {Tsunetsugu}},\ }\href@noop {} {\bibfield  {journal} {\bibinfo  {journal}
  {Nature Physics}\ }\textbf {\bibinfo {volume} {4}},\ \bibinfo {pages} {639}
  (\bibinfo {year} {2008})}\BibitemShut {NoStop}%
\bibitem [{\citenamefont {Cho}\ \emph {et~al.}(2015)\citenamefont {Cho},
  \citenamefont {Hsieh}, \citenamefont {Morimoto},\ and\ \citenamefont
  {Ryu}}]{cho2015topological}%
  \BibitemOpen
  \bibfield  {author} {\bibinfo {author} {\bibfnamefont {G.~Y.}\ \bibnamefont
  {Cho}}, \bibinfo {author} {\bibfnamefont {C.-T.}\ \bibnamefont {Hsieh}},
  \bibinfo {author} {\bibfnamefont {T.}~\bibnamefont {Morimoto}}, \ and\
  \bibinfo {author} {\bibfnamefont {S.}~\bibnamefont {Ryu}},\ }\href@noop {}
  {\bibfield  {journal} {\bibinfo  {journal} {Physical Review B}\ }\textbf
  {\bibinfo {volume} {91}},\ \bibinfo {pages} {195142} (\bibinfo {year}
  {2015})}\BibitemShut {NoStop}%
\bibitem [{\citenamefont {{Dubinkin}}\ and\ \citenamefont
  {{Hughes}}(2018)}]{2018arXiv180709781D}%
  \BibitemOpen
  \bibfield  {author} {\bibinfo {author} {\bibfnamefont {O.}~\bibnamefont
  {{Dubinkin}}}\ and\ \bibinfo {author} {\bibfnamefont {T.~L.}\ \bibnamefont
  {{Hughes}}},\ }\href@noop {} {\bibfield  {journal} {\bibinfo  {journal}
  {ArXiv e-prints}\ } (\bibinfo {year} {2018})},\ \Eprint
  {http://arxiv.org/abs/1807.09781} {arXiv:1807.09781 [cond-mat.str-el]}
  \BibitemShut {NoStop}%
\end{thebibliography}
\end{document}